\documentclass[usenatbib,useAMS]{mnras}

\usepackage[T1]{fontenc}
\usepackage{ae,aecompl}
\usepackage[dvipdfmx]{graphicx} 
\usepackage{amsmath}    
\usepackage{amsfonts}
\usepackage{multicol}        
\usepackage{bm}         
\usepackage{pdflscape}  
\usepackage{epstopdf}
\usepackage{comment}
\usepackage{xcolor}

\newcommand{\yp}[1]{{\color{black}{#1}}}

\newcommand{\mtrv}[1]{{\color{black}{#1}}}

\newcommand{\beq}{\begin{equation}}
\newcommand{\beqa}{\begin{eqnarray}}
\newcommand{\eeq}{\end{equation}}
\newcommand{\eeqa}{\end{eqnarray}}

\newcommand{\pmemj}{p_{{\rm mem},j}}

\newcommand{\lo}{\lambda_\mathrm{obs}}
\newcommand{\lt}{\lambda_\mathrm{true}}

\newcommand{\ds}{\Delta\!\Sigma(R)}
\newcommand{\wpr}{w_\mathrm{p}\!(R)}

\newcommand{\mpch}{h^{-1}{\rm Mpc}}

\newcommand{\simgt}{\lower.5ex\hbox{$\; \buildrel > \over \sim \;$}}
\newcommand{\simlt}{\lower.5ex\hbox{$\; \buildrel < \over \sim \;$}}

\def\avrg#1{\left\langle #1 \right\rangle}

\title[Projection effects on cluster observables]{The impact of projection effects on cluster observables: stacked lensing and projected clustering}

\author[T.~Sunayama et al.]{Tomomi Sunayama\thanks{E-mail: tomomi.sunayama@ipmu.jp}$^{1}$, 
Youngsoo~Park\thanks{E-mail: youngsoo.park@ipmu.jp}$^{1}$, 
Masahiro~Takada\thanks{E-mail: masahiro.takada@ipmu.jp}$^{1}$, 
Yosuke~Kobayashi$^{2,1}$, 
\newauthor
Takahiro~Nishimichi$^{3,1}$,
Toshiki~Kurita$^{2,1}$, 
Surhud~More$^{4,1}$,
Masamune~Oguri$^{5,2,1}$, and
\newauthor
Ken~Osato$^{6}$
\\
$^{1}$Kavli Institute for the Physics and Mathematics of the Universe (WPI), The University of Tokyo Institutes for Advanced Study (UTIAS),\\
The University of Tokyo, 5-1-5 Kashiwanoha, Kashiwa-shi, Chiba, 277-8583, Japan\\
$^{2}$Physics Department, The University of Tokyo, Bunkyo, Tokyo 113-0031, Japan\\
$^{3}$Center for Gravitational Physics, Yukawa Institute for Theoretical Physics, Kyoto University, Kyoto 606-8502, Japan\\
$^{4}$The Inter-University Center for Astronomy and Astrophysics, Post bag 4, Ganeshkhind,
Pune, 411007, India\\
$^{5}$Research Center for the Early Universe, The University of Tokyo, Tokyo 113-0033, Japan\\
$^{6}$Institut d'Astrophysique de Paris, Sorbonne Universit\'e, CNRS, UMR 7095, 75014 Paris, France
}

\begin{document}
\label{firstpage}
\pagerange{\pageref{firstpage}--\pageref{lastpage}}
\maketitle

\begin{abstract}
An optical cluster finder inevitably suffers from {\it projection effects}, where it misidentifies a superposition of galaxies in multiple halos along the line-of-sight as a single cluster. Using mock cluster catalogs built from cosmological $N$-body simulations, we quantify the impact of these projection effects with a particular focus on the observables of interest for cluster cosmology, namely the cluster lensing and the cluster clustering signals. We find that ``observed'' clusters, i.e. clusters identified by our cluster finder algorithm, exhibit lensing and clustering signals that deviate from expectations based on a statistically isotropic halo model -- while both signals agree with halo model expectations on small scales, they show unexpected boosts on large scales, by up to a factor of 1.2 or 1.4 respectively.
We identify the origin of these boosts as the inherent selection bias of optical cluster finders for clusters embedded within filaments aligned with the line-of-sight, and show that a minority ($\sim 30\%$) of such clusters within the entire sample is responsible for this observed boost. We discuss the implications of our results on previous studies of optical cluster, as well as prospects for identifying and mitigating projection effects in future cluster cosmology analyses.
\end{abstract}




\section{Introduction} 
\label{sec:intro}

Galaxy clusters are the most massive gravitationally self-bound objects in the Universe. These clusters form at the rare high peaks of the primordial density fluctuations, and they subsequently trace the growth of structure in the Universe as they grow in mass and abundance. As such, clusters constitute a natural cosmological probe for constraining the properties of the primordial fluctuations  \citep{1986ApJ...304...15B,2008ApJ...687...12D,2006ApJ...652...71W}
as well as cosmological parameters including the nature of dark energy \citep{Whieetal:93,Haiman:2001,Vikhlininetal:09,Rozoetal:10,TakadaBridle:07,OguriTakada:2011} \citep[also see][for a review]{Weinberg:2013}. 
Many current and future surveys, such as the Hyper Suprime-Cam (HSC) survey \citep{HSCOverview:17}, the Dark Energy Survey\footnote{\url{ https://www.darkenergysurvey.org}} (DES) \citep{DES2005}, the Kilo Degree Survey\footnote{\url{http://kids.strw.leidenuniv.nl/}} (KiDS) \citep{KiDs2015}, the Rubin Observatory Legacy Survey of Space and Time\footnote{\url{https://www.lsst.org}} (LSST) \citep{LSST2009}, \textit{Euclid}\footnote{\url{ https://sci.esa.int/web/euclid}} \citep{euclid2018}, and the Nancy Grace Roman Telescope\footnote{\url{https://wfirst.gsfc.nasa.gov}} \citep{WFIRST2019}, are aiming to capitalize on this prospect by utilizing clusters as one of their key cosmological probes. We expect these surveys to find galaxy clusters in unprecedented numbers and carry out cluster-based cosmology analyses with great statistical precision if systematic effects are well under control.

Photometric surveys such as HSC, KiDS and DES are expected to lead the immediate next generation of cluster cosmology with their catalogs of optically identified galaxy clusters. Generally, these surveys employ cluster finding algorithms \citep{2000AJ....120.2148G,2007ApJ...655..128G,2010ApJS..191..254H,Rykoff_etal2014,Rozo2014,Rozo2015,Rozo2015_2,Ogurietal:17} that look for small overdense regions of red galaxies with similar observed colors, using multi-wavelength imaging data. Red galaxies are known to have smaller scatters in their estimated photometric redshifts due to their characteristic 4000\AA~break in flux that moves through the different optical wavelength bands with varying redshifts. In the absence of spectroscopic redshifts, this photometric feature helps to distinguish galaxies at different redshifts occupying the same position on the sky. 

Identified clusters are often characterized by their richnesses, i.e. \yp{the number counts of red galaxies weighted by their membership probabilities,} which are identified as cluster members, commonly denoted as $\lambda$. Analyses typically subsample the identified clusters into bins of richness and redshift, from which they measure the observables of interest such as the number of clusters in each bin (cluster abundance), the stacked shapes of background galaxies around clusters (cluster lensing), and the two-point correlation function of clusters (cluster clustering). The key challenge in these analyses is the accurate calibration of the richness-mass relation. While theoretical predictions for the measured observables depend on the masses of the clusters in a given sample, these masses cannot be measured directly. Instead, a relationship between a directly observed property (richness) and cluster mass must be inferred. Joint analyses of multiple observables, especially combining cluster abundances with cluster lensing \citep{dePutterWhite2005,Johnston:2007,Okabe:2010,Hoekstraetal:15,vanUitertetal.16,Simet:2016,Melchior:2016,2018ApJ...854..120M,Murataetal:19,Allen_etal2011}, provides a promising path for estimating cluster masses and subsequently calibrating the richness-mass relation, as they allow for simultaneous constraints on both cosmological and richness-mass relation parameters, i.e. a self-calibration of the richness-mass relation \citep{Lima:2005,OguriTakada:2011}.

Even such joint analysis schemes, however, can yield inaccurate calibrations and ultimately biased cosmological constraints 
if systematic effects are not under control \citep{White_etal2010,NohCohn2012,Henson_etal2017}.
Despite the better-than-average photometric redshift uncertainties of red galaxies, the line-of-sight resolution of photometric surveys for cluster members is still significantly poorer relative to the angular resolution. Consequently, the effective window of a cluster region identified by an optical cluster finder extends much broader along the line-of-sight than across. This incurs a characteristic systematic effect, commonly referred to as \textit{projection effects,} where interloper galaxies along the line-of-sight to a cluster are mistaken as genuine members of the cluster. Some degree of projection effect is inevitable for any optical cluster finder \citep[see][and references therein]{2005MNRAS.359..272C,Cohn:2007,costanzietalprojection,2018arXiv181009456C}. Even for the redMaPPer cluster catalog \citep{Rykoff_etal2014}, 
constructed from the Sloan Digital Sky Survey (SDSS) DR8 data \citep{Aihara:2011} and considered as the most homogeneous and well-calibrated catalog of optical clusters, there are several hints of projection effects affecting the sample. The most well-known consequence of projection effects is the misidentification of cluster richnesses, where the inclusion of interlopers as cluster members causes a general up-scatter in the estimated richnesses \citep[e.g.][]{costanzietalprojection}. 

More interestingly, there are hints suggesting that the impact of projection effects extend beyond richness misidentification. First, when \citet{Miyatake:2016} reported a possible detection of assembly bias for subsamples of clusters divided based on the concentration of member galaxies, the signal appeared too large compared to theoretical predictions \citep{2006ApJ...652...71W,2008ApJ...687...12D} for the $\Lambda$CDM model. Subsequent works found that this large apparent signal might be due to projection effects \citep{Zu:2016,BuschWhite:17,SunayamaMore}. 
Second, when \citet{More:2016} reported a detection of the so-called ``splashback'' radius, a physically motivated boundary of cluster halos, for redMaPPer clusters, the location of the splashback radius was found to be smaller than theoretical expectations \citep[also see][for a similar detection]{2018ApJ...864...83C}. Follow-up studies \citep{2019ApJ...874..184Z,2019MNRAS.487.2900S} used samples of clusters selected based on the Sunyaev-Zel'dovich effect and found a different location of the splashback radius consistent with theoretical expectations, albeit with larger errors, suggesting that the original location may have been impacted by projection effects. The recent analysis in \citet{Murataetal:20} further indicated that previous analyses of the splashback radius for optically selected clusters may suffer from projection effects. Third, \citet{2018ApJ...854..120M} developed a forward modeling approach to calibrate the richness-mass relation from a joint measurement of cluster abundances and cluster lensing. However, they found that a population of less massive halos, down to $10^{12}h^{-1}M_\odot$, had to be introduced in order for the $\Lambda$CDM model prediction to match the observations. Overall, these studies indicate that projection effects may impact not only cluster richnesses but also other cluster observables such as cluster lensing, which, if true, can render problematic the standard approach for cluster cosmology employed by photometric surveys.

Hence the purpose of this paper is to perform a quantitative study on the impact of projection effects with a particular focus on cluster observables beyond richness.
To this end, we use a large set of cosmological $N$-body simulations in \citet{Nishimichi2018}; (i) we populate mock red galaxies into individual halos in $N$-body simulations, (ii) run a cluster finder algorithm with line-of-sight projection in each mock catalog to identify a sample of ``observed'' clusters, and (iii) measure cluster observables -- abundance, lensing, and clustering -- across multiple richness bins. Furthermore, to obtain physical insights on the results, we compare the cluster observables measured from the mock cluster catalogs with predictions obtained from the same simulations assuming statistical isotropy \citep[more exactly we will use the Dark Emulator developed in][]{Nishimichi2018}. This comparison illuminates the characteristic features arising from the line-of-sight projection inherent in cluste finders.

This paper is structured as follows. In Section~\ref{sec:methods}, we discuss the conventional approach for modeling cluster observables based on the halo model and the richness-mass relation. We then describe the details of our $N$-body simulations as well as our methods for generating mock cluster catalogs, measurements of cluster observables, and theoretical predictions. In Sections~\ref{sec:signal} and \ref{sec:signal2}, we respectively discuss how projection effects impact the cluster observables, and then explore the physical insights on the identified impacts from various tests. Section~\ref{sec:summary} is devoted to discussions and conclusions. Throughout this paper we employ the units convention $c=1$ for the speed of light.

\section{Methods}
\label{sec:methods}

\subsection{Cluster Observables}
\label{subsec:cluster_observables}

Cluster-based cosmology analyses rely on prescriptions for identifying and characterizing clusters in an observed catalog of galaxies, commonly referred to as cluster finders. While details would vary between different approaches \citep[e.g.][]{Rykoff_etal2014,Rozo2014,Oguri:2014,Ogurietal:17}, a core concept is shared by all cluster finders: there exists a \textit{fiducial definition} of a cluster, and the properties of a selected sample of clusters \textit{depend on this definition.} This suggests that any cluster finder will naturally come with its own set of distinct selection biases, and consequently that the characteristic selection of clusters induced by a cluster finder must be properly folded into the modeling of the cosmological observables. 

Let us begin by discussing the conventional approach for modeling this selection. Suppose that each cluster in a survey volume is identified and selected according to some selection observable. We will consider the observed optical richness $\lo$, a weighted count of red member galaxies, as a working example for the selection observable, but note that the following discussion can be applied to any other selection observable such as the X-ray brightness or the Sunyaev-Zeldovich (SZ) signal. To model the connection between the selection observable and various cosmological probes, we begin by defining a few key quantities:
\begin{itemize}
\item $P(\lo|M)$: the probability distribution of the selection observable, i.e. the observed richness ($\lo$), for halos of mass $M$. $P(\lo|M)$ encapsulates the connection from our selection observable to the underlying physical quantity of interest. 
\item $P(\lt|M)$: the probability distribution of ``true'' richness ($\lt$) for halos of mass $M$. By ``true'' we mean that $\lt$ is unaffected by biases from e.g. measurement errors or systematic effects. 
\item $P(\lo|\lt)$: the probability distribution of the observed richness ($\lo$) for clusters of true richness $\lt$. This relationship can be biased and scattered by both measurement errors as well as selection biases from physical effects.
\end{itemize}
Throughout this section, we will use these quantities to represent the selection of clusters based on observed richnesses.

We begin by applying the above formalism to the modeling of cluster abundances.
The expected (ensemble-averaged) number density of clusters in a given richness bin $i$ is
\begin{align}
\bar{n}_{{\rm c},i}&=\int_{\lambda_{{\rm obs},i}}\mathrm{d}\lo \int\!\!\mathrm{d}M~\frac{\mathrm{d}n_{\rm h}}{\mathrm{d}M}
P(\lo|M)\nonumber\\
&=\int_{\lambda_{{\rm obs},i}}\mathrm{d}\lo
\!
\int\!\mathrm{d}\lt~
P(\lo|\lt)\int\!\!\mathrm{d}M~\frac{\mathrm{d}n_{\rm h}}{\mathrm{d}M}
P(\lt|M), 
\label{eq:ni}
\end{align}
where $\mathrm{d}n_{\rm h}/\mathrm{d}M$ is the halo mass function in the mass range $[M,M+\mathrm{d}M]$ and the integral $\int_{\lambda_{{\rm obs},i}}\mathrm{d}\lo$ is taken over the range of observed richnesses for the $i$-th bin. The cluster number count for this bin is then
\begin{align}
N_i&=V_{\rm S}\bar{n}_{{\rm c},i}, 
\label{eq:Ni}
\end{align}
where $V_{\rm S}$ is the survey volume. Note that $V_{\rm S}$ depends on an assumed cosmological model, as it must be estimated from the solid angle and the redshift interval 
of a survey using the angular diameter and radial distances under the assumed cosmology:
\beq
V_{\rm S}\equiv \Omega_{\rm S}\int_{z_{\rm l}}^{z_{\rm u}}\!\frac{\mathrm{d}z}{H(z)}~ \chi(z)^2~.
\label{eq:VS}
\eeq
Here, $\chi(z)$ is the angular comoving distance to redshift $z$. In contrast to Eq.~(\ref{eq:VS}) above, we did not take any redshift dependence into account in Eq.~(\ref{eq:ni}) and did not include effects from survey masks, e.g. due to bright stars, for simplicity.
Finally, note that the following identity holds:
\begin{align}
P(\lo|M)&=
\int\!\!\mathrm{d}\lt~
P(\lo|\lt)P(\lt|M). 
\end{align}
In order to make the formulation in Eq.~(\ref{eq:Ni}) valid, the probability distributions need to satisfy the normalization condition: $\int_0^\infty\!\mathrm{d}\lo~P(\lo|M)=\int_{0}^\infty\!\mathrm{d}\lt~P(\lt|M)=\int_{0}^\infty\!\mathrm{d}\lo~ P(\lo|\lt)=1$. This means that every cluster or halo needs to be counted once and only once in computing relevant statistics.

Next, let us consider cluster lensing. By measuring the statistically averaged shapes of background galaxies around clusters as a function of cluster-centric radius, we can probe the average matter distribution around clusters. 
The physical quantity underlying cluster lensing, i.e. the projected (``surface'') mass density profile around clusters, is given by
\begin{align}
\Sigma_{i}(R)=\bar{\rho}_{\rm m0}\int_{-\infty}^{\infty}\!\!\mathrm{d}\pi~
\left[1+\xi_{{\rm cm},i}\!\left(\sqrt{R^2+\pi^2}\right)\right],
\label{eq:Sigma}
\end{align}
where $\bar{\rho}_{\rm m0}$ is the mean matter density today, $\pi$ and $R$ are the separations parallel and perpendicular to the line-of-sight direction from the cluster center, respectively,
and 
$\xi_{{\rm cm},i}(r)$ is the cluster-matter correlation function for the $i$-th richness bin, in turn given by
\begin{align}
\xi_{{\rm cm},i}(r)&=\frac{1}{\bar{n}_{{\rm c},i}}
\int_{\lambda_{{\rm obs},i}}\!\mathrm{d}\lo\int\!\mathrm{d}\lt~P(\lo|\lt)\nonumber\\
&\hspace{2em}\times 
\int\!\!\mathrm{d}M~\frac{\mathrm{d}n_{\rm h}}{\mathrm{d}M} P(\lt|M)
 \xi_{\rm hm}(r; M),
 \label{eq:xi_cm}
\end{align}
with $\xi_{\rm hm}(r; M)$ being the halo-matter correlation function for halos of mass $M$.
The above equation together with Eq.~(\ref{eq:Ni})
shows that the effects of the selection function, $P(\lo|\lt)$ or $P(\lt|M)$, cancel out to some extent between the numerator and the denominator in $\xi_{\rm cm}$, implying that $\Sigma(R)$ may be less sensitive to selection effects than number counts
\citep[also see][for a similar discussion]{2018ApJ...854..120M}. 
The excess surface mass density profile $\Delta\!\Sigma$, which is 
a direct observable of cluster lensing,
is given in terms of the surface mass density profile $\Sigma$ as
\begin{align}
\Delta\! \Sigma_i(R)&\equiv \avrg{\Sigma_i}\!(<R)-\Sigma_{i}(R)\nonumber\\
&= \frac{2}{R^2} \int_0^R\! \mathrm{d}R' R' \Sigma_i(R') - \Sigma_i(R)\nonumber \\
&= \bar{\rho}_{\rm m0}\int\!\!\frac{k\mathrm{d}k}{2\pi}~P_{{\rm cm},i}(k)J_2(kR), 
\label{eq:DeltaSigma}
\end{align}
where $\avrg{\Sigma_i}\!(<R)$ denotes the average surface mass density within a circle of radius $R$, 
$P_{{\rm cm},i}(k)$ is the cluster-matter power spectrum for the $i$-th richness bin, i.e. the Fourier transform of Eq.~(\ref{eq:xi_cm}),
and $J_2(x)$ is the second-order Bessel function. 

Cluster clustering can be modeled similar to the above formalism for cluster lensing. The direct cluster clustering observable is the projected cluster two-point function $w_{\rm{p}} (R)$, which represents the excess probability of finding a second cluster around a given cluster as a function of the projected separation between the two clusters in the two-dimensional plane perpendicular to the line-of-sight:
\begin{align}
w_{{\rm p},ij}(R)&=2 \int_{0}^{\pi_{\rm max}}\!\!\mathrm{d}\pi
~ 
\xi_{{\rm cc},ij}\!\!\left(\sqrt{R^2+\pi^2}\right).
\label{eq:wp_def}
\end{align}
Here $\pi_{\rm max}$ is the projection length along the line-of-sight direction in the $w_{{\rm cc},ij}$ computation, and $\xi_{{\rm cc},ij}(r)$ is the three-dimensional cluster auto-correlation function between richness bins $i$ and $j$ defined as
\begin{align}
\xi_{{\rm cc},ij}(r)&=\frac{1}{\bar{n}_{{\rm c},i}\bar{n}_{{\rm c},j}}
\nonumber\\
&\times \int_{\lambda_{{\rm obs},i}}\!\!\!\mathrm{d}\lo\!\!\int\!\!\mathrm{d}\lt P(\lo|\lt)\!\!
\int\!\!\mathrm{d}M\!\frac{\mathrm{d}n_{\rm h}}{\mathrm{d}M} P(\lt|M)\nonumber \\
&\times
\int_{\lambda^\prime_{{\rm obs},j}}\!\!\!\mathrm{d}\lambda'_{\rm obs}\!\!\int\!\!\mathrm{d}{\lt^\prime} P(\lambda^\prime_{\rm obs}|\lt^\prime)\!\!
\int\!\!\mathrm{d}M'\!\frac{\mathrm{d}n_{\rm h}}{\mathrm{d}M'} P(\lt^\prime|M')
\nonumber\\
&\hspace{1em}\times
\xi_{{\rm hh}}(r;M,M'),
\end{align}
where $\xi_{{\rm hh}}(r; M,M')$ is the three-dimensional halo auto-correlation function between two halos of masses $M$ and $M'$.  Similar to the cluster lensing case, the above equation shows that selection effects may cancel out to some extent for cluster clustering.

Note that this modeling approach contains a number of assumptions. First, it assumes that observed clusters constitute a fair sample of halos for a given halo mass. Halos of different masses will have different probabilities of being included in a cluster sample, as specified by $P(\lo|M)$ relation and the  range of $\lo$
used for selection, but halos with equal masses will have equal probabilities of being included regardless of their additional characteristics. Put differently, it assumes that the cluster finding algorithm does not introduce a selection bias among halos of equal masses. \yp{Second, it assumes that observed clusters have statistically isotropic clustering properties determined solely by their masses; more specifically, it assumes that projected statistics such as $\Delta\Sigma$ or $w_\mathrm{p}$ do not depend on the choice of projection direction, as shown in Eqs.~(\ref{eq:Sigma}) and~(\ref{eq:wp_def}), and that there are no assembly bias effects inducing relationships between clustering properties and cluster characteristics other than mass.} A major focus of this paper is to study whether these assumptions are valid in modeling optical clusters.

\subsection{$N$-body Simulations and Halo Catalogs}
\label{sec:signal:sim}

Now that we have defined the cluster observables of interest, we need to build mock cluster catalogs and measure these observables. We begin that process with the $N$-body simulations and halo catalogs from \citet{Nishimichi2018}. Briefly, all $N$-body simulations were performed with $2048^3$ particles in a comoving cubic box with side lengths of either 1 or 2~$h^{-1}{\rm Gpc}$, assuming the best-fit flat $\Lambda$CDM model\footnote{$\{\omega_{\rm b},\omega_{\rm c},\Omega_{\rm \Lambda},\ln(10^{10}A_{\rm s}),n_{\rm s}\} = \{0.02225,0.1198,0.6844,3.094,0.9645\}$} from \textit{Planck} Data Release 2 \citep{Planck:2015}. The initial displacement vector and the initial velocity of each $N$-body particle was set by second-order Lagrangian perturbation theory \citep{scoccimarro98,crocce06a,crocce06b,nishimichi09} with an input linear matter power spectrum computed from the publicly available Boltzmann code {\tt CAMB} \citep{camb}, and the subsequent time evolution of the particle distribution was simulated using the parallel Tree-Particle Mesh code {\tt Gadget2} \citep{Springel:2005}. We use a single species of particles to represent the total matter distribution, i.e. we do not consider the distinct evolution of massive neutrinos in these simulations, but we do set $\omega_\nu=0.00064$ for the physical density of massive neutrinos to set up the initial conditions. 
The {\it Planck} model has, as derived parameters, $\Omega_{\rm m}=0.3156$
(the present-day matter density parameter), $\sigma_8=0.831$ (the present-day RMS linear mass density fluctuations within a top-hat sphere of radius $8~h^{-1}{\rm Mpc}$) and $h=0.672$ for the Hubble parameter. The particle mass is $1.02$ or $8.16\times 10^{10}~h^{-1}M_\odot$ for the 1 or 2~$h^{-1}{\rm Gpc}$ box simulations, respectively. 

To generate halo catalogs, we first take simulation snapshots at redshift $z=0.251$ -- chosen to be close to the mean redshift of SDSS redMaPPer clusters -- and identify halos using the Friends-of-Friends (FoF) halo finder \texttt{Rockstar} developed in \citet{Behroozi:2013} \citep[also see][for details]{Nishimichi2018}. We use the ``200m'' halo definition, defining halo masses as $M\equiv M_{\rm 200m}=(4\pi/3)(r_{\rm 200m})^3(200\bar{\rho}_{\rm m0})$ where $r_{\rm 200m}$ is the spherical halo boundary radius within which the mean mass density is $200\times \bar{\rho}_{\rm m0}$. Note that the use of the present-day mean mass density $\bar{\rho}_{\rm m0}$ is due to our use of comoving coordinates, meaning that $r_{\rm 200m}$ is also in comoving length units. We employ the default settings of the {\tt Rockstar} algorithm and define the center of each halo as the center-of-mass of a subset of member particles in the inner part of that halo, which is considered as a proxy for the gravitational potential minimum.
Our definition of halo mass includes all particles within the radius $r_{\rm 200m}$ from the halo center, i.e. includes particles even if they are not gravitationally bound to the halo.
After identifying halos, we classify them into central and satellite halos; when the separation between the centers of two halos is closer than $r_{\rm 200m}$ of the more massive halo, we mark the less massive halo as a satellite. We only keep central halos with masses above $10^{12}\,h^{-1}M_\odot$ in the final halo catalog we use in this paper. The ``minimum halo'' at
$M=10^{12}~h^{-1}M_\odot$ consists of 12 or 100 $N$-body particles for the 2 or 1~$h^{-1}$Gpc box, respectively.

\subsection{Mock Catalogs of Red-Sequence Galaxies}
\label{subsec:mocks}

Most optical cluster finders operate by identifying a concentration of ``red-sequence'' galaxies, i.e. passively evolving early-type galaxies with no active star formation, in a small spatial region. These red galaxies serve as good tracers of clusters due to their preferential formation in overdense regions. In addition, due to their passive evolution in color, they allow for an accurate redshift estimation.
For instance, the redMaPPer cluster finder uses a catalog of relatively bright red galaxies with $L\simgt 0.2L_*$, as implemented for SDSS as well as DES data \citep{Rykoff_etal2014,Rykoff:2016} \citep[also see][for a similar method applied to the Subaru HSC data/KiDS data]{Ogurietal:17,Vakili2019}.

Thus, in order to study the behavior of such cluster finders from simulations, we must construct a mock catalog of red galaxies from the $N$-body simulations discussed in Section~\ref{sec:signal:sim}. For this we use the halo occupation distribution (HOD) formulation \citep{1998ApJ...494....1J,seljak:2000uq,peacock:2000qy,2005ApJ...633..791Z} to populate mock red galaxies in halos. Our HOD model gives the expected numbers of central and satellite galaxies, $N_{\rm cen}(M)$ and $N_{\rm sat}(M)$, as functions of halo mass $M$:
\begin{align}
\left < N_{\rm cen} \right > (M)=
\frac{1}{2}\left[
1+{\rm erf}\left(\frac{\log M-\log M_{\rm cut}}{\sigma_{\log M}}\right)
\right]
\end{align}
and
\begin{align}
\left < N_{\rm sat} \right >(M)=N_{\rm cen}(M)\left(\frac{M-\kappa M_{\rm cut}}{M_1}\right)^\alpha ,
\label{eq:HOD_sat}
\end{align}
where $M_{\rm cut}, M_1, \sigma_{\log M}, \kappa,$ and $\alpha$ are model parameters. Following \citet{2018arXiv181009456C}, we assume fiducial parameter values of $M_{\rm cut}=10^{11.7}~h^{-1}M_\odot$, $M_1=10^{12.9}~h^{-1}M_\odot$, $\sigma_{\log M}=0.1$, $\kappa=1.0$, and $\alpha=1.0$. This parameter configuration implies $N_{\rm cen}(M)= 1 \mbox{ for } M\ge 10^{12}h^{-1}M_\odot$, i.e. all identified halos in our halo catalogs receive a central galaxy.

With the HOD prescription in hand, we populate galaxies into halos as follows \citep[also see][]{Kobayashi:2019jrn}:
\begin{itemize}
\item[(i)] {\it Central galaxies} -- We populate a central galaxy at the center of each halo with $M\ge 10^{12}~h^{-1}M_\odot$. We do not consider any off-centering between central galaxies and halo centers in this work for simplicity.
\item[(ii)] {\it Satellite galaxies} --  For each halo with $M\ge 10^{12}~h^{-1}M_\odot$, we first determine the number of satellite galaxies $N_{\rm sat}$ from a Poisson random draw with mean given by Eq.~(\ref{eq:HOD_sat}). Once $N_{\rm sat}$ is determined, we populate each satellite galaxy according to a Navarro-Frenk-White \citep[][hereafter NFW]{nfw97} profile specified by the halo mass and the \citet{2015ApJ...799..108D} mass-concentration relation. We limit the extent of the NFW profile to within the $r_{\rm 200m}$ boundary.
\end{itemize}
The resulting galaxy number density of our mock galaxy catalogs is about $7.4\times 10^{-3}~(h^{-1}{\rm Mpc})^{-3}$ on average, which is roughly consistent with the number density of red galaxies used in the SDSS redMaPPer catalog. 

We perform this procedure on 20 independent realizations of the $1~{h^{-1}{\rm Gpc}}$ box simulations, as well as on 14 realizations of the 2~$h^{-1}{\rm Gpc}$ box simulations, for a total of 34 independent halo catalogs. We will use the former ($1~{h^{-1}{\rm Gpc}}$) catalogs to study cluster lensing and the latter ($2~{h^{-1}{\rm Gpc}}$) catalogs to achieve sufficiently accurate statistical errors for cluster clustering measurements.

Finally, in addition to these baseline catalogs, we prepare a second set of catalogs where we populate satellite galaxies in each halo using the distribution of FoF member particles provided in the {\tt Rockstar} output for each halo instead of the NFW profile. This allows for satellite galaxies to follow the aspherical shapes of their host halos, and comparing our baseline results to those from the ``shape-dependent'' catalogs allows us to study the impact of halo shapes, or halo triaxiality, on our main results (see Appendix~\ref{app:shape}).

\subsection{Cluster Finder and Mock Cluster Catalogs}
\label{sec:cluster_finder}

Our cluster finder algorithm is a modified version of the algorithm used in \citet{SunayamaMore}, which in turn is based on the redMaPPer red-sequence cluster finder \citep{Rykoff_etal2014,Rozo2014, Rozo2015, Rozo2015_2}. The key difference between the cluster finder in this work and \citet{SunayamaMore} is that we differentiate between central and satellite galaxies in creating the mock galaxy catalog and consider only the central galaxies as potential cluster centers. This also implies that we eliminate (on purpose) potential miscentering effects from our cluster finding process. Note that while the redMaPPer algorithm uses multi-wavelength imaging data to identify overdensities of red galaxies as clusters, our mock catalog of galaxies does not have color information. Thus, instead of a color filter, we employ a spatial top-hat filter extending $d_{\rm cyl}^{\rm max} = 60~h^{-1}{\rm Mpc}$ from each cluster center on both directions along the line-of-sight
to identify member galaxies and estimate optical richnesses. This choice of projection distance corresponds to a redshift width of $\pm \Delta z=0.023$ at $z=0.24$, respectively corresponding to the typical photometric redshift uncertainty and the typical redshift of SDSS redMaPPer clusters. Note that we also tried different projection lengths (more specifically $d_{\rm proj}$ of $30h^{-1}{\rm Mpc}$ and $120h^{-1}{\rm Mpc}$) of our top-hat filter and found only insignificant differences among the resulting cluster catalogs.

To initialize the cluster finder, we first consider all central galaxies in the mock galaxy catalog as potential cluster centers and rank order the center candidates by their host halo masses. We then descend the ranked list, assigning all galaxies (both central and satellite) within a cylinder of radius $0.5\mpch$ and length $|\pi| <
d_{\rm cyl}^{\rm max}$ around a center candidate as members of that candidate.
In this first step, the membership probability for all galaxies is set to
unity if it is within the above defined cylinder.
At the end, each potential cluster is assigned a richness $\lambda$ equal to its total number of member galaxies, and we eliminate all clusters with $\lambda<3$ from the list of center candidates.  The purpose of this first step is to find a broad set of overdense regions that are potential clusters. 

Once the cluster finder is initialized, we perform a set of iterative percolation steps. We remake the list of cluster candidates, this time rank ordering by the preliminary richnesses from the initialization step. Note that we simplify the rank-ordering by using $\lambda$ alone while the actual redMaPPer algorithm implemented in \citet{Rykoff_etal2014} takes into account an absolute luminosity filter for candidate member galaxiesas well as the preliminary richnesses. In addition we set $p_\mathrm{free}=1$ for all galaxies at the beginning of each iteration, where $p_\mathrm{free}$ is the probability that a given galaxy does not belong to other clusters.

Each iteration of the percolation step descends the ranked list and performs the following operations on each potential cluster:
\begin{enumerate}
\item Given the $j$-th cluster in the list, recompute the richness
and the membership probability for each member galaxy based on the galaxy catalog. The cluster richness and membership probabilities are given by
\begin{equation}
\lambda = \sum_{
R_j
<R_{\rm c}(\lambda)} 
p_{{\rm free},j} 
\pmemj(R_j|\lambda)\,
\label{eq:lambda}
\end{equation}
and
\begin{equation}
p_{{\rm mem}, j}(R_j|\lambda) = \frac{\lambda u(R_j|\lambda)}{\lambda u(R_j|\lambda) + b}\,,
\label{eq:pmem}
\end{equation}
where $R_j$ is the projected cluster-centric distance of 
the $j$-th
member galaxy,
$R_{\rm c}(\lambda)$ is the projected fiducial ``cluster boundary'' given by 
\begin{equation}
R_{\rm c}(\lambda) = R_0 (\lambda/100.0)^{\beta}\,
\label{eq:radius}
\end{equation}
with $R_0=1.0\mpch$ and $\beta=0.2$ as in the redMaPPer algorithm, $u(R_j|\lambda)$ is the projected radial profile of 
member galaxies around the cluster center, and $b$ is the background galaxy density.

Following \citet{Rykoff_etal2014}, we use the projected NFW profile for $u(R)$, assuming that the spatial distribution of satellite galaxies follows the profile, and employ a fixed NFW scale radius of $r_{\rm s}=0.15~h^{-1}{\rm Mpc}$ for all clusters. Note that the profile $u(R|\lambda)$ is truncated at $R_{\rm c}$, i.e. $u(R)=0$ at $R>R_{\rm c}$, and is normalized such that $\int_0^{R_{\rm c}}\!2\pi R\mathrm{d}R~u(R)=1$. This gives 
$u(R|\lambda)$ 
the dimensions of projected number density, i.e. $[({\rm Mpc}/h)^{-2}]$. The background density $b$ is assumed to be a constant to model the uncorrelated galaxies in the foreground and the background; we employ
\begin{align}
b= 2 \bar{n}_{\rm g}d_{\rm cyl}^{\rm max},
\end{align}
where $\bar{n}_{\rm g}$ is the mean number density of mock galaxies, and $d_{\rm cyl}^{\rm max}=60~h^{-1}{\rm Mpc}$ in our case. Similar to $u(R)$, $b$ has the dimensions of number of galaxies per unit area, i.e. $[({\rm Mpc}/h)^{-2}]$.

\item Numerically solve Eqs.~(\ref{eq:lambda}) and (\ref{eq:pmem}) to obtain membership probabilities and richness. If there is a central galaxy within $R_{\rm c}(\lambda)$ whose host halo mass is larger than that for the current cluster center, consider the central galaxy with the most massive halo as the new center.

\item If a new central galaxy was identified, recompute $p_{\rm mem}$ and $\lambda$ with respect to the new central galaxy. After converging on a central galaxy, update $p_{\rm free}$ \citep[see Section~9.3 in][]{Rykoff_etal2014}:
\beq
p_{\rm free} \longrightarrow p_{\rm free} ( 1 - p_{\rm mem})\,.
\eeq
This is to take into account situations where potential member galaxies around lower-ranked cluster candidates already have non-zero probabilities of belonging to higher-ranked candidates.
 If $p_{\rm free}<0.5$ for a central galaxy, that galaxy is eliminated from the list of center candidates.

\item Move down to the next cluster candidate in the list and repeat steps (i)--(iii).
\end{enumerate}
We perform these steps until the results, i.e. the assigned richnesses for clusters, converge. In the following, we refer to the converged richness values obtained from the above procedure as ``observed'' richnesses or ``$\lo$''.
Note that, for every iteration, we reset $p_{\rm free}=1$ for all galaxies.

For comparison, we also define the ``true'' richness of each cluster in the mock catalog as follows.
In the presence of projection effects, multiple halos or galaxies residing in different halos can be identified as a single cluster if those are aligned along the line-of-sight direction. We thus define the most massive halo in each identified cluster as the ``primary'' halo of the cluster, such that we can establish a one-to-one correspondence between the identified clusters and the primary halos in each simulation realization to avoid any double counting of halos or clusters in the statistics. We then define the ``true'' richness, $\lambda_{\rm true}$, for each cluster based on its primary halo, i.e.
\begin{align}
\lambda_{\rm true}\equiv\sum_{i; r_i<r_{\rm 200m}^{\rm primary}} 1~, 
\label{eq:lt}
\end{align}
where the summation runs over all true member galaxies of the primary halo within $r_{\rm 200m}$. We use $r$ to describe a three-dimensional radius while $R$ is used for the projected radius. Note that even if a member galaxy is located at $r>R_{\rm c}$, we include it in the richness definition if $r<r_{\rm 200m}$. The true richness defined this way is simply the total number of central and satellite galaxies in the primary halo\footnote{When we consider the mock cluster catalog taking into account the aspherical shape of halo in Appendix~\ref{app:shape}, we also use the total number of central and satellite galaxies in the primary halo even if a satellite galaxy is at $R>r_{\rm 200m}$.}, and is an idealized definition for cluster richness. 
Note that $\lambda_{\rm true}$ is an integer number by definition. Furtheremore, there are a few cases that $\lambda_{\rm obs}=0$ even when $\lambda_{\rm true}>20$, because these halos are considered as a member of more massive halos. Also, clusters with $\lo<\lambda_{\rm true}$ can occur if some of satellite galaxies in the primary halo are outside the cutoff radius $R_{\rm c}$ in the cluster finding process.

\subsection{Measurements and {\tt Dark Emulator} Predictions}
\label{sec:emulator}

Based on the true and observed richnesses defined above, we construct our true and observed mock cluster samples with the richness binning $\lambda_\mathrm{true/obs} \in [20,30),~[30,40),~[40,55),~[55,200)$. We then make measurements of the cluster lensing and cluster clustering signals for each bin, following \citet{2011A&A...527A..87V}. 
For cluster lensing, we first project the dark matter particle and cluster distributions along the $z$-axis, our fiducial choice for the line-of-sight direction, to define the 2D fields in each realization of the $1(h^{-1}{\rm Gpc})^3$ volume. We use a Nearest Grid Point (NGP) interpolation to create grid-based fields of dark matter and clusters, and then compute the power spectrum, $P^{\rm 2D}_{\rm cm}(k_\perp)$, using Fast Fourier Transforms (FFT) \citep[also see][for details]{Kobayashi:2019jrn}. Finally, we perform inverse FFTs on the computed power spectra to obtain the projected cluster-matter cross-correlation functions and subsequently the $\ds$ profiles according to Eq.~(\ref{eq:DeltaSigma}). To increase the spatial resolution in the FFT computations, we ``fold'' the projected 2D fields up to 5 times, i.e. to a factor of $2^5$, and adopt a $46332^2$ FFT grid at each folding step. This FFT resolution 
is sufficient to probe the lensing profiles down to the spatial resolution limit of $N$-body simulations (we will show the results down to $R=0.02~h^{-1}{\rm Mpc}$ in the following).
Note that while we are projecting over $\pi_{\rm max}=500~h^{-1}{\rm Mpc}$ in our 1$h^{-1}{\rm Gpc}$ boxes, this is equivalent to projecting over an infinite integration length, i.e. $\pi_{\rm max} \to \infty$, as we employ periodic boundary conditions in our $N$-body simulations.

For cluster clustering, we similarly use the NGP interpolation to compute 3D grid-based fields of cluster number densities with $1024^3$ grid points. We again compute the Fourier coefficients $\tilde{\delta}_{{\rm c},{\bf k}}$ at each grid point using FFT, and then perform inverse FFT to obtain the 3D cluster auto-correlation function $\xi_{\rm cc}({\bf r}_\perp,\pi)$ according to the definition 
$\xi_{\rm cc}({\bf r})=\int\!\!\mathrm{d}^3{\bf k}/(2\pi)^3 |\tilde{\delta}_{{\rm c}, \bf k}|^2 \exp[i{\bf k}\cdot{\bf r}]$, where ${\bf r}$ is the three-dimensional separation vector. We then compute the the 2D correlation function, $\xi_{\rm cc}(R,\pi)$, from the azimuthal average over $\varphi$ in the $xy$-plane, where ${\bf r}=(R\cos\varphi,R\sin\varphi,\pi)$. Finally, projecting $\xi_{\rm cc}(R,\pi)$ along the $\pi$ direction, we arrive at $w_{\rm p}(R)$ following Eq.~(\ref{eq:wp_def}). We employ $\pi_{\rm max}=500~h^{-1}{\rm Mpc}$ as our fiducial choice of projection length for consistency with the lensing measurements. However, as cluster clustering arises from correlations between distinct clusters by definition, the length scales of interest here are larger than those for cluster lensing and the $1024^3$ grid provides sufficient resolution for our purposes.

To compute theoretical predictions for comparison against these measurements, we rely on the \texttt{Dark Emulator} developed in \citet{Nishimichi2018}. In addition to simulations assuming the \textit{Planck} cosmology, \citet{Nishimichi2018} also generated an ensemble of $N$-body simulations and corresponding halo catalogs for 101 distinct $w$CDM cosmological models. These models span over a broad range of cosmological parameter space, including our fiducial {\it Planck} cosmology as well as the range of cosmological models inferred from large-scale structure probes such as the Subaru HSC cosmic shear analysis \citep[][]{2019PASJ...71...43H}. They then used the simulations to build an emulator for halo statistics, i.e. the {\tt Dark Emulator}, that allows for fast and accurate computations of the halo mass function, the halo-matter correlation function, and the halo auto-correlation function in terms of halo masses, redshifts, separations, and cosmological models. The {\tt Dark Emulator} outputs are typically accurate to better than a few per cent within the sampled $w$CDM cosmologies over the range of separations and mass scales we are interested in, including the quasilinear and the fully nonlinear regimes of matter clustering, but assumes statistical isotropy for the computed halo statistics. This makes the {\tt Dark Emulator} particularly useful for our purposes, as we can compare the cluster properties measured from mock catalogs with projection effects against the isotropic emulator predictions and isolate out the specific effect of line-of-sight projections while remaining independent of analytic modeling assumptions. This is a notable feature of our study.

\begin{figure*}
    \includegraphics[width=0.43\textwidth]{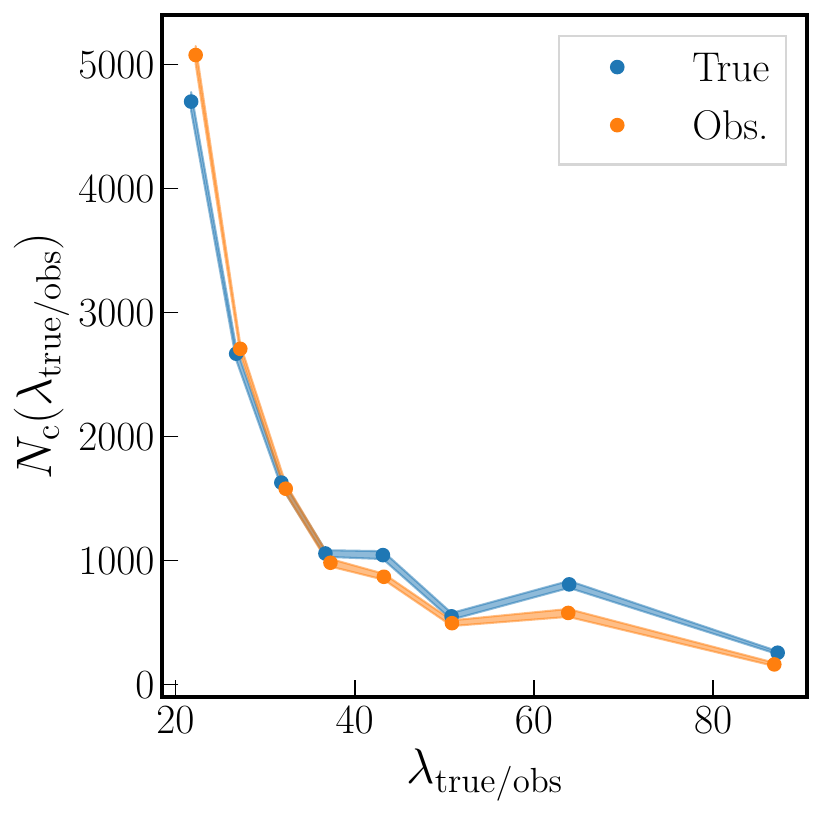}
    \includegraphics[width=0.43\textwidth]{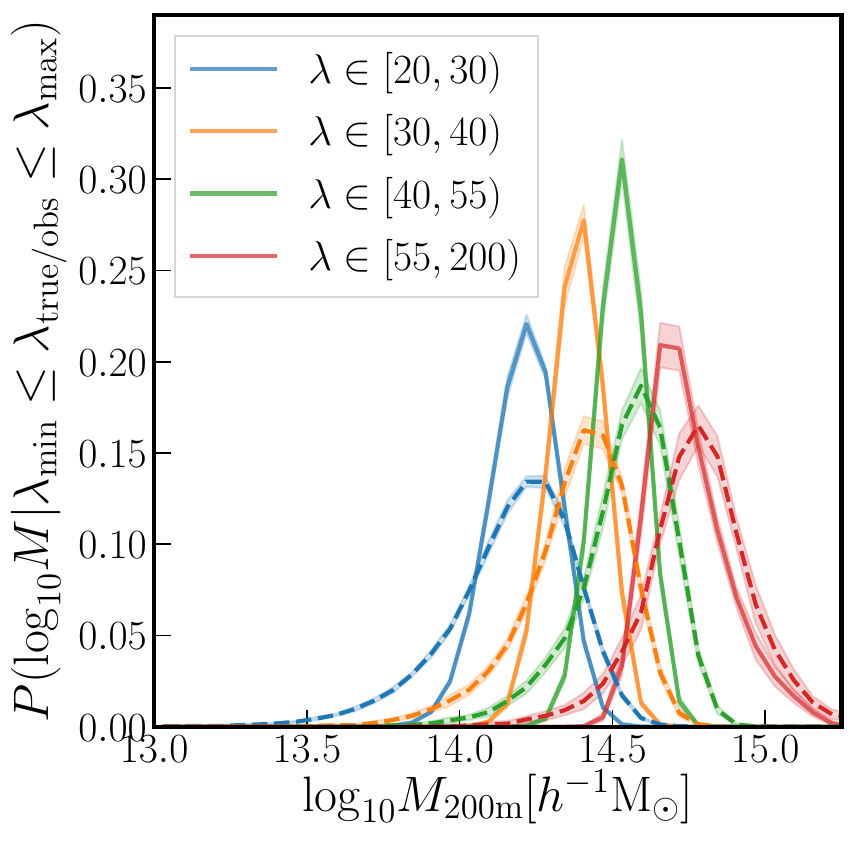}
    \caption{\label{fig:abundance}
    {\it Left}: 
    The number of galaxy clusters in a 
    $1~(h^{-1}{\rm Gpc})^3$ 
    volume
    for different richness bins, measured from the true (blue) and the observed (orange) cluster samples. The fluctuations in the two distributions are due to finite binning effects. 
    {\it Right}:
    The halo mass distributions for different richness bins (colors), measured from the true (solid) and the observed (dashed) cluster samples. Mass distributions for the observed sample show a low-mass end tail from projection effects, as well as some shifts towards higher masses from finite aperture effects in higher richness bins.
    Throughout this paper, we plot the mean of 20 independent realizations as markers, and use the sample variance estimated from the same set of independent realizations to show uncertainties plotted as shaded regions. 
    }
\end{figure*}
\begin{figure}
    \centering
     \includegraphics[width=0.45\textwidth]{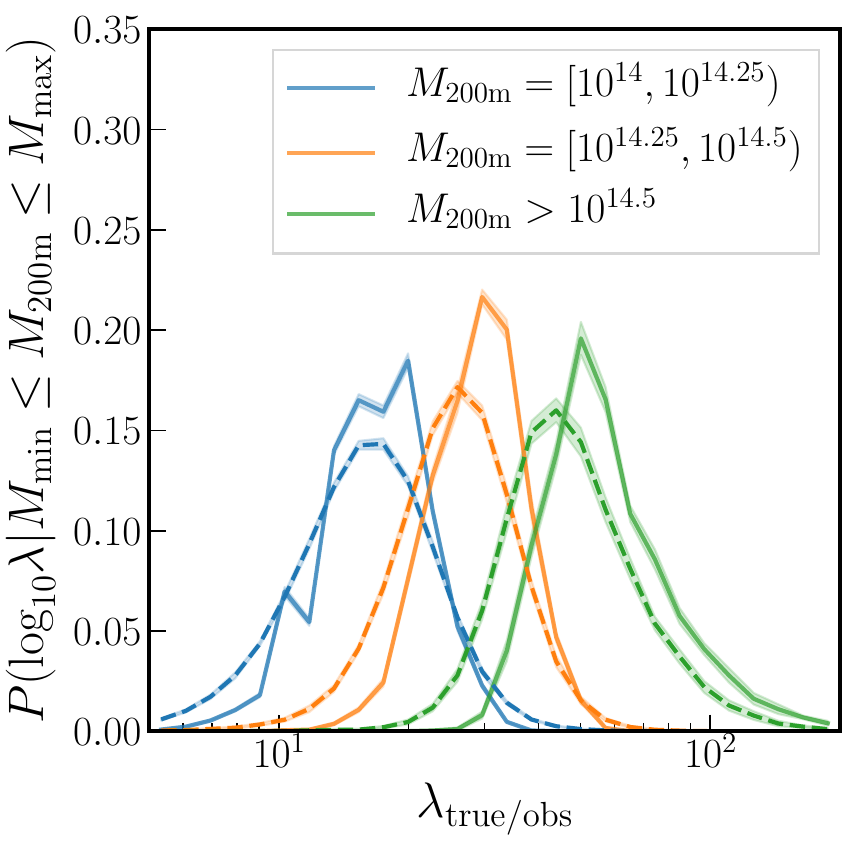}
    \caption{The distribution of true (solid) and observed (dashed) richnesses for halos of different mass bins. Again, we note in the observed distributions a high-richness tail from projection effects for lower mass halos, as well as a shift towards lower richnesses from finite aperture effects for higher mass halos. The fluctuations in the true richness distributions are due to finite binning effects, i.e. from taking logarithmically spaced bins of integer-valued $\lt$.
}
    \label{fig:richness_mass}
\end{figure}
\begin{figure}
    \centering
     \includegraphics[width=0.45\textwidth]{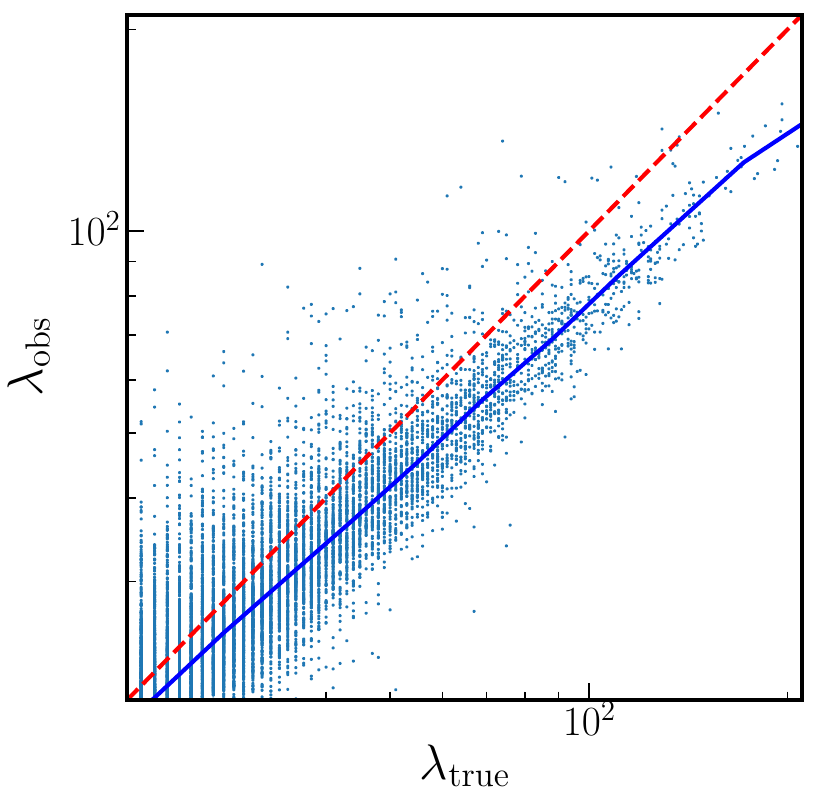}

    \caption{Comparison of the true richness $(\lt)$ and the observed richness ($\lo$) for individual clusters with $\lo\ge 20$. The scatter is larger for low-richness clusters, and $\lo$ tends to be underestimated compared to $\lt$ for high-richness clusters. \yp{The solid line shows the median of $\lo$-distribution in each bin of $\lt$. }
    The dashed line represents the $\lo=\lt$ relation.}
    \label{fig:pheno-lobs}
\end{figure}

\section{Quantifying the impact of projection effects}
\label{sec:signal}
We now present our main results based on measurements of cluster observables from the mock cluster catalogs. Recall that we have defined above the following two sets of cluster samples:
\begin{itemize}
    \item The ``true'' cluster sample, selected according to the true richness ($\lt$) defined in Eq.~(\ref{eq:lt}).
    \item The ``observed'' cluster sample, selected according to the observed richness ($\lo$) obtained from the cluster finder algorithm described in Section~\ref{sec:cluster_finder}.
\end{itemize}
The key difference between the true and observed samples is the existence of projection effects from the characteristic selection effects introduced by our cluster finder algorithm, which affects the observed sample but not its true counterpart. Our goal is to identify how projection effects propagate into the cosmological observables (cluster abundances, $\ds$ and $\wpr$) as well as the selection observable $\lo$ by comparing their measurements between the true and observed samples.

\subsection{Cluster Abundances and the Richness-Mass Relation}
\label{subsec:richness-mass}

Optical cluster finders are known to introduce a non-trivial relationship between the observed richness $\lo$ and the true richness $\lambda_{\rm true}$. On the one hand, line-of-sight projections of (un)correlated interlopers lead to an overestimation of $\lo$. On the other hand, the percolation of members and background subtraction lead to an underestimation of $\lo$. 
These additional complications must be calibrated within the context of a richness-mass relation in order to properly estimate the underlying halo mass distribution and the resulting cluster abundance in a cluster sample selected by $\lo$. In fact, the term ``projection effect'' has often been used to refer to this exact phenomenon, i.e. the misidentification of $\lo$ and the resulting complexities in modeling the $\lt$--$\lo$ relation and cluster abundances. Thus, we begin the comparison of our true and observed results by exploring how cluster abundances and the richness-mass relation are altered by our cluster finder algorithm. 

The left panel of Fig.~\ref{fig:abundance} shows cluster abundances as a function of richness. The observed cluster abundances show small discrepancies against measurements from the true sample, but these discrepancies alone do not fully reveal the impact of projection effects. The right panel, which compares the halo mass distributions of the true and observed samples across different richness bins, provides more information for our purposes. Note that for the observed sample we define cluster mass as the mass of the primary (most massive) halo in the identified cluster (see~Section~\ref{sec:cluster_finder}). For all richness bins, the observed sample exhibits a low-mass end tail compared to the true sample. This trend is consistent with previous results from the actual SDSS redMaPPer catalog, e.g. \citet{2018ApJ...854..120M} (Fig.~7 in their paper). The observed sample also shows a high-mass end tail that is more pronounced for higher richness bins. For massive clusters, the finite aperture radius $R_{\rm c}(\lambda)$ we use to define the cluster can be smaller than $r_{\rm 200m}$ of the host halo, which leads to an underestimation of $\lo$ and the subsequent inclusion of such clusters in lower richness bins than naively expected. 

Fig.~\ref{fig:richness_mass} shows the distributions of richness ($\lambda_{\rm true/obs}$) for a given halo mass bin, i.e. $P(\lambda|M)$. The richness distributions for the observed sample are shifted lower from those for the true sample, due to the finite aperture effect discussed above. For the lowest mass bin, however, the observed distribution shows a high-richness tail arising from projection effects. Finally, in Fig.~\ref{fig:pheno-lobs}, we directly show the $\lt-\lo$ relationship. At lower richnesses, $\lo$ shows a large scatter at a given $\lt$ and shows a tendency for overestimation, i.e. $\lo>\lt$. At higher richnesses, on the other hand, $\lo$ tends to be underestimated compared to $\lt$, and the $\lt$--$\lo$ scatter becomes much smaller.

The results in Figs.~\ref{fig:abundance}--\ref{fig:pheno-lobs} are in good agreement with the two main effects we expect from the cluster finder, i.e. projection effects and finite aperture effects, and are also in qualitative agreement with earlier studies, e.g. \citet{costanzietalprojection}. While the true sample shows straightforward mass and richness distributions arising from the intrinsic Poisson scatter in $\lt$ we introduced (see Section~\ref{subsec:mocks}), the observed sample exhibits richness-dependent shifts and additional scatters in comparison, complicating the modeling of underlying cluster mass distributions. 
For cluster cosmology analyses, the hope is to build models for underlying mass distributions that capture these complexities and calibrate them using additional cluster observables that we discuss next.

\subsection{Cluster Lensing and Cluster Clustering}

\label{subsec:lensing}

\begin{figure}
    \centering
    \includegraphics[width=0.45\textwidth]{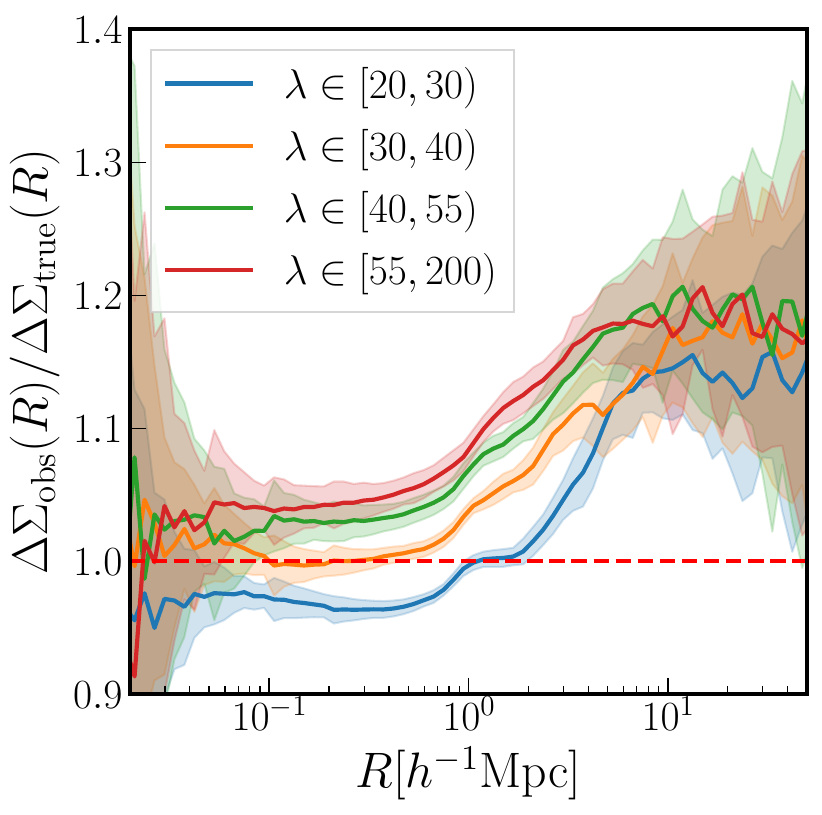}
    \caption{\label{fig:lensing1} Ratios of lensing profiles between the observed and the true cluster samples across different richness bins. The suppression/boost in the one-halo term ($R<1$--$2h^{-1}{\rm Mpc}$) for lower/higher richness bins is explained by the changes in the underlying mass distributions between the true and the observed samples, but the boosts on large scales -- up to a factor of 1.2 -- are not explained by cluster masses. Again, the lines and the shaded regions respectively represent the mean and the sample variance for a volume of $1~(h^{-1}{\rm Gpc})^3$ estimated from 20 independent realizations.}
\end{figure}
\begin{figure*}
    \centering
    \includegraphics[width=0.32\textwidth]{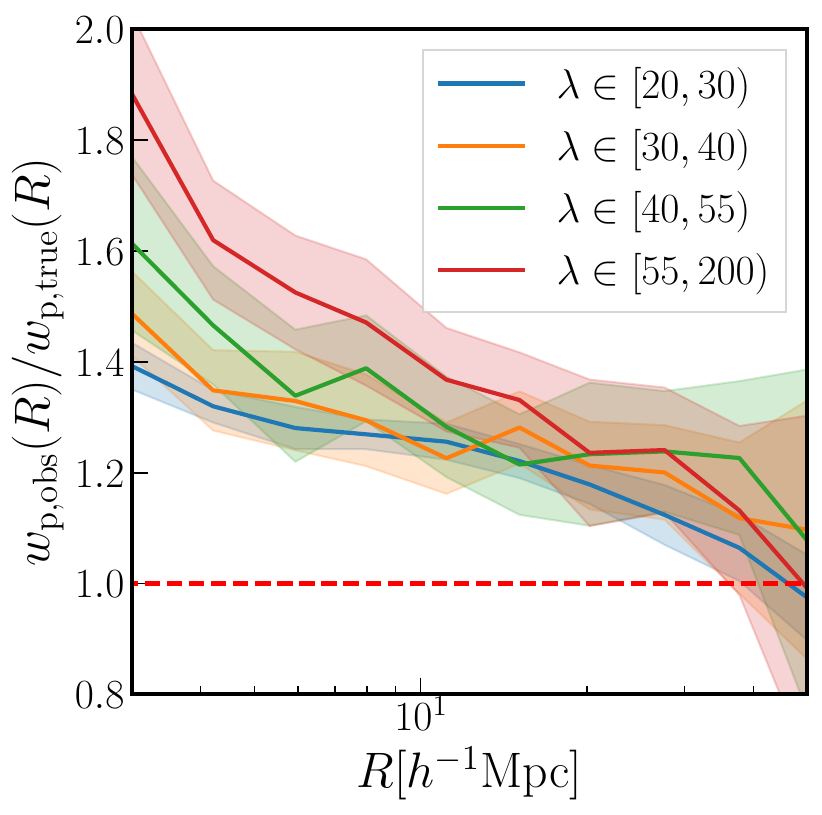}
    \includegraphics[width=0.32\textwidth]{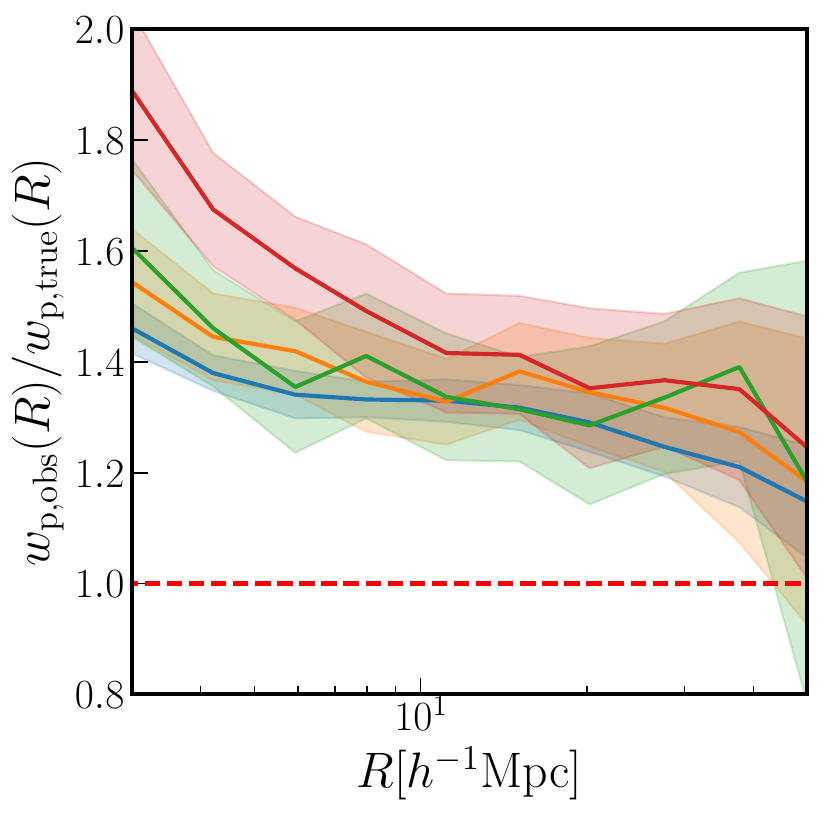}
    \includegraphics[width=0.32\textwidth]{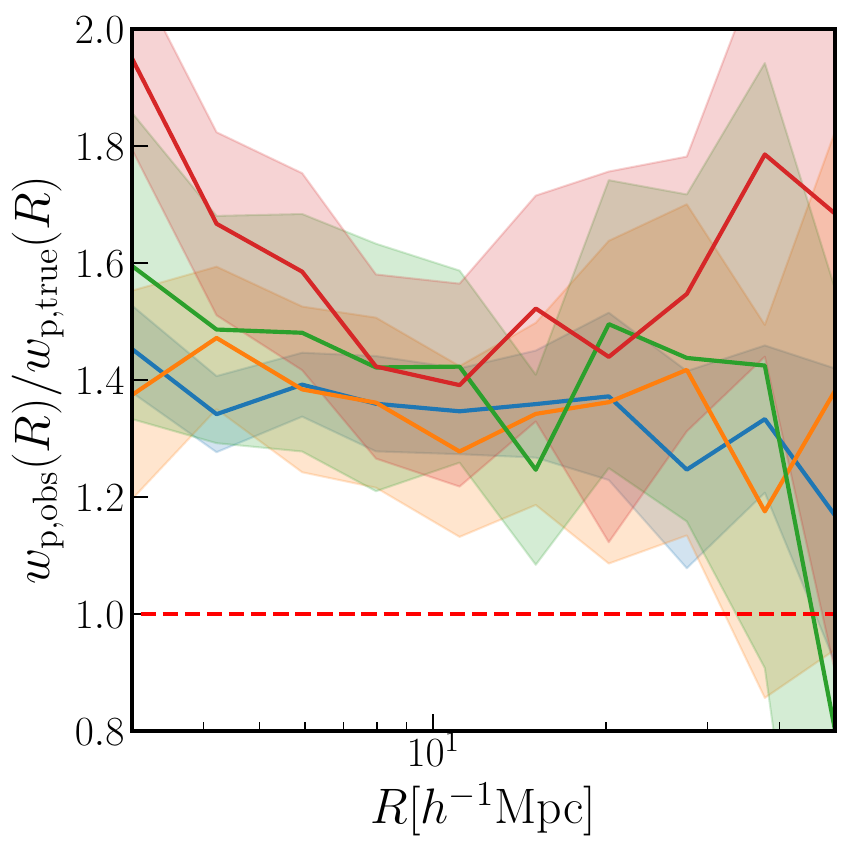}
    \caption{\label{fig:auto1} Ratios of projected auto-correlation functions between the true and the observed samples for different richness bins. The left, middle, and right panels respectively correspond to projection lengths of $\pi_{\rm max}=50, 100$, and $500 h^{-1}{\rm Mpc}$ in computing the the projected correlation function (Eq.~\ref{eq:wp_def}).
    All richness bins exhibit a boosted clustering signal for the observed sample compared to the true sample, by up to a factor of 1.4, which is about double the boost in the two-halo term of the lensing profile in Fig.~\ref{fig:lensing1}. Similar to  Fig. \ref{fig:lensing1}, the shaded regions represent the sample variance for a volume of $1~(h^{-1}{\rm Gpc})^3$.}
\end{figure*}

We now turn our attention to the comparison of the cluster lensing and the cluster clustering signals between the true and the observed cluster samples. More specifically, we are interested in studying whether these signals measured from the observed sample match expectations based on our earlier findings on the underlying cluster mass distributions. Starting with cluster lensing, we show in Fig.~\ref{fig:lensing1} the ratios of $\Delta\!\Sigma(R)$ measurements between the true and observed samples across different richness bins. On small scales, the lensing profiles correspond to the one-halo term in the halo model picture, and thus are expected to probe the average halo mass of the sample. The measured ratios agree with this expectation; the observed sample shows suppressed/boosted small-scale lensing signals for lower/higher richness bins when compared against the true sample, consistent with the shifts in the underlying mass distributions presented in Fig.~\ref{fig:abundance}.

The more interesting aspect of this comparison, however, appears on large scales, i.e. in the two-halo regime. On these scales, say beyond a few Mpc$/h$, the observed sample displays a significant boost in the amplitudes of the lensing signal -- up to $\sim$20\% -- when compared against the true sample. In a standard
halo model picture, the two-halo lensing signal is expected to follow $\xi_\mathrm{hm} \sim b_\mathrm{h}(M)\xi_\mathrm{mm}$, where $b_\mathrm{h}$ is the linear halo bias. As the $b_\mathrm{h}(M)$ relation is a monotonically increasing function of halo mass, we would naively expect the two-halo lensing signals for the observed sample to follow the trends in the one-halo regime, i.e. to show suppression for lower richness bins and boosts for higher richness bins. The unanimous boost on large scales for all richness bins we observe, therefore, indicates that the differences in $\ds$ between the true and observed samples cannot be explained entirely by changes in the underlying halo mass distribution. Rather, it points to a potential failure of the 
standard halo model, or more specifically of a discrepancy between the one- and two-halo lensing signals in the observed cluster sample.

We further confirm this discrepancy in the comparison of the cluster clustering signal $w_{\rm p}\!(R)$. In Fig.~\ref{fig:auto1}, we show the ratio of the cluster clustering signal measured from the true and observed samples. Since $w_{\rm p}\!(R)$ is measured only between distinct halos, they are by construction ``on large scales'' or entirely in the two-halo regime, and we find that the observed cluster clustering signal is boosted on all scales. This is in agreement with the discrepancy we find in the comparison of lensing signals, and in disagreement with the halo model expectation that the lower richness bins of the observed sample, with lower overall halo masses compared to their true counterparts, should show suppressed clustering signals. The different panels in Fig.~\ref{fig:auto1} also shows that the amount of the observed boosts depend on the projection length $\pi_{\rm max}$ used in measuring $w_{\rm p}\!(R)$. As previously discussed, we employ $\pi_{\rm max}=500~h^{-1}{\rm Mpc}$ as our fiducial setting, from which we find that the boost in clustering is about double of that in lensing. This has meaningful implications, which we will discuss further in later sections. 

\subsection{Comparisons against \texttt{Dark Emulator} Predictions}
The two-halo term boosts in $\ds$ and $w_{\rm p}\!(R)$ discussed above are both significant and somewhat surprising, as they would invalidate the usual halo model connection between the one- and the two-halo regimes. To ensure that our findings are genuine, we further test our measurements against predictions using the \texttt{Dark Emulator} described in Section~\ref{sec:emulator}. Emulator predictions are particularly useful for our purposes, as they are (i) independent of modeling assumptions, e.g. for halo bias or halo density profiles; (ii) able to capture the full non-linearity of the dark matter and halo distributions; and most importantly (iii) statistically isotropic. To maximize the utility of these predictions, we use the actual underlying halo mass distributions for both the true and the observed samples in their computation. That is, we define our emulator predictions as
\begin{align}
\Delta\!\Sigma^\mathrm{pred,true/obs}(R) & =  \int_{M\in {\rm bin}}\!\!\mathrm{d}M~ P^\mathrm{true/obs}\!(M)\Delta\!\Sigma^\mathrm{emu}\!(R|M), \nonumber\\
w_{\rm p}^\mathrm{pred,true/obs}(R) & = 
\int_{M_1\in{\rm bin}} \int_{M_2\in{\rm bin}}\!\!\mathrm{d}M_1\mathrm{d}M_2~  P^\mathrm{true/obs}\!(M_1) \nonumber\\
&\times P^\mathrm{true/obs}\!(M_2) w_{\rm p}^\mathrm{emu}\!(R|M_1,M_2).
\end{align}
Here $P^\mathrm{true/obs}\!(M)$ is the normalized halo mass distribution drawn directly from the truth information in the halo catalogs for the true/observed cluster samples, while $\Delta\!\Sigma^\mathrm{emu}(R|M)$ and $w_{\rm p}^\mathrm{emu}\!(R|M_1,M_2)$ are generated from the emulator. Using the truth information for $P^\mathrm{true/obs}\!(M)$ is equivalent to assuming a perfect knowledge of $P(\lambda_\mathrm{true/obs}|M)$. Thus, any mismatch between the emulator predictions and the measurements would point towards systematic effects that cannot be captured by the formalism in Section~\ref{subsec:cluster_observables}.

\begin{figure*}
    \centering
    \includegraphics[width=0.45\textwidth]{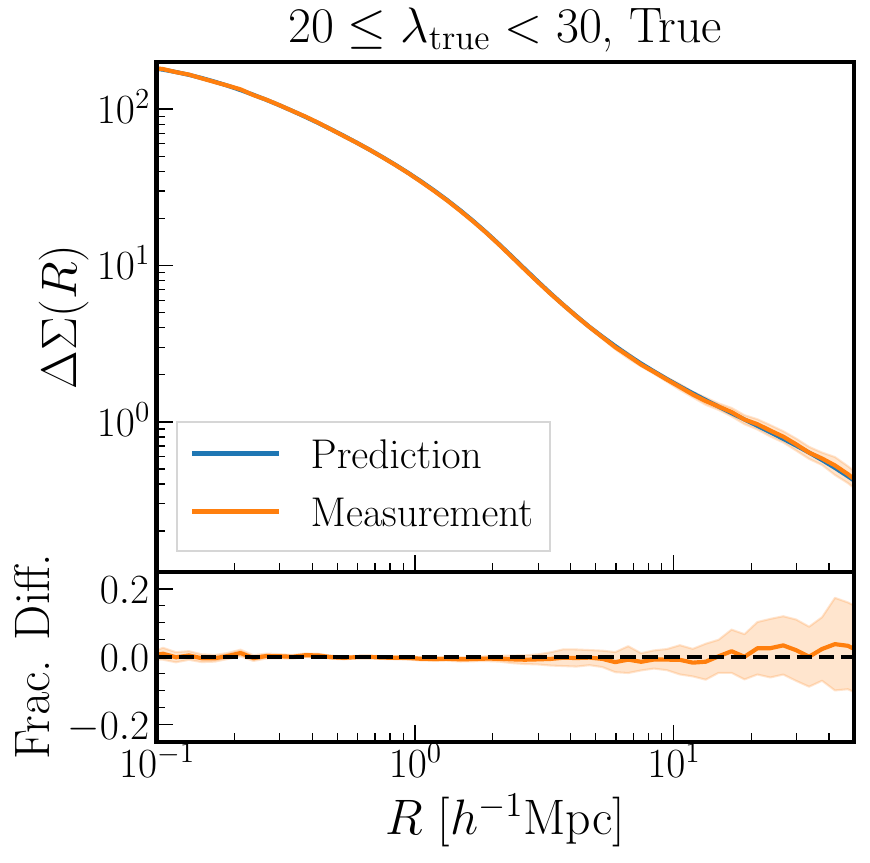}
    \includegraphics[width=0.45\textwidth]{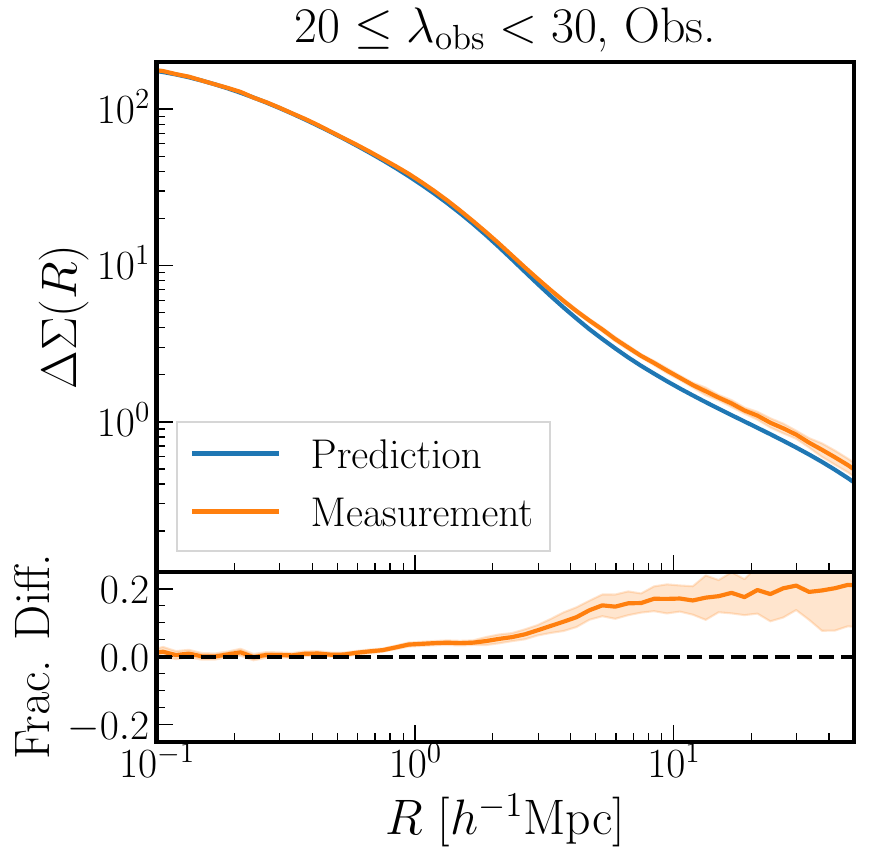}
    \caption{\label{fig:lensing_pred1} 
    {\it Left}: 
    Comparison of the measured cluster lensing signal for the true sample (blue) against the lensing signal predicted by the emulator 
    using
    the underlying halo mass distribution of the true sample (orange). 
    Results for the $20<\lambda<30$ richness bin are shown here.
    {\it Right}:
    Same as the left panel, but for the observed cluster sample. Note that the emulator predictions for the observed sample use the masses of ``primary'' halos for each identified cluster (see text for details). While the measured lensing profile for the true sample agrees fairly well with the result from the emulator, the measured lensing profile for the observed sample is larger than the emulator prediction on scales $R\simgt 1~h^{-1}{\rm Mpc}$, by up to a factor of 1.2.
}
\end{figure*}

\begin{figure*}
    \includegraphics[width=0.45\textwidth]{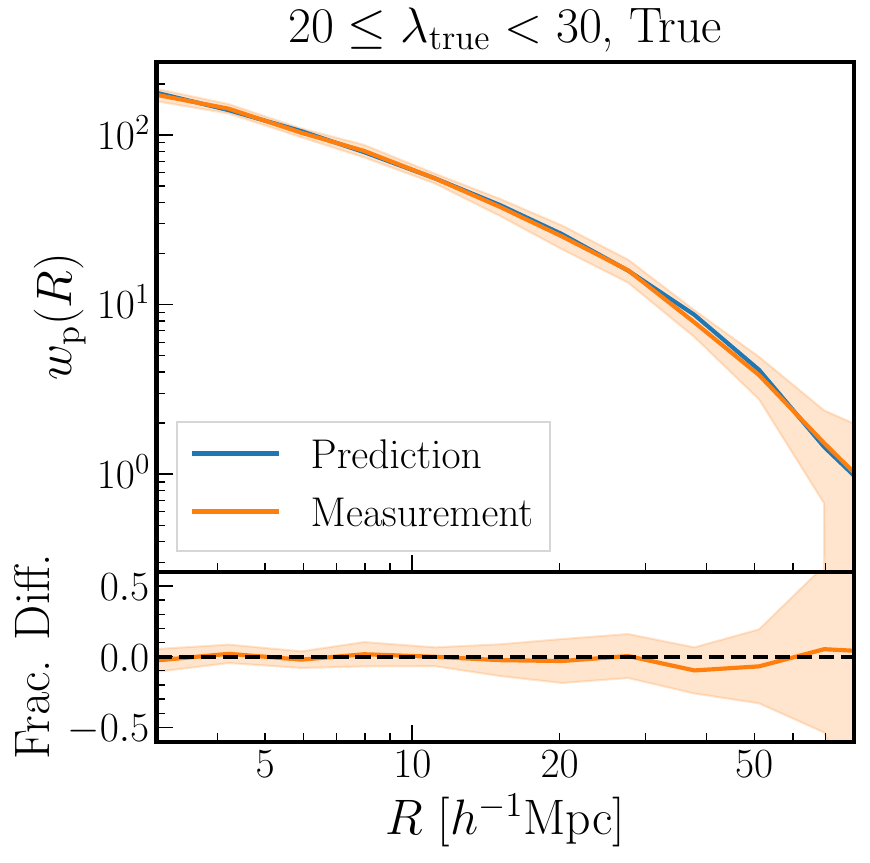}
    \includegraphics[width=0.45\textwidth]{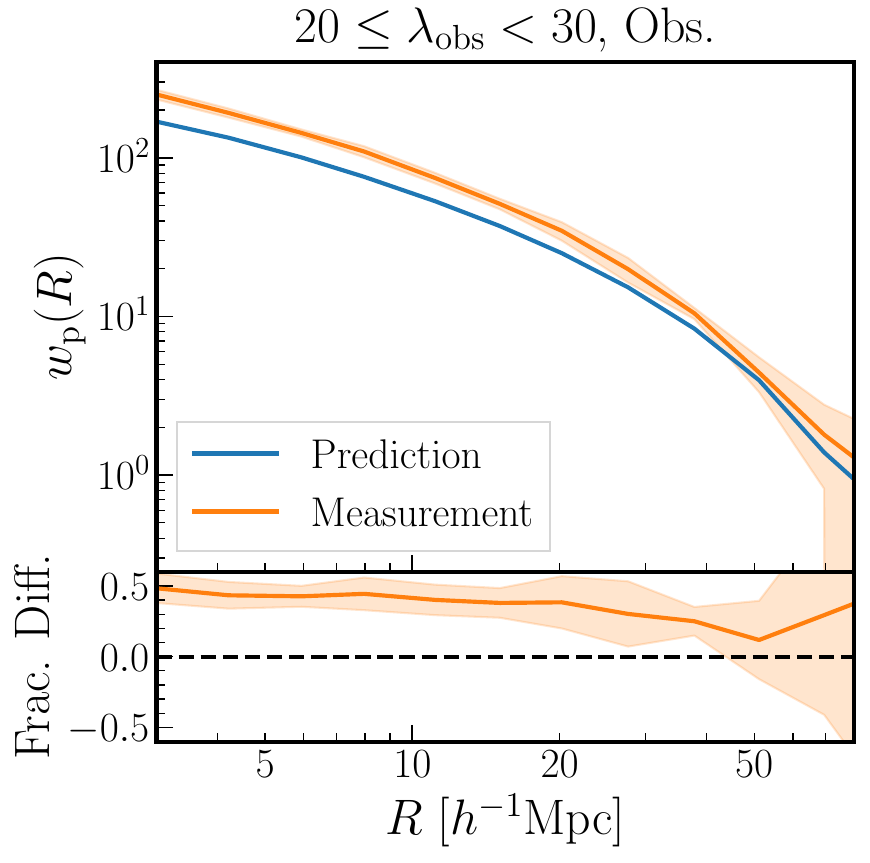}
    \caption{\label{fig:auto_pred1}
    {\it Left}: Similar to the previous figure, but for the
    comparison of the measured cluster clustering signal for the true sample (blue) against the clustering signal predicted by the emulator and the underlying halo mass distribution of the true sample (orange). 
    {\it Right}:
    Same as the left panel, but for the observed cluster sample.  While the measured clustering signal for the true sample agrees well with the emulator prediction, the observed sample shows a boosted cluster signal on all scales compared to the emulator result, by a factor of 1.4.}
\end{figure*}

In Fig.~\ref{fig:lensing_pred1}, we compare the lensing profiles measured from the true/observed samples against the corresponding emulator predictions. We again note that, for the observed sample, the cluster mass is given by the mass of the primary (most massive) halo within the cluster region. For the true sample, the measurements and the predictions agree well on all scales. This is expected by design, and shows that our emulator predictions are working well. However, for the observed sample, the measured signals agree with the emulator predictions only up to $R\sim 1~h^{-1}{\rm Mpc}$, roughly corresponding to the virial radii of cluster-scale halos, and displays boosts up to 20\% on large scales. Note that we only show the results for the $20<\lambda<30$ bin here, but the trends are consistent for other richness bins (see Appendix~\ref{app:all_richness}). This confirms our earlier finding and suspicion that the large-scale boosts in the cluster lensing signal for the observed sample cannot be explained by changes in the underlying cluster mass distribution. We find similar results for the cluster clustering signal, as shown in Fig.~\ref{fig:auto_pred1}. Measurements from the true sample again matches well with emulator predictions, but the observed sample shows a significant boost on all scales compared to the emulator output. The size of these boosts is up to 40\%, again roughly double the boost observed from the lensing side.


The above comparisons confirm our previous finding that the lensing and clustering signals of the observed cluster samples deviate from expectations based on the mass distribution of the sample. On the one hand, the reasonably good agreement between measurements and emulator predictions in the one-halo lensing signal is encouraging for cluster cosmology, as this suggests that we may use the cluster lensing signal on small scales to infer cluster masses or the richness-mass relation. However, the boosted lensing signal on large (two-halo) scales, i.e. beyond $1~h^{-1}{\rm Mpc}$, as well as the boosted clustering signal, will lead to overestimations on the inferred halo masses \citep[also see][for the similar discussion]{Osato_etal2018} if included. Furthermore, we cannot ascertain that the small-scale lensing signal will always be well-behaved, as the impact of projection effects may vary with respect to e.g. cosmology, HOD, or the details of the cluster finder algorithm. Thus, our findings ultimately imply that projection effects not only impact the richness-mass relation and the mass distributions of cluster samples, but in addition introduce characteristic modifications to the cluster lensing and cluster clustering signals of optically identified clusters. These modifications must be understood and modeled to perform an unbiased cluster cosmology analysis including two-halo scales.

\subsection{Interpretation}
\label{sec:interpretation}

\begin{figure*}
    \includegraphics[width=0.45\textwidth]{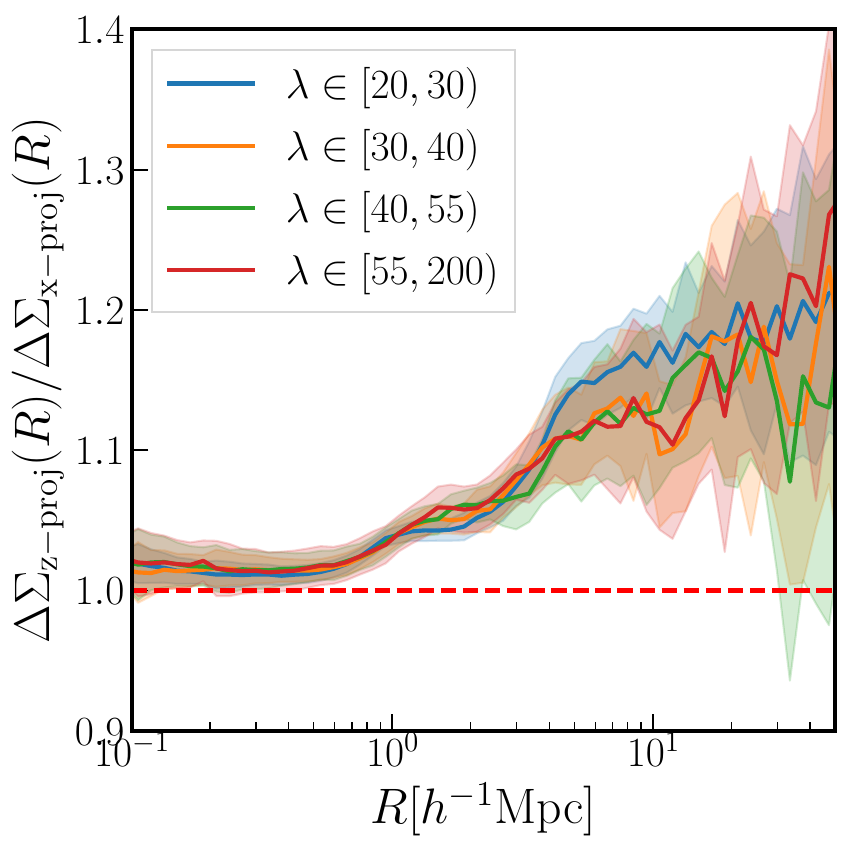}
    \includegraphics[width=0.45\textwidth]{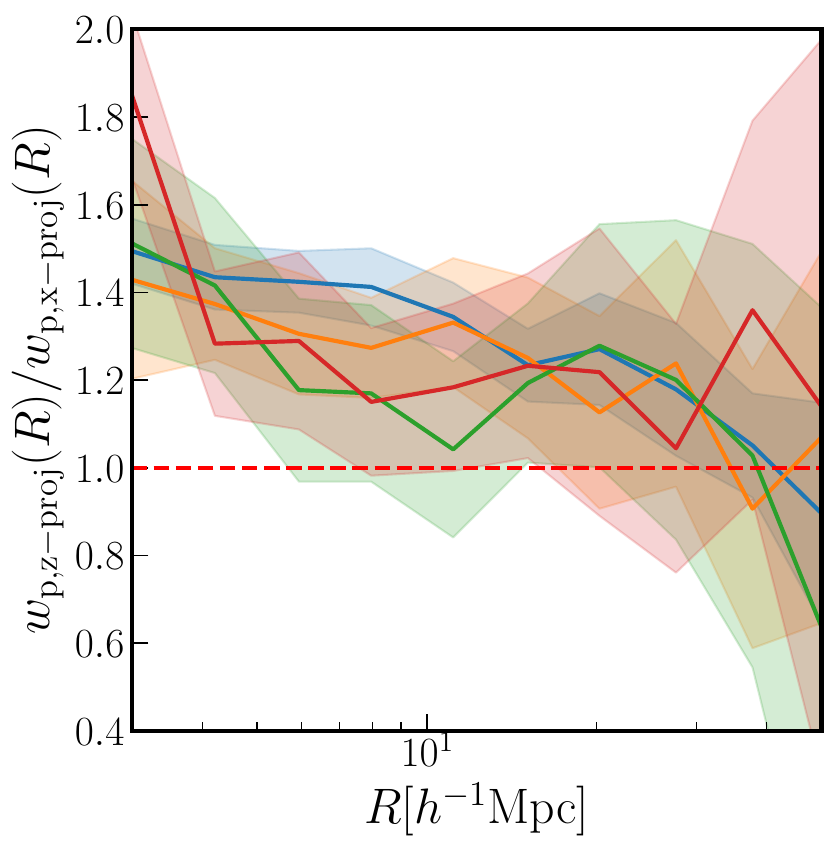}
 
    \caption{\label{fig:orientation} Comparison of the lensing profiles (left) and auto-correlation functions (right) for the observed cluster sample with different projection axes. 
    Note that we choose the $z$-axis as the fiducial line-of-sight (LOS) direction when we run our cluster finder algorithm in each mock catalog. Here we compare our default measurements against a second set of measurements using the $x$-axis as the LOS direction. The $z$-projected measurements show characteristic large-scale boosts when compared to the $x$-projected measurements, implying that the disagreement between measurements and predictions in Fig.~\ref{fig:lensing_pred1} depends on the choice of the LOS direction.}
\end{figure*}

Where are the two-halo boosts in lensing and clustering coming from? At first sight, our results seem comparable to some degree with the results in \citet{Osato_etal2018}, which studied the dependence of halo surface mass density profiles on the orientation of halo shapes. They showed that halos with their major axes aligned with the line-of-sight exhibit boosted lensing profiles; for cluster-scale halos, the shape orientation dependence induced boosts up to 25\% in the lensing signals, on both one- and two-halo regimes. However, our default mock galaxy catalog \textit{does not include} any information on the shapes of individual clusters or their host halos, as we populate galaxies in halos following a spherically symmetric NFW profile around the true halo center. This implies that our cluster finder algorithm has no explicit knowledge of halo shapes or their orientations, and consequently that there is no direct causal link between the boosts that we find above and the effect of shape orientations discussed in \citet{Osato_etal2018}.

We can, however, postulate a common underlying cause for both results. \yp{Projection effects can cause cluster finders to preferentially select clusters embedded within filaments that are aligned with the line-of-sight; multiple halos and their member galaxies within a filament will be projected together along the line-of-sight, which then contributes to a high concentration of galaxies in the projected 2D plane that cluster finders look for.} This selection bias, if it does exist, implies that identified clusters will have their lensing and clustering signals modified by the presence of the filament and associated correlated large-scale structure  \citep[also see][for the similar discussion]{Cohn:2007,Okumuraetal:17}.  
In addition, as halo shape orientations are correlated with the surrounding matter distribution, we expect halos with their major axes aligned with the line-of-sight to have higher chances of being embedded within aligned filaments, which would give rise to boosted lensing and clustering signals similar to Fig.~\ref{fig:lensing1}. In other words, our results and results from \citet{Osato_etal2018} can both be explained by considering the impact of anisotropic matter distributions around halos, in our case manifesting through a preferential selection of clusters in aligned filaments by our cluster finder.

To further investigate this hypothesis, we test the anisotropic nature of the observed boosts. In Fig.~\ref{fig:orientation} we compare the lensing and clustering signals from the observed sample with two different line-of-sight directions. For this comparison, we make a new set of measurements for cluster lensing and cluster clustering; we use the exact same sample of observed clusters, but we now use the $x$-axis of the simulation box as the line-of-sight direction or the projection axis for making measurements, as opposed to our fiducial choice of the $z$-axis as the line-of-sight direction. 
Recall that the 2D observables, i.e. $\Delta\!\Sigma$ and $w_{\rm p}$, are sensitive to Fourier modes in the two-dimensional plane perpendicular to the projection direction. Thus, the $x$-projected measurements study modes in the $yz$ plane, while the fiducial $z$-projected measurements correspond to the modes in the $xy$ plane.

On the lensing side, the left panel of Fig.~\ref{fig:orientation} shows that the default, $z$-projected lensing profile is boosted on large scales compared to the $x$-projected profile. The scale dependence of the boost is similar to that from Fig.~\ref{fig:lensing_pred1}, implying that the $x$-projected measurements are closely following the emulator predictions. 
This can be explained as follows: aligned filaments in the $z$-projected measurements would lie perpendicular to the line-of-sight in the $x$-projected measurements, and thus their impact would mostly disappear in the azimuthally averaged statistic \mtrv{in the $yz$-plane},
$\ds$. The right panel of Fig.~\ref{fig:orientation} similarly compares the auto-correlation function projected along the $z$- and  $x$-axes. Again, we find that the $z$-projected measurements are boosted compared to the $x$-projected measurements, and that the profiles of the boosts are consistent with those from Fig.~\ref{fig:auto_pred1}.
These results confirm that the large-scale boosts in $\Delta\!\Sigma$ and $w_{\rm p}$ are fundamentally anisotropic, further strengthening the hypothesis that aligned filaments give rise to the large-scale boosts in the lensing and clustering signals of the observed cluster sample.

\begin{figure*}
    \centering
    \includegraphics[width=0.32\textwidth]{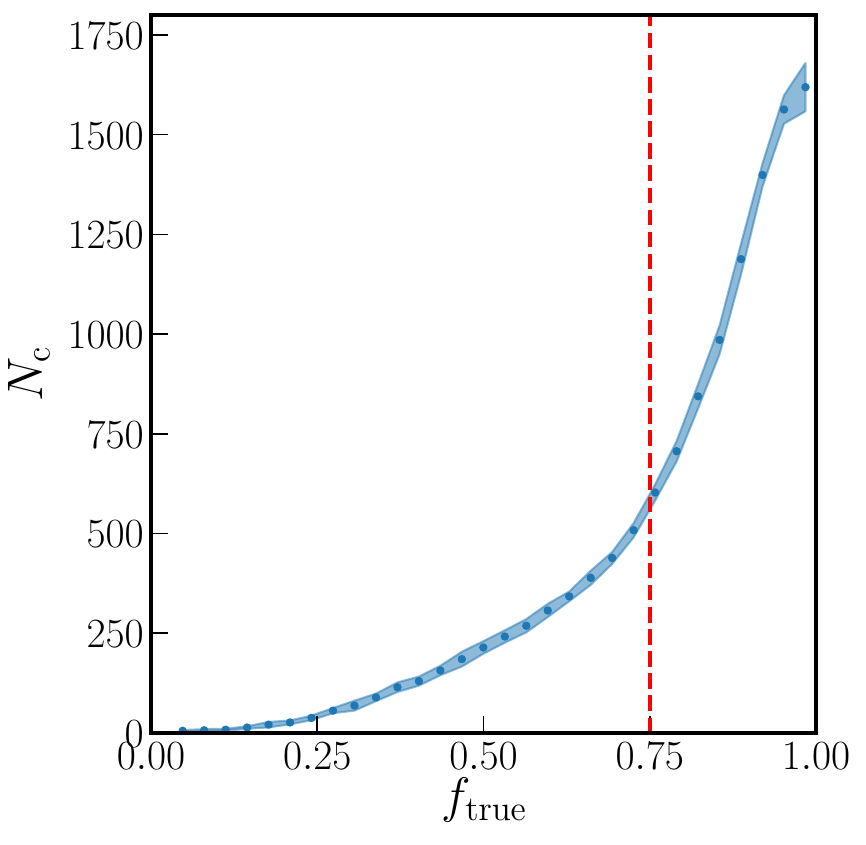}
    \includegraphics[width=0.32\textwidth]{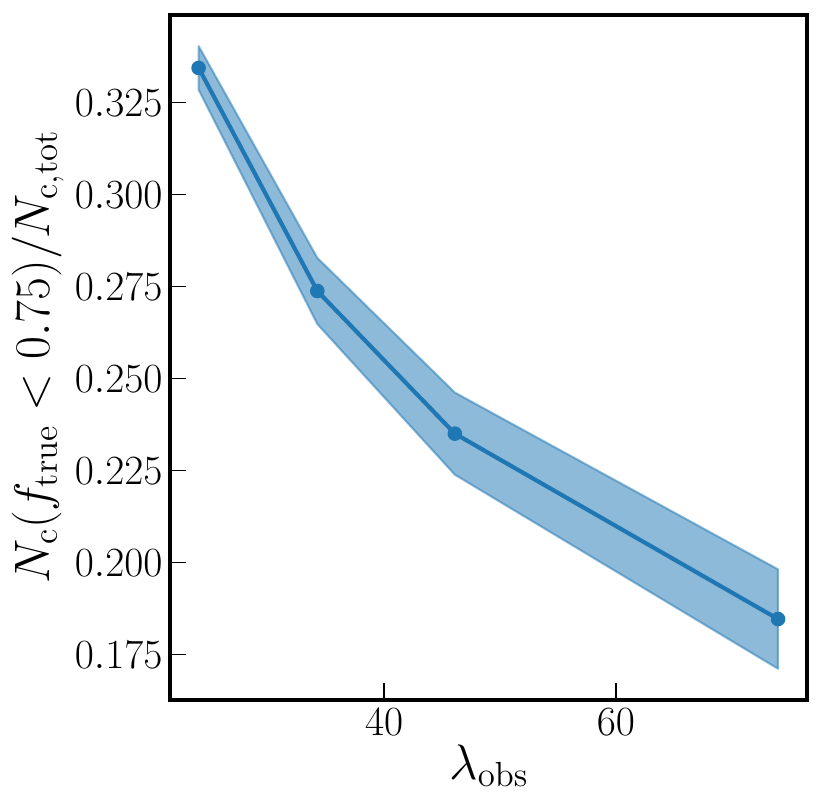}
    \includegraphics[width=0.32\textwidth]{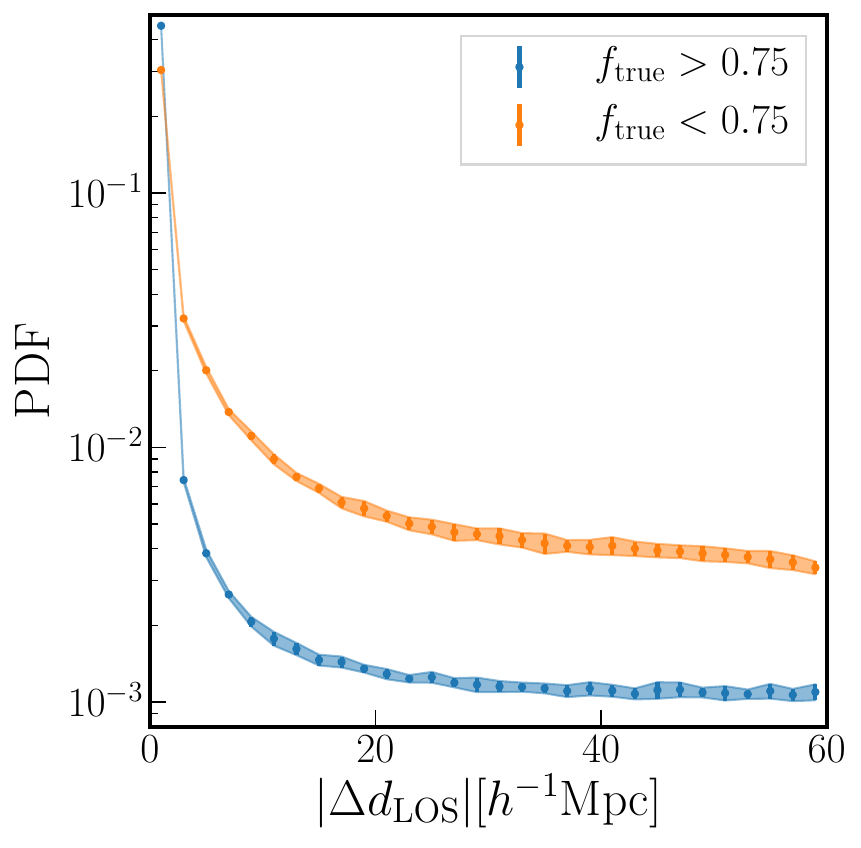}
    \caption{\label{fig:ftrue_frac} Exploratory statistics for the $f_\mathrm{true}$-based sample splits. \textit{Left:} The distribution of true member fraction in the observed cluster sample with $\lambda_{\rm obs}>20$. The red line shows the location of the $f_{\rm true}=0.75$ cut 
    that divides the pure and projected subsamples. \textit{Middle:} The fraction of clusters with $f_{\rm true}<0.75$ 
    for different richness bins. The fraction tends to be higher for smaller richness clusters. \textit{Right:} The \mtrv{normalized distribution of} line-of-sight (LOS)
    distances of 
    member galaxies from the cluster center \mtrv{for clusters with $\lambda_{\rm obs}>20$.} Member galaxies in $f_{\rm true}<0.75$ clusters tend to be more scattered across the projection length than the ones in $f_{\rm true}>0.75$ clusters.}
\end{figure*}
\begin{figure*}
    \centering
    \includegraphics[width=0.45\textwidth]{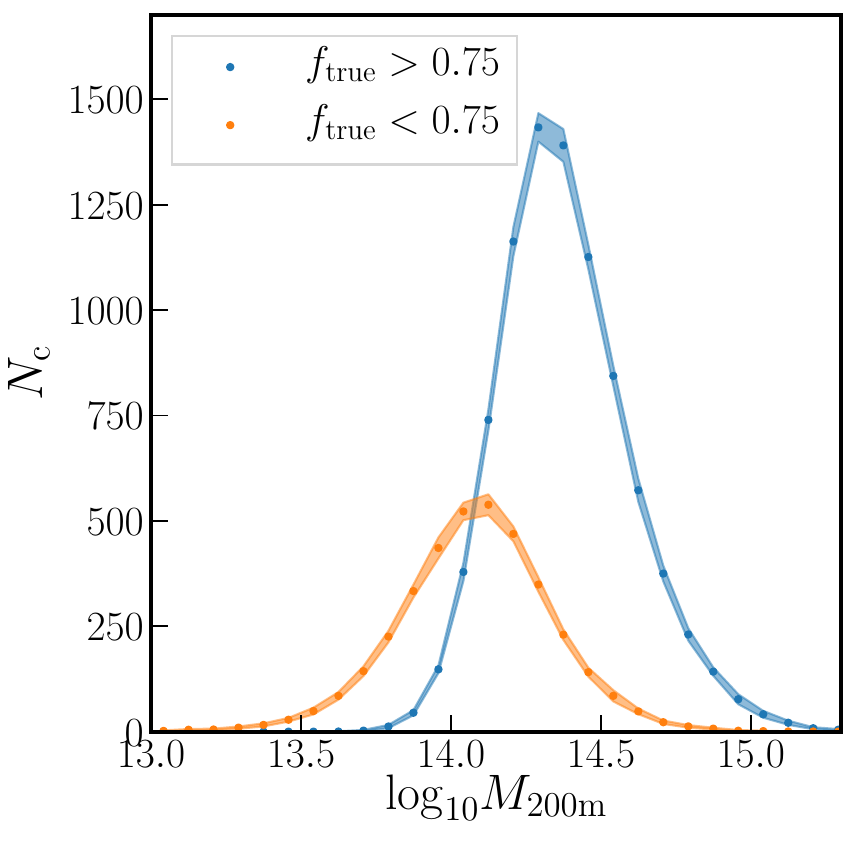} 
    \includegraphics[width=0.45\textwidth]{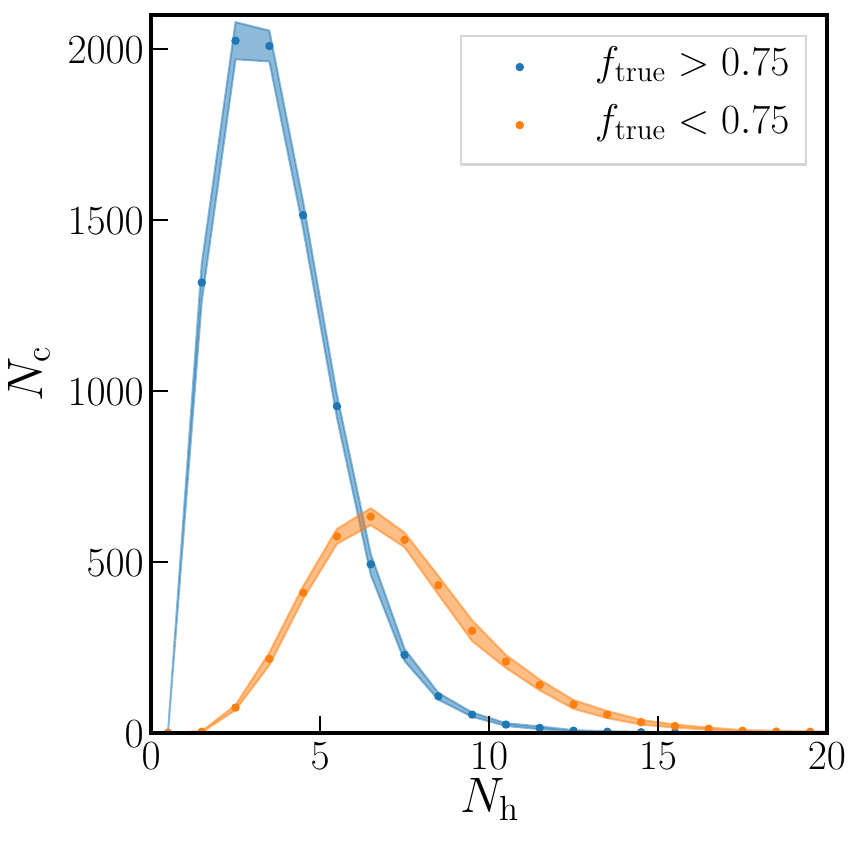}
    \caption{\label{fig:ftrue_mass} 
    {\it Left}:
    The mass distributions for primary halos in individual clusters with $f_{\rm true}>0.75$ (blue; ``pure'' subsample) and $f_{\rm true}<0.75$ (orange; ``projected'' subsample). See Eq.~(\ref{eq:frtrue}) for the definition of $f_{\rm true}$. The projected clusters tend to have lower masses than the 
    \mtrv{pure} clusters. 
    {\it Right}: The distribution of number of included halos \mtrv{with $M\ge 10^{12}~h^{-1}M_\odot$} for each cluster. A halo is considered to be ``included'' in a cluster if at least one galaxy in the halo is identified as a member of the cluster. As expected, projected clusters include member galaxies from a larger number of halos, and the distributions for the pure and projected subsamples are qualitatively different. Results for \mtrv{clusters with $\lambda_{\rm obs}>20$} are shown.}
\end{figure*}

\section{Projected Clusters, Aligned Filaments, and Boosts}
\label{sec:signal2}

So far, we have confirmed that the observed large-scale boosts in $\Delta\!\Sigma$ and $w_{\rm p}$ are unexplained by the underlying halo mass distribution and highly anisotropic. We then have argued based on these findings that the origin of these boosts is the preference for clusters embedded within aligned filaments inherent in our cluster finder. Now, we test the validity of our argument. Our working hypothesis implies that (i) there should be clusters with such aligned filaments in our cluster sample, and (ii) such clusters would exhibit boosted signals in the two-halo regimes when compared to isotropic emulator predictions. In this section, we describe how we test for these implications and what our subsequent findings are.

\subsection{Projected Clusters and True Member Fractions}
\label{subsec:member}

We begin by defining a metric to identify clusters with aligned filaments. We do not explicitly identify filaments from our $N$-body simulations, but exploit the fact that filaments generally present themselves as a ``string'' of low-mass halos. This implies that clusters with aligned filaments will have member galaxies from lower-mass secondary halos in the projected cluster region in addition to those from the primary halo, while clusters without such filaments will largely consist of galaxies from a single primary halo. In other words, we may identify clusters with aligned filaments by measuring the contribution from secondary, line-of-sight projected halos to the total richness. We thus define the \textit{true member fraction} $f_\mathrm{true}$ as
\begin{equation}
\label{eq:frtrue}
f_{\rm true}\equiv
\frac{{\displaystyle \sum_{{i}\in {\rm primary}; R_i\le R_{\rm c}}}~ p_{{\rm mem},i}}{\lo}.
\end{equation}
For a given cluster, the sum in the numerator is taken only over the true member galaxies of the primary halo. In addition, we only include galaxies within the cluster boundary $R_{\rm c}$ in the sum (as opposed to the halo boundary $R_\mathrm{200m}$) to enforce $0\le f_{\rm true}\le 1$. This definition is comparable to the definition of $w^{\rm halo}$ given in Eq.~(6) of \citet{SunayamaMore}\footnote{However, \citet{SunayamaMore} used the virial radius $R_{\rm vir}$ for the definition of halo boundary, instead of $r_{\rm 200m}$ in our paper. We found that, due to the fact of $R_{\rm vir}<r_{\rm 200m}$ at the redshift of interest, the different boundary definitions of each cluster lead to differences in the fraction of true members of primary halo to the total richness.}.

The left panel of Fig.~\ref{fig:ftrue_frac} shows the average distribution of $f_\mathrm{true}$ in the entire observed cluster sample. While the majority of identified clusters show values of $f_\mathrm{true}$ close to unity, we note that there is a non-negligible tail of clusters showing high degrees of secondary halo contributions. Based on this distribution, we proceed to split the observed sample into two subsamples: a ``pure'' subsample with $f_{\rm true}>0.75$ and a ``projected'' subsample with $f_{\rm true}<0.75$. This results in a roughly 70--30 percentile split of the observed cluster sample into the pure and projected subsamples, respectively. The middle panel of Fig.~\ref{fig:ftrue_frac} shows the fraction of projected clusters with $f_{\rm true}<0.75$ in each richness bin. As expected, the contamination of projected clusters is greater for lower richness bins.

Finally, the right panel of Fig.~\ref{fig:ftrue_frac} shows the distribution of the line-of-sight separation of cluster members from their centers. Member galaxies in the pure sample are highly concentrated around the center, with their distribution truncating sharply after the halo virial radius of a few $h^{-1}$Mpc. Member galaxies in the projected sample, on the other hand, show a significant presence beyond the halo virial radius, extending all the way out to the maximum separation allowed by the line-of-sight window ($60h^{-1}{\rm Mpc}$). Further comparisons between the two subsamples, shown in Fig.~\ref{fig:ftrue_mass}, indicate that the projected subsample has lower primary halo masses and higher total number of included halos compared to the pure subsample, both of which are expected outcomes for contaminations from aligned filaments. These exploratory statistics suggest that $f_\mathrm{true}$ can be used to identify projected clusters, and more importantly that projected clusters do exist in our observed cluster sample, satisfying the first requirement we have given ourselves above.

\subsection{Boosted Signals from Projected Clusters}

\begin{figure*}
    \centering
    \includegraphics[width=0.45\textwidth]{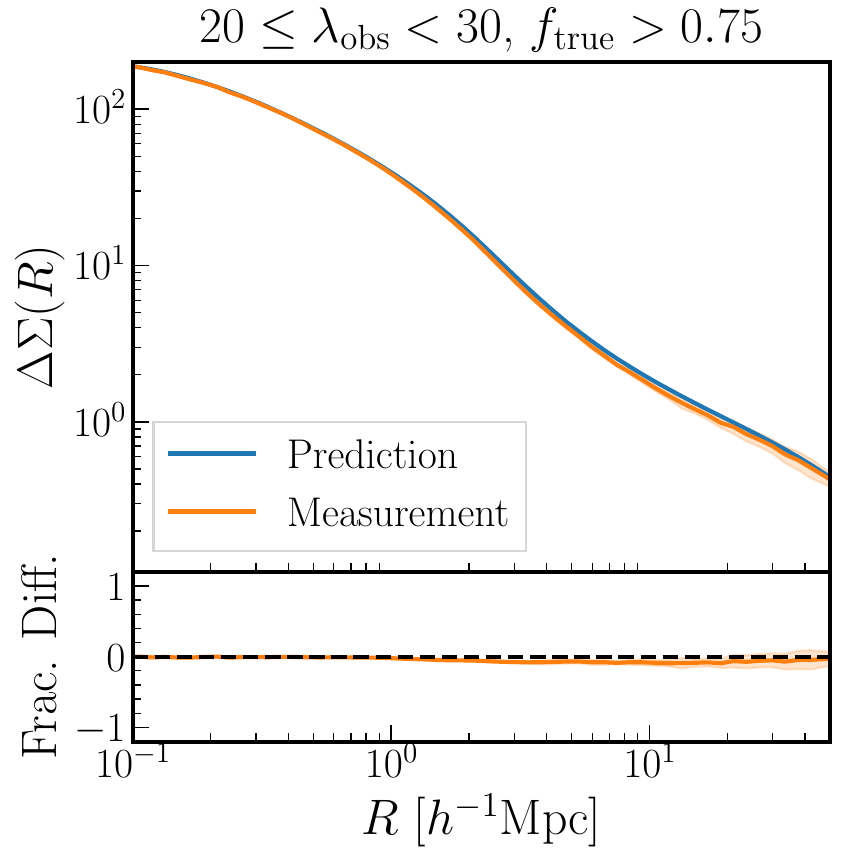} 
    \includegraphics[width=0.45\textwidth]{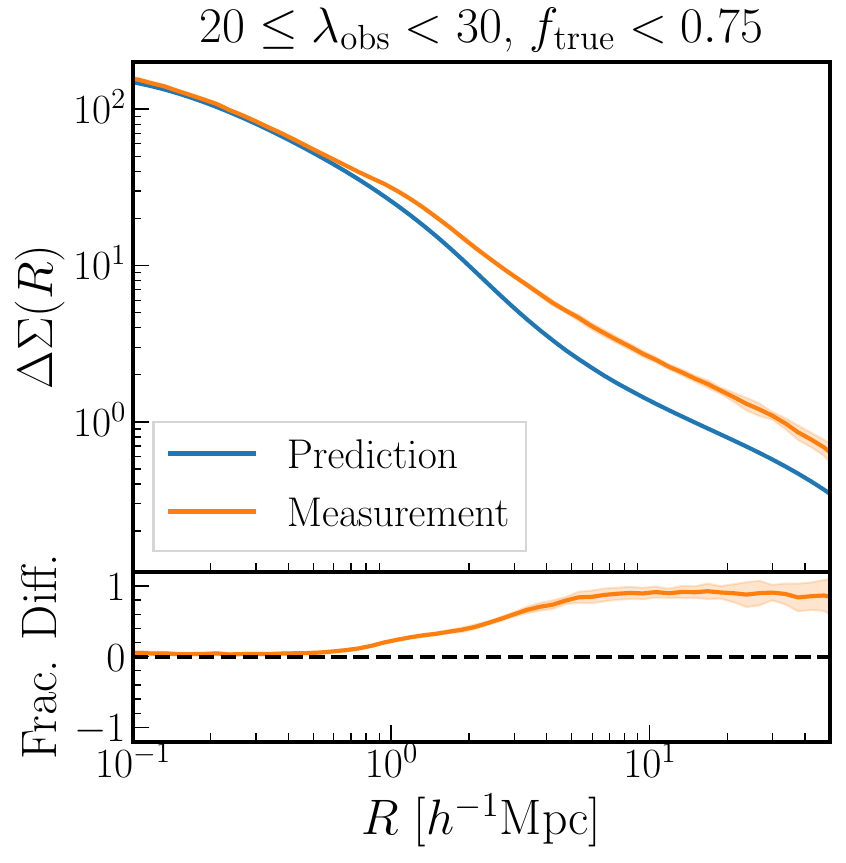}
    \caption{\label{fig:ftrue_lensing} {\it Left panel}: 
    Measured lensing profiles of the clusters with $f_{\rm true}>0.75$ compared to predictions from the emulator. The results agree well on the one-halo term as well as on the two-halo term. 
    {\it Right}: 
    The same as the left figure, but for the clusters with $f_{\rm true}<0.75$. The lensing profile on the one-halo term agrees with the prediction. However, the amplitude on the two-halo term is larger than the prediction by up to a factor of 1.8. Note that the results here are for the clusters with $20<\lambda<30$.}
\end{figure*}
\begin{figure*}
    \centering
    \includegraphics[width=0.32\textwidth]{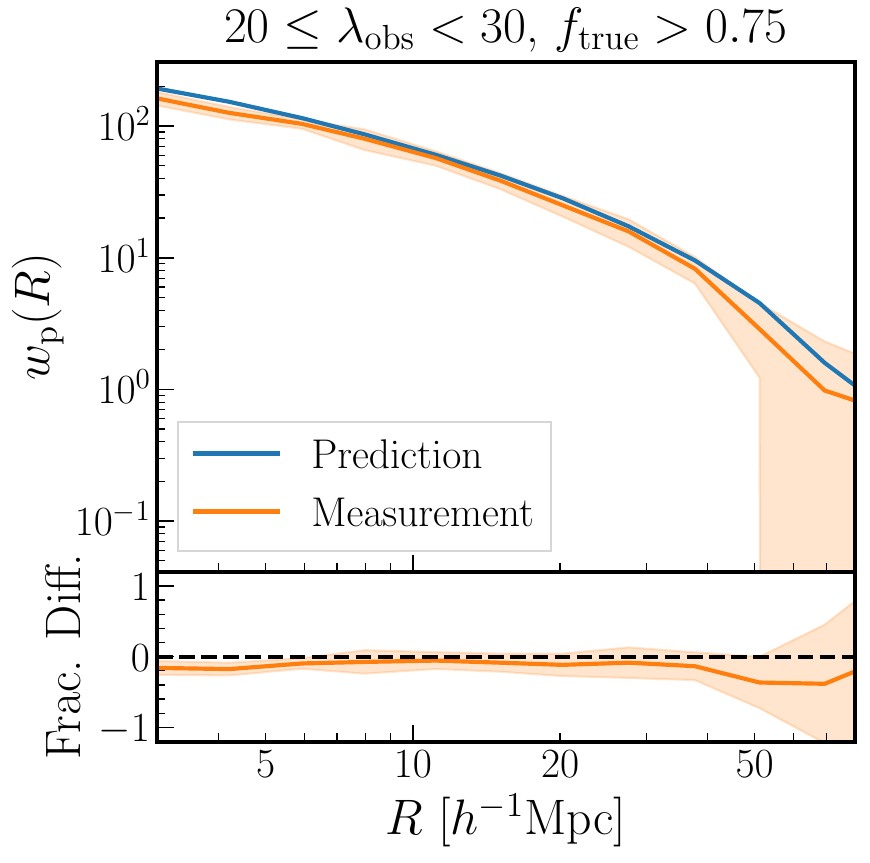}     \includegraphics[width=0.32\textwidth]{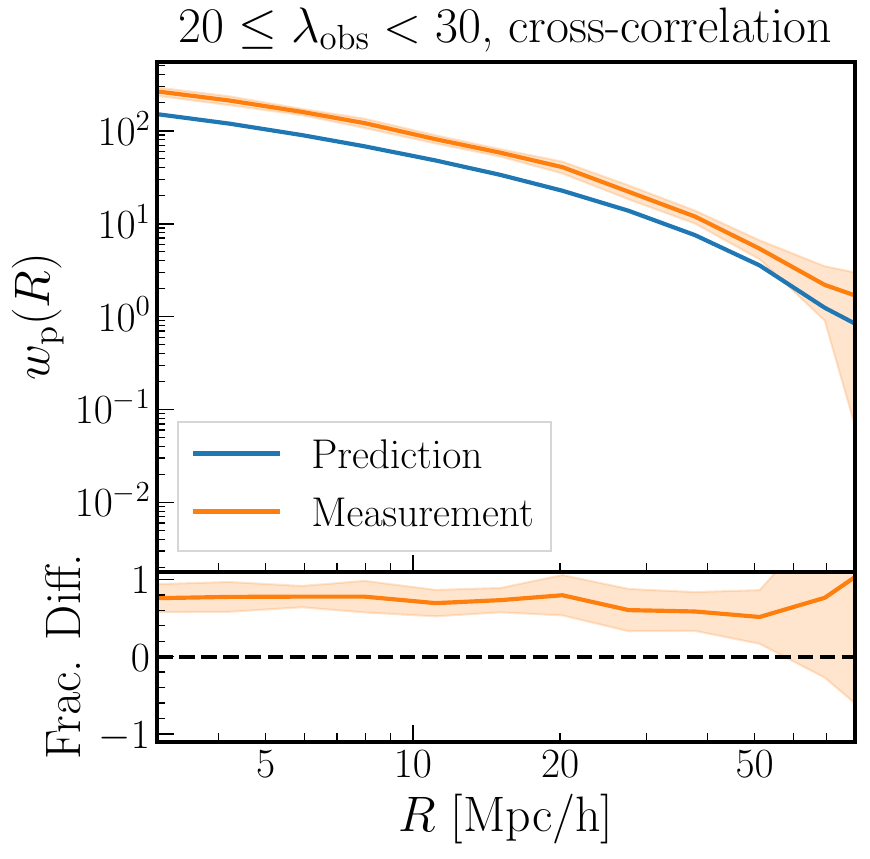} 
    \includegraphics[width=0.32\textwidth]{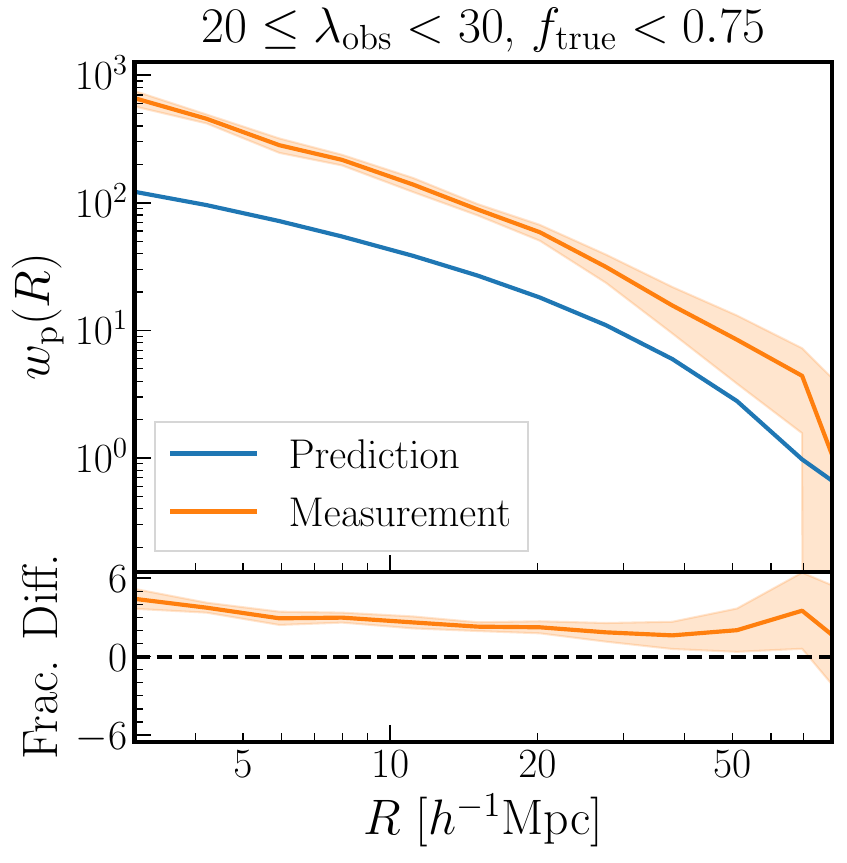}
    \caption{\label{fig:ftrue_auto} {\it Left}: Auto-correlation functions of the clusters with $f_{\rm true}>0.75$ compared to the prediction from the emulator. The results agree fairly well with each other. Note that the auto-correlation function is always in the two-halo regime by definition.
    {\it Middle}: The same as the left panel, but for the cross-correlation between pure clusters ($f_{\rm true}>0.75$) and projected clusters ($f_{\rm true}<0.75$). The amount of boost is similar to that of the boost in the two-halo term regime of lensing profile in Fig.~\ref{fig:ftrue_lensing}.
    {\it Right}: Same comparison for clusters with $f_{\rm true}<0.75$. 
    The amplitude is larger than the prediction by a factor of 3, which is roughly the square of 1.8, i.e. the boost observed from the cross-correlation in the middle panel.}
\end{figure*}

With our subsamples defined and the projected clusters identified, we proceed to test whether the boosted signals indeed arise from the projected subsample. In Fig.~\ref{fig:ftrue_lensing}, we show comparisons between the emulator predictions and the measurements from the mock catalogs for the lensing profiles of each subsample. The predictions and measurements are in relatively good agreement for the pure subsample, although the measurements show a slight suppression on large scales. On the other hand, for the projected sample, the measurements show significant boosts up to a factor of 1.8 when compared against the emulator predictions. 

In Fig.~\ref{fig:ftrue_auto}, we show similar comparisons for cluster clustering, this time also including the cross-correlation between the pure and projected subsamples. The pure subsample again agrees well with emulator predictions, while the 
pure-projected cross-correlation function and the auto-correlation function for the projected subsample shows boosts up to factors of 1.8 and 3, respectively.

\begin{figure*}
    \centering
    \includegraphics[width=0.32\textwidth]{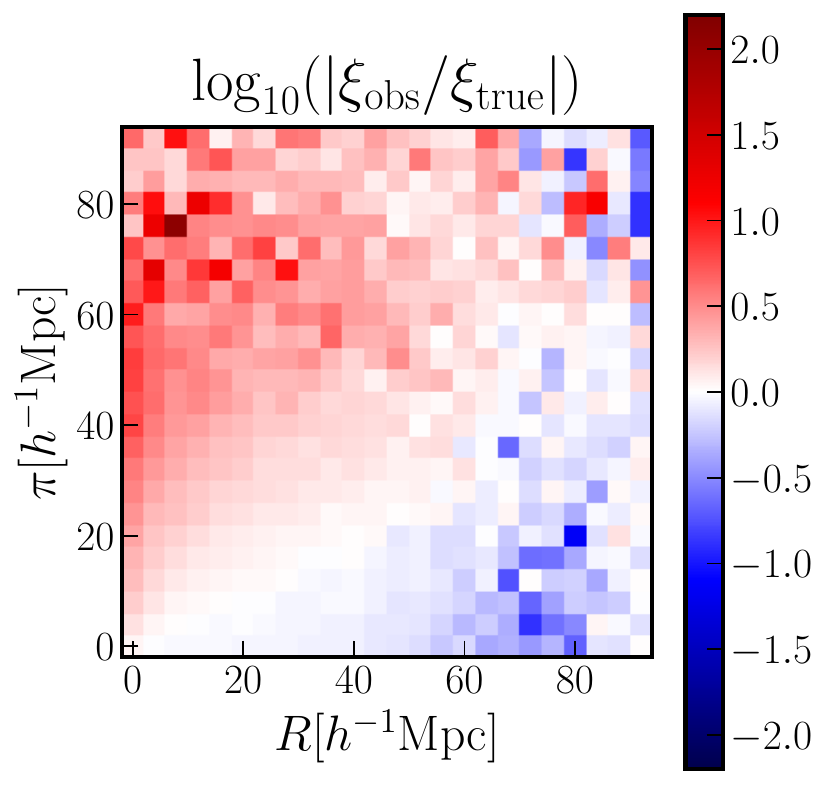}
    \includegraphics[width=0.32\textwidth]{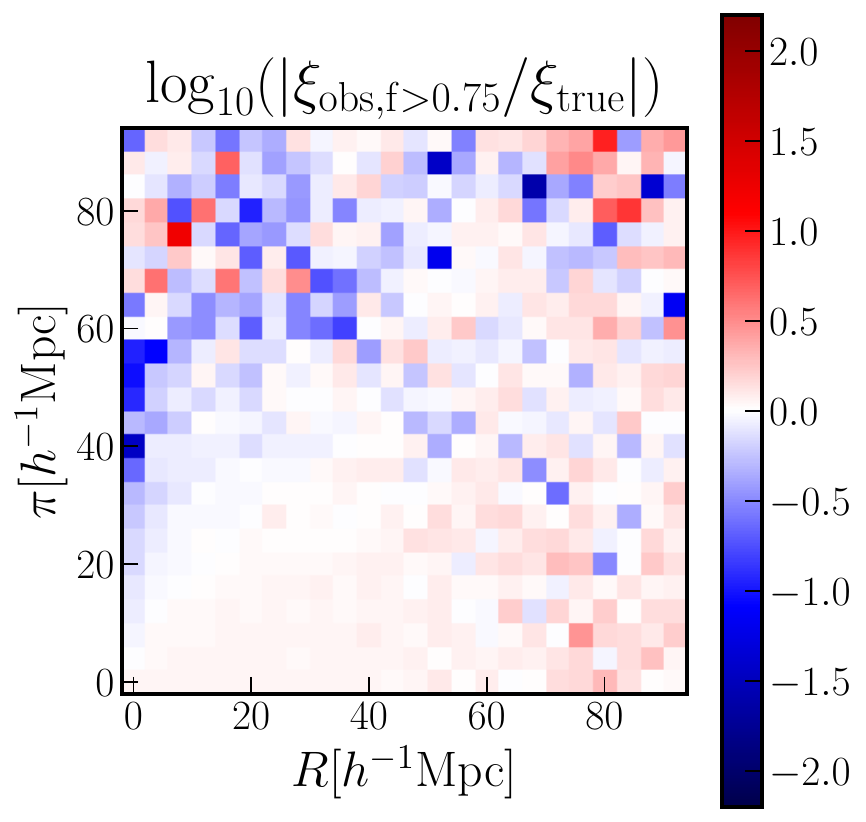}
    \includegraphics[width=0.32\textwidth]{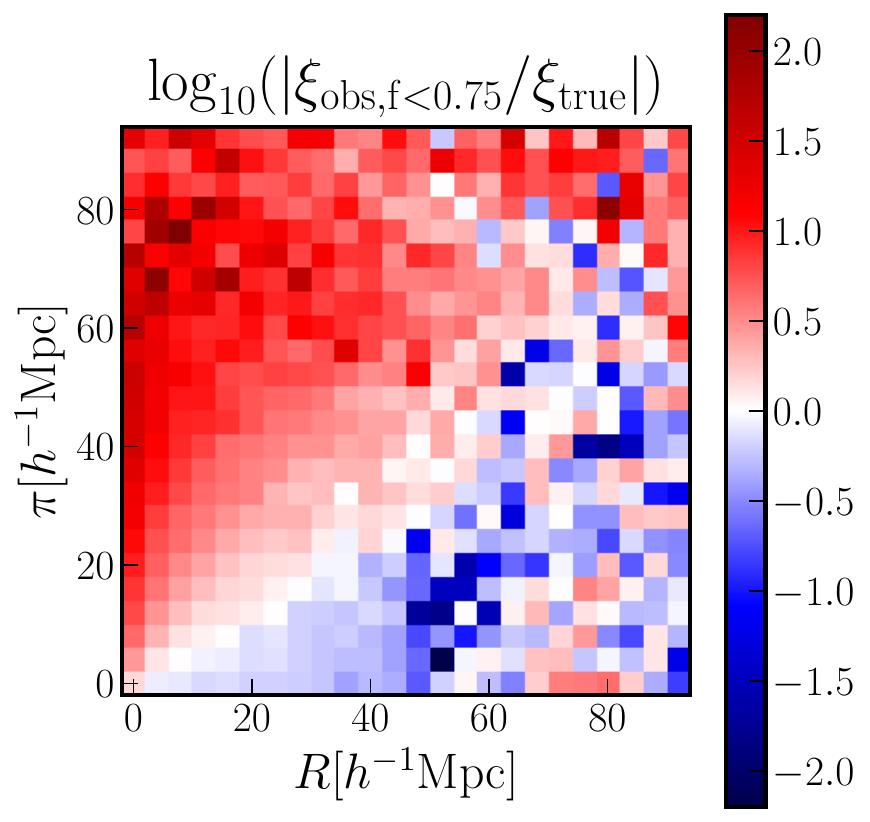}
    \caption{\label{fig:auto2D_ftrue}\textit{Left:} Ratio of 2D auto-correlation functions between the observed and the true samples, respectively for $\lt\ge20$ and $\lo\ge 20$. Note that one cluster is at the origin while the location of the other cluster is a specified by $R$ and $\pi$, respectively the separations perpendicular and parallel to the line-of-sight. \textit{Middle:} Same comparison, but between the pure ($f_\mathrm{true}>0.75$) subsample and the true sample. \textit{Right:} Again, same comparison, but between the projected ($f_\mathrm{true}<0.75$) subsample and the true sample.}
\end{figure*}

These findings clearly indicate that clusters containing large numbers of secondary halos projected along the line-of-sight show significant boosts in their measured lensing and clustering signals on large scales, satisfying our second requirement above. Combined with our conclusion from the previous section, this completes our description for the origin of the observed projection effects: cluster finders preferentially select clusters embedded within aligned filaments, and these filaments give rise to the boosted large-scale lensing and clustering signals from aligned filaments. We find further support for this hypothesis from studying the anisotropic 2D cluster clustering signal in Fig.~\ref{fig:auto2D_ftrue}, where we respectively show the ratios of the 2D clustering signal for the observed, pure, and projected samples to that for the true sample. While the pure subsample shows little anisotropy, the observed and the projected samples display a significant quadrupole pattern in the 2D correlation function, similar to Figs.~6 and 7 in \citet{Osato_etal2018} \citep[also see][for discussion on the impact of large-scale tidal field on small-scale structures]{2017PhRvD..95h3522A,2018PhRvD..97f3527A}. Note that the line-of-sight projection of $\xi(R, \pi)$ along $\pi$ gives the projected correlation functions we have studied so far. This explains the $\pi_{\rm max}$ dependence of the observed boosts in the cluster clustering signal shown in Fig.~\ref{fig:auto1}.

For comprehensiveness of our discussion, we show in Appendix~\ref{app:shape} how our results change if we include the effects of halo shapes. Also, in Appendix~\ref{app:all_richness}, we show the comparisons between measurements and emulator predictions, i.e. results similar to those presented in Figs.~\ref{fig:lensing_pred1}, \ref{fig:auto_pred1}, \ref{fig:ftrue_lensing} and \ref{fig:ftrue_auto}, but for all richness bins.

\subsection{Implications}

Our findings have immediate implications for several previous studies with potential contaminations from projection effects. First, our findings could explain the apparent assembly bias signal discussed in \citet{Miyatake:2016}. For a given $\lo$, pure clusters would largely consist of a single massive primary halo while projected clusters would likely contain multiple lower-mass halos aligned with the line-of sight. This implies that projected clusters likely have a higher concentration of member galaxies than pure clusters, and consequently boosted lensing/clustering signals on large scales that \citet{Miyatake:2016} observed \citep[also see][for the similar discussion]{SunayamaMore}. 
In addition, the boosts we observe amplify the two-halo term while leaving intact the one-halo term. This means that the location of the splashback radius, roughly corresponding to the transition between the one- and the two-halo regimes, would be pushed to smaller scales. This could explain the smaller-than-expected splashback radii of SDSS redMaPPer clusters found in \citet{More:2016} \citep[also see][]{SunayamaMore,Murataetal:20}.

On a different note, the distinct radial behavior of the observed boosts, i.e. their tendency for asymptoting to a constant factor on large scales, may help mitigate their impact on cluster cosmology. This is particuarly interesting because the boost in the lensing signal and the pure-projected cross-correlation clustering signal are roughly equal, at a factor of 1.8, while the boost in the auto-correlation of the projected sample (factor of 3) is roughly square of that. This motivates a simple modeling approach where we introduce a multiplicative factor to the large-scale lensing and clustering signals of the projected sample, i.e. 
\begin{eqnarray}
\Delta\!\Sigma_\mathrm{proj} & = & (1+\alpha) \Delta\!\Sigma_\mathrm{pure},\\
w_\mathrm{p,proj-pure} & = &  (1+\alpha) w_\mathrm{p,pure}, \\
w_\mathrm{p,proj} & = &  (1+\alpha)^2 w_\mathrm{p,pure}.
\end{eqnarray}
This ansatz is equivalent to introducing a multiplicative factor to the large-scale bias in the halo-matter correlation function around projected clusters. Propagating this model to the net observed signals, we obtain
\begin{eqnarray}
\Delta\!\Sigma_\mathrm{obs} & = & 
(1-f_\mathrm{proj})\Delta\!\Sigma_\mathrm{pure} + f_\mathrm{proj}\Delta\!\Sigma_\mathrm{proj} \nonumber \\
& = & (1+\alpha f_\mathrm{proj}) \Delta\!\Sigma_\mathrm{pure}
\label{eq:dsfac}
\end{eqnarray}
for the lensing signal, and
\begin{eqnarray}
w_\mathrm{p,obs} & = & 
(1-f_\mathrm{proj})^2 w_\mathrm{pure} 
+ 2f_\mathrm{proj}(1-f_\mathrm{proj}) w_\mathrm{pure-proj}\nonumber \\
& & + f_\mathrm{proj}^2 w_\mathrm{proj,obs} \nonumber \\ 
& = & (1+2\alpha f_\mathrm{proj} + \alpha^2f_\mathrm{proj}^2) w_\mathrm{pure,obs}
\label{eq:wpfac}
\end{eqnarray}
for the clustering signal. If we consider the $\lambda \in [20,30)$ bin and use $f_\mathrm{proj} = 0.33$ (the middle panel of 
Fig.~\ref{fig:ftrue_frac}), $\alpha \simeq 0.6$ reproduces the factors of 1.2 and 1.4 we observed in Figs.~\ref{fig:lensing1} and~\ref{fig:auto1}, i.e. the large-scale boosts in the lensing and clustering signals of the observed cluster sample relative to the true sample, assuming $\Delta\!\Sigma_\mathrm{true} \simeq \Delta\!\Sigma_\mathrm{pure}$ and $w_\mathrm{p,true} \simeq w_\mathrm{p,pure}$. \yp{In addition, noting that both Eqs.~\ref{eq:dsfac} and \ref{eq:wpfac} only exhibit the multiplicative combination $\alpha f_\mathrm{proj}$ in their expressions, we can also consolidate them as
\begin{eqnarray}
\Delta\!\Sigma_\mathrm{obs} & = & (1+f)\Delta\!\Sigma_\mathrm{pure} \\
w_\mathrm{p,obs} & = & (1+f)^2 w_\mathrm{p,pure},
\end{eqnarray}
where $f=\alpha f_\mathrm{proj}$, resulting in a one-parameter model for the observed large-scale boosts. 

While these discussions are only exploratory, and constructing a concrete systematics model for projection effects is beyond the scope of this paper, they do offer encouraging prospects for an eventual mitigation strategy based on simple parametrizations inspired by physical arguments. We plan to further pursue these possibilities with follow-up papers in the near future.}

\section{Summary and Discussion}
\label{sec:summary}

In this paper, we have investigated the origin and the extent of projection effects on the cluster observables of interest for cosmological analyses. We summarize our conclusions as follows:
\begin{itemize}
    \item The richness-mass relations of observed cluster samples show complex shifts and scatters arising from two main sources: projection effects and finite aperture effects. This is in agreement with previous findings from e.g. \citet{costanzietalprojection}.
    \item However, even when the changes in the underlying richness-mass relation are fully taken into account, the measurements from the observed cluster samples show a significant boost in both the cluster lensing and the cluster clustering observables on large scales, respectively up to $20\%$ and $40\%$, when compared to emulator predictions. 
    \item This suggests that changes in the richness-mass relation cannot fully explain the observed projection effects. We hypothesize that the observed boosts are caused by the preferential selection of certain anisotropic matter distributions around clusters, i.e. aligned filaments, by our cluster finder.
    \item We find support for the aligned filaments hypothesis from our analysis of projected clusters with $f_\mathrm{true}<0.75$. We find that these clusters, with high degrees of richness contribution from secondary halos along the line-of-sight, exhibit 
    significant
    large-scale boosts in the lensing and clustering signals. While they constitute only about $30\%$ of the entire observed sample of clusters, but exhibit large-scale boosts up to factors of 1.8 and 3, respectively for the lensing and clustering observables, strong enough to significantly drive up the mean signal for the overall sample.
\end{itemize}

Our findings have a number of significant implications for the modeling of projection effects in cluster cosmology analyses. First, we find a strong connection between the observed impact of projection effects on cluster observables and an underlying physical phenomenon, namely that of anisotropic large-scale structure around optically selected clusters. \yp{This implies that we may be able to restrict the freedom in future mitigation schemes for projection effects based on these physical arguments.} The overall impact of projection effects, of course, is tightly coupled to the selection properties of the cluster finder being used, but modeling the underlying physical effect that couples to the selection properties will lead to a more accurate characterization of projection effects that can retain more cosmological information. For instance, we presented a rough empirical approach to model the impact of projection effects at the end of Section~\ref{sec:signal2}, which could be used to identify projection effects and extract unbiased cosmological constraints in optical cluster samples, especially with the combined information from cluster lensing and cluster clustering. Also, cross-correlations of cluster samples with neighboring spectroscopic galaxies from e.g. SDSS/BOSS data could also be used to mitigate projection effects \citep{2019MNRAS.tmp.2563S}. Finally, with a better understanding of the physical nature of the aligned filaments surrounding projected clusters, we may be able to physically model the impact of projection effects with minimal amounts of additional parameters involved. We plan to make further studies on these possibilities immediately following this paper.

There are several simplifications we employed in this paper. We employed a sharp truncation in the line-of-sight window to identify clusters from the mock galaxy catalogs, and in reality a more complex line-of-sight kernel arising from the photometric redshift uncertainties of cluster member galaxies must be taken into account. We have also ignored the redshift evolution of cluster observables, and further studies on the redshift dependence of large-scale boosts would give us a better handle for distinguishing between projection effects and cosmological signals. In addition, as we are able to describe in some detail the properties of the clusters mainly responsible for giving rise to the observed projection effects, we can begin formulating strategies for ``cleaning'' optical cluster samples with e.g. follow-up observations in X-ray or SZ to mitigate the impact of projection effects. These are important and exciting research questions for the ultimate goal of carrying out precision cosmology with clusters, and we plan to explore them in the near future.

\section*{Acknowledgements}
We would like to thank Eduardo Rozo for useful discussion. 
This work was supported in part by World Premier International Research Center Initiative (WPI Initiative), MEXT, Japan, JSPS KAKENHI Grant Numbers JP15H03654, JP15H05887, JP15H05892, JP15H05893, JP15H05896, JP15K21733, JP17K14273, JP18K03693, and JP19H00677, and Japan Science and Technology Agency CREST JPMHCR1414. KO is supported by JSPS Overseas Research Fellowships. Numerical simulations presented in this paper were carried out on Cray XC30 and XC50 at Centre for Computational Astrophysics, National Astronomical Observatory of Japan.

\appendix

\section{The effect of halo shapes on the projection effect}
\label{app:shape}

For our fiducial mock catalogs of galaxies, and thus for the results presented above, we ignored the effects of halo shapes/triaxiality by populating galaxies into halos with a spherically symmetric NFW profile (see Section~\ref{subsec:mocks}). In this Appendix, we discuss how our results correlate with halo shapes and their orientations, as well as how our results shift if we take halo shapes into account. 

We first examine the indirect correlation between halo shape orientations and projection effects or equivalently aligned filaments we discussed in Sec.\ref{sec:interpretation} by studying the shape orientations of halos in our default, i.e. \textit{shape-agnostic}, catalogs. To measure shape orientations, we first compute the inertia tensor of the dark matter distribution for each halo from its member particles in the simulations as
\begin{align}
I_{ij}&=\sum_{p=1}^{N} 
r_{p,i}r_{p,j} , 
\label{eq:inertia}
\end{align}
where $r_p$ is the separation of the $p$-th member particle from the halo center in ellipsoidal coordinates following the method in \citet{Osato_etal2018}, and $i,j=1,2$ or 3 corresponds to each axis in the coordinate system.
The eigenvectors of the inertia tensor correspond to the directions of the principal axes of the particle distribution within each halo. We define halo orientation as the direction of the major axis.

\begin{figure*}
 \centering
   \includegraphics[width=0.4\textwidth]{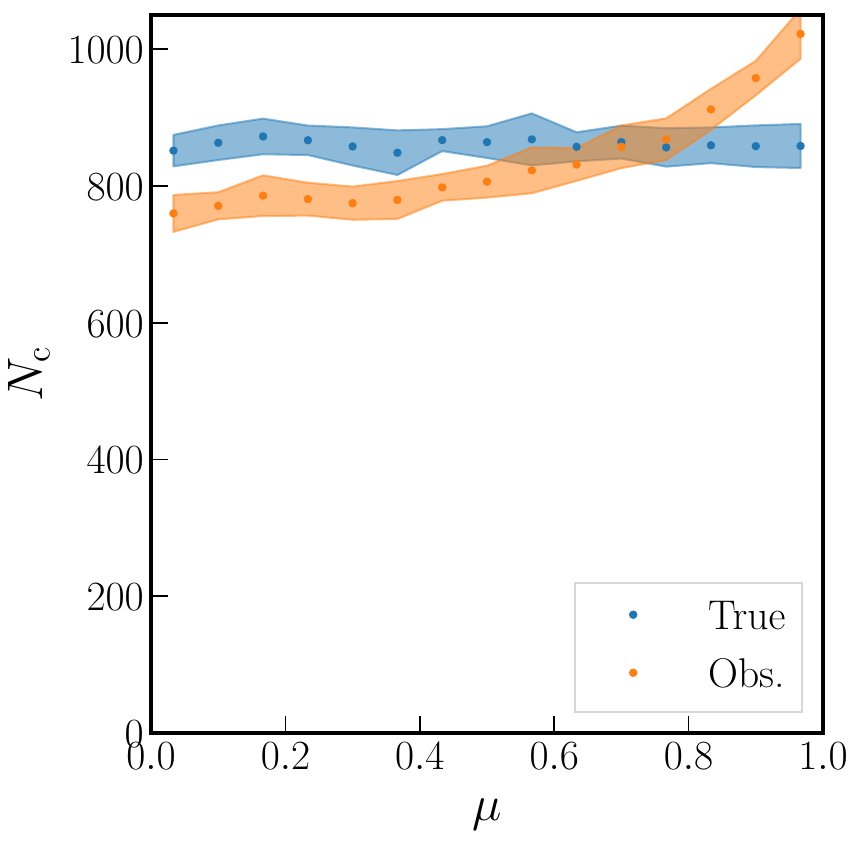}\includegraphics[width=0.4\textwidth]{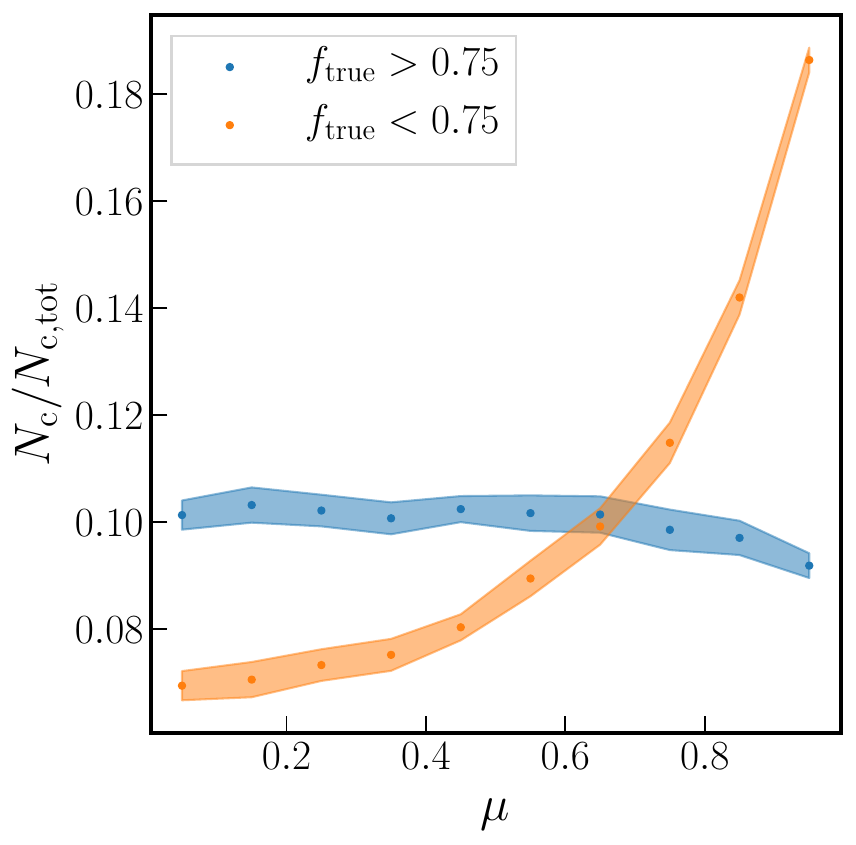}
  \caption{{\it Left}: Distributions of $\mu={\rm cos} i$, where $i$ is the angle between the major axis of a halo and the line-of-sight, for the true/observed cluster samples. While halo orientations in the true sample are randomly distributed, clusters oriented along the line-of-sight, i.e. those with large $\mu$, are preferentially selected by the cluster finder for the observed sample. We plot results for all clusters with $\lambda_{\rm true/obs}>20$. {\it Right}: Similar to the left panel, but for the pure/projected subsamples. The orientations are again randomly distributed in the pure subsample, but they show a strong selection bias in the projected subsample, indicating that the overall selection bias in the entire observed sample is mainly driven by the projected subsample.}
     \label{fig:orientation_hist}
\end{figure*}

The left panel of Fig.~\ref{fig:orientation_hist} shows the distribution of halo orientation for the true/observed samples from our fiducial catalog. Note that we only use the primary halo to define the halo orientation for each cluster. There is no orientation dependence for the clusters in the true sample, while clusters elongated along the line-of-sight are preferentially selected in the observed sample. Since these results come from our fiducial, shape-agnostic catalogs, they imply that there indeed is an indirect correlation between projection effects and halo orientations, likely arising from the existence of aligned filaments. The right panel similarly shows the distribution of halo orientation for the pure/projected subsamples. The clusters in the pure subsample show little dependence on halo orientation, and most of the preferential selection of clusters elongated along the line-of-sight happens in the projected subsample. Overall, we find that our results are consistent with the picture that projected clusters are strongly correlated with aligned filaments.

\begin{figure*}
    \includegraphics[width=0.4\textwidth]{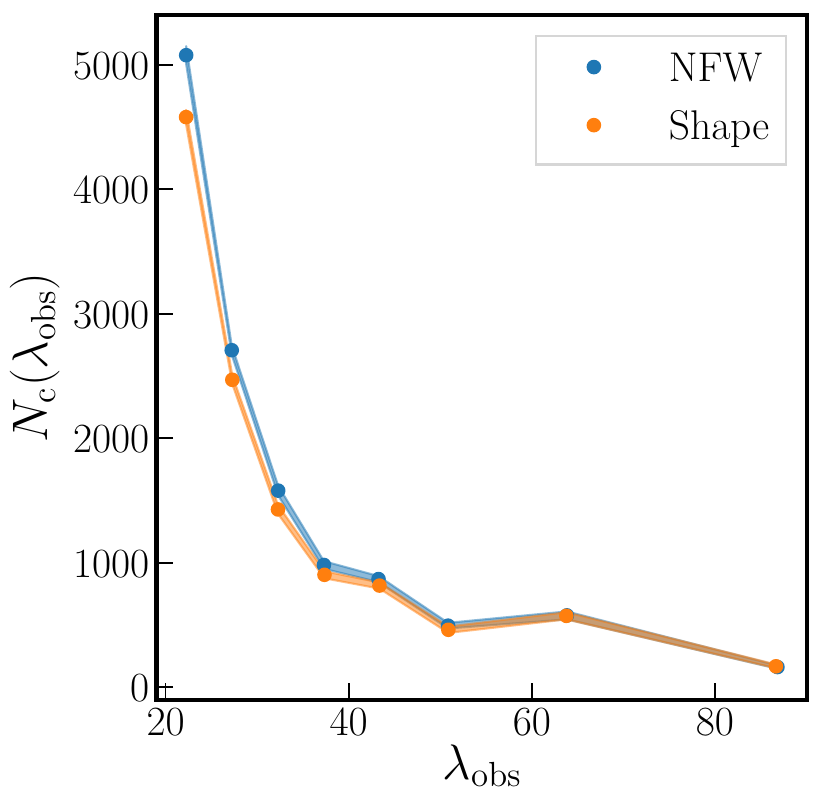}\includegraphics[width=0.4\textwidth]{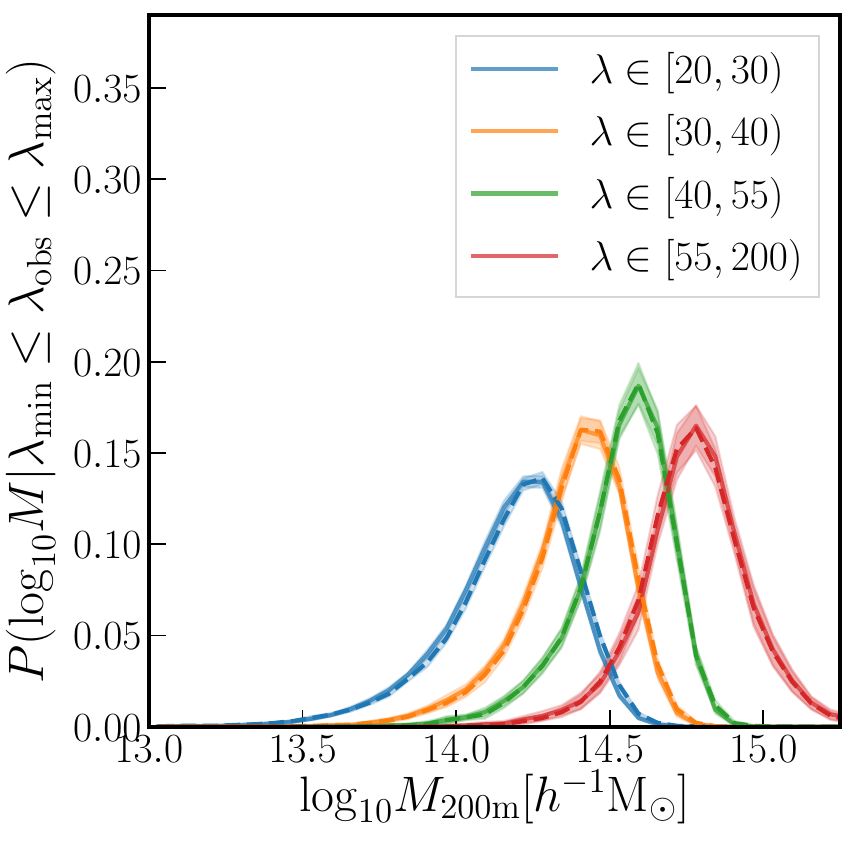}
    \caption{\label{fig:abundance_shape}
    {\it Left}: The comparison of cluster abundances as a function of richness between the ``NFW'' and the ``shape'' catalogs. We populate galaxies into halos either based on NFW profiles (blue: labeled as ``NFW'') or by using member particles of each halo (orange: labeled as ``shape''). Clusters are separately identified by our cluster finder in the two different mock galaxy catalogs. {\it Right}: The mass distributions of primary halos for clusters identified in the two different galaxy catalogs. 
    The solid line shows the mass distribution of the ``NFW'' cluster sample, while the dashed line is for the ``shape'' sample. Different colors correspond to the different richness bins. We observe that the underlying mass distributions are almost identical between the two catalogs.}
\end{figure*}
\begin{figure}
    \centering
    \includegraphics[width=0.4\textwidth]{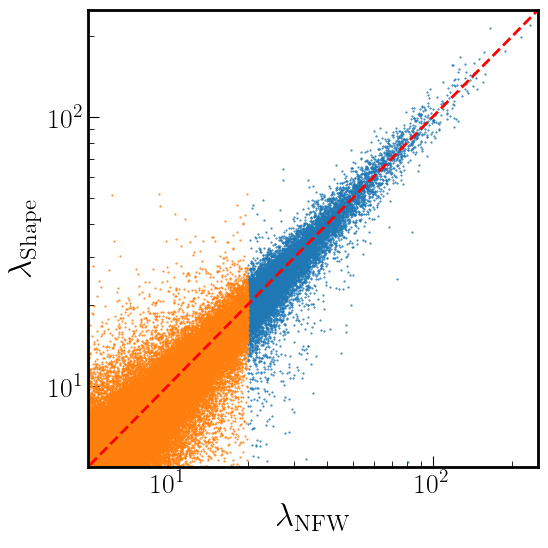}
    \caption{\label{fig:richness_shape} The relationship between the observed richnesses obtained from the ``NFW'' catalog ($\lambda_\mathrm{NFW}$) and the ``shape'' catalog ($\lambda_\mathrm{shape}$) for clusters identified in both catalogs.}
\end{figure}
\begin{figure}
    \centering
    \includegraphics[width=0.4\textwidth]{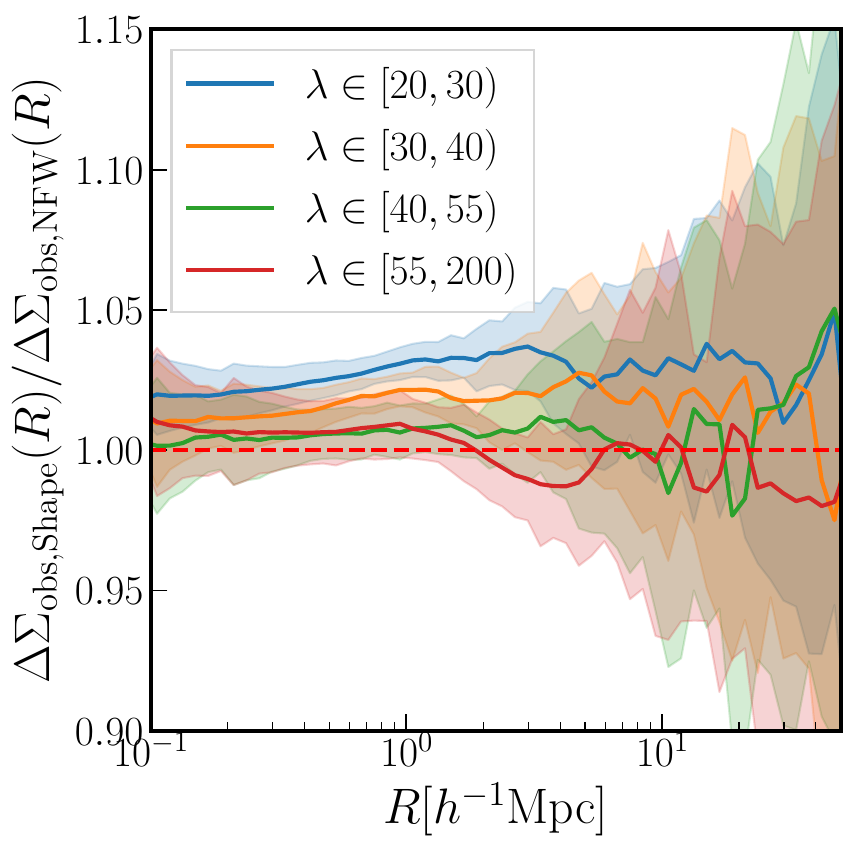}

    \caption{\label{fig:cross_shape} Ratios of lensing profiles for observed clusters, between the ``NFW'' and the ``shape'' catalogs. The ratios are consistent with unity for all richness bins, implying that the clusters identified in the ``shape'' catalog exhibit similar characteristics as those in the default ``NFW'' catalog. 
    }
\end{figure}
\begin{figure}
    \includegraphics[width=0.4\textwidth]{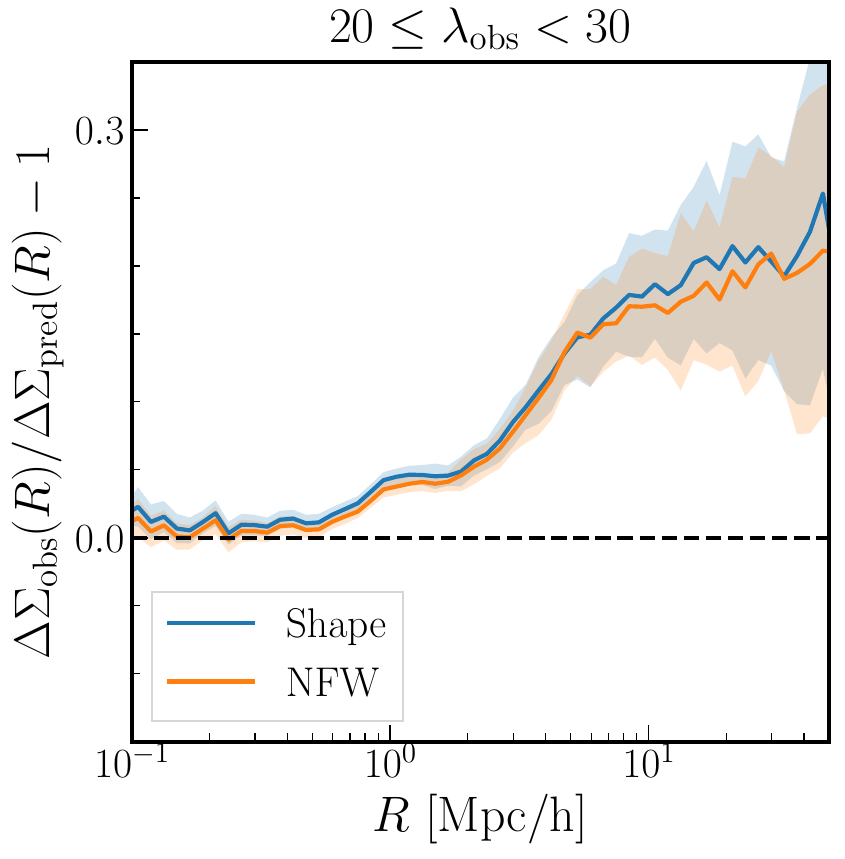}
    \caption{\label{fig:shape_nfw_comp} 
    The comparison of fractional differences between measurements and emulator predictions for the observed samples from ``NFW'' (orange) and ``shape'' (blue) catalogs. The ``NFW'' result shows the default large-scale boost in lensing that we have discussed above, and the ``shape'' result shows very similar trends, implying that the inclusion of halo shapes in the mock galaxy catalog generation process does not significantly impact the extent of projection effects. 
    }
\end{figure}

Next, we study the impact of halo shapes on cluster identification. To this end, we build a new set of galaxy mock catalogs, taking into account the shape of each halo in the simulations. Namely, instead of using an NFW profile, we randomly select dark matter particles in each halo to determine the positions of mock galaxies when populating them. Note that we maintain the same number of populated galaxies for each halo between the new (``shape'') and the default (``NFW'') catalogs. We then run the cluster finder on the new catalog to identify a new set of observed cluster samples.

Fig.~\ref{fig:abundance_shape} shows the abundances and the richness-mass relations for clusters identified from the two different mock catalogs. Cluster abundances in lower richness bins are smaller for the ``shape'' mock catalogs; while halos oriented along the line-of-sight would retain all member galaxies, halos oriented perpendicular to the line-of-sight would lose a portion of their members as they may lie outside of $R_c(\lambda)$. This implies that halo shapes tend to cause underestimations of richnesses. However, despite this effect, the underlying mass distributions are almost identical between the ``NFW'' and the ``shape'' catalogs.
Fig.~\ref{fig:richness_shape} compares the richness for identical halos between the ``NFW'' and ``shape'' mock catalogs. At lower richness, more clusters in the ``shape'' mock catalogs tend to have smaller richness values than the clusters in the ``NFW'' mock catalogs. This means that a larger number of clusters with $\lambda_{\rm NFW}=20$ have $\lambda_{\rm Shape}<20$ and are excluded from the cluster catalog than vice versa, which explains why the abundance at low richness bin for the ``shape'' mock catalog is smaller.

Finally, in Fig.~\ref{fig:cross_shape}, we compare the lensing profiles at each richness bin between the ``NFW'' and 
the ``shape'' catalogs. Note that the observed samples for the two catalogs are not exactly the same, as the galaxy distribution in each catalog is different and thus the cluster finder identifies slightly different samples of clusters between the two catalogs. Nevertheless, the ratios are consistent with unity, implying that the lensing profiles are very similar between the two catalogs. We identify the subtle differences as being caused by slight differences in the underlying halo mass distributions, as indicated by the right panel of Fig.~\ref{fig:abundance_shape}. Fig.~\ref{fig:shape_nfw_comp} compares the measurements with the emulator predictions for the observed sample of NFW/shaped catalogs. The levels of the boost on the large scales are almost the same for both catalogs. This implies that the difference shown in Fig.~\ref{fig:cross_shape} is due to the difference in cluster mass of the samples.

In summary, we conclude that the shape/triaxiality of halos, or their orientations, does not play a major role in the observed projection effects, but rather should be considered as correlated quantities that share the same origin of aligned filaments as projection effects.

\section{Results for All Richness Bins}
\label{app:all_richness}

\begin{figure*}
    \centering
    \includegraphics[width=0.24\textwidth]{meas_pred_true_20to30.pdf}
    \includegraphics[width=0.24\textwidth]{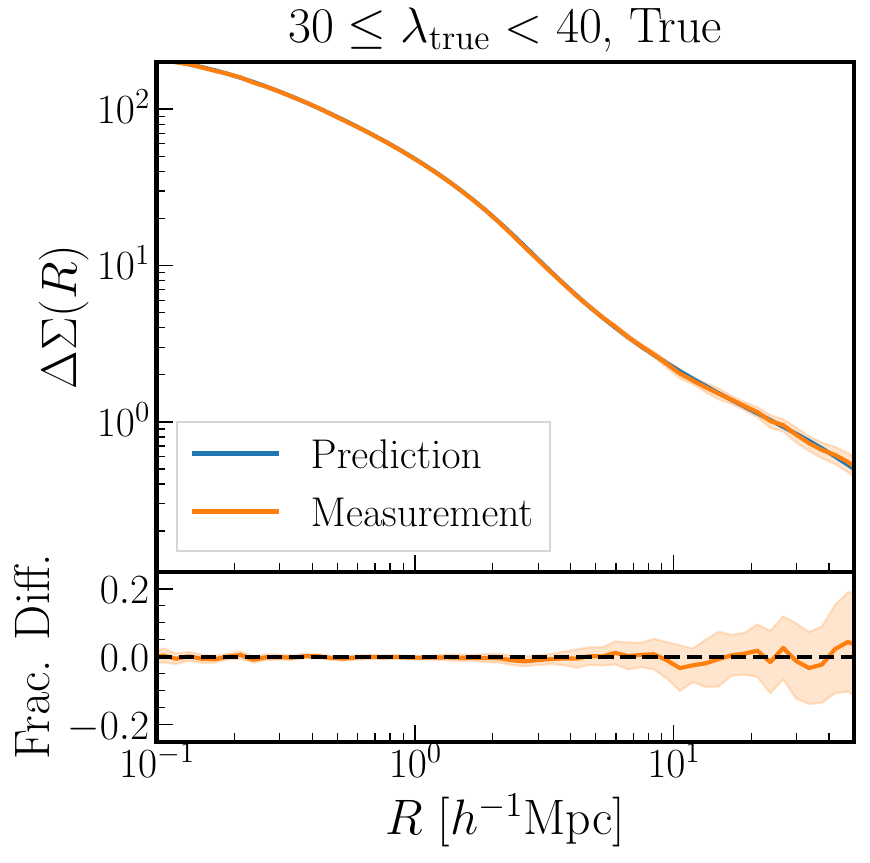}
    \includegraphics[width=0.24\textwidth]{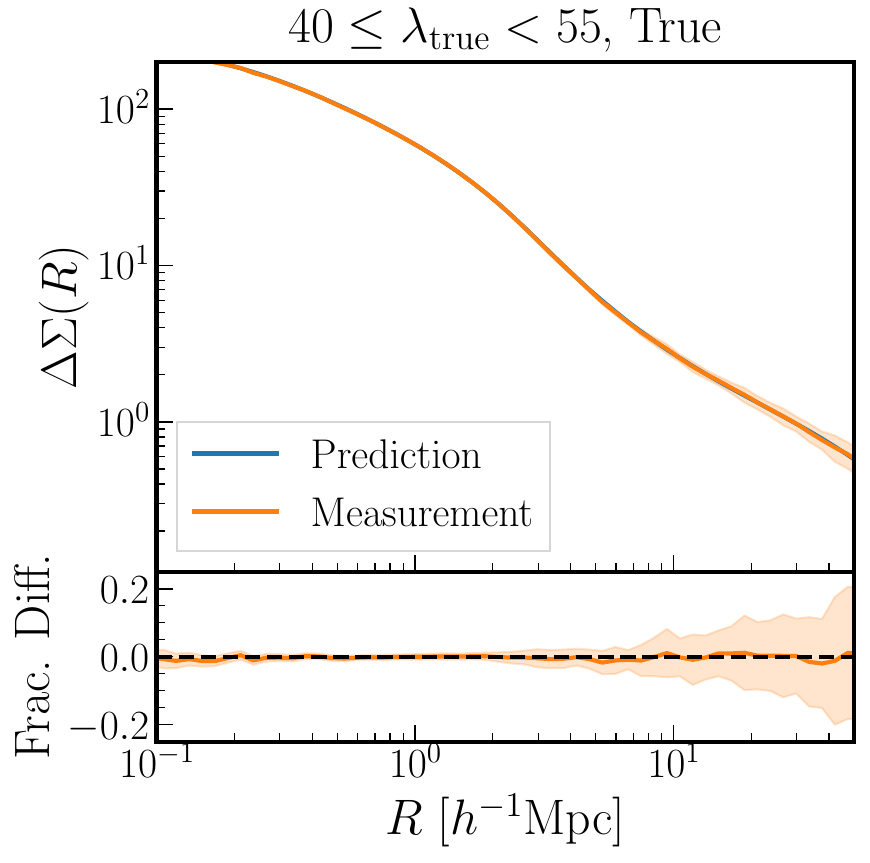}
    \includegraphics[width=0.24\textwidth]{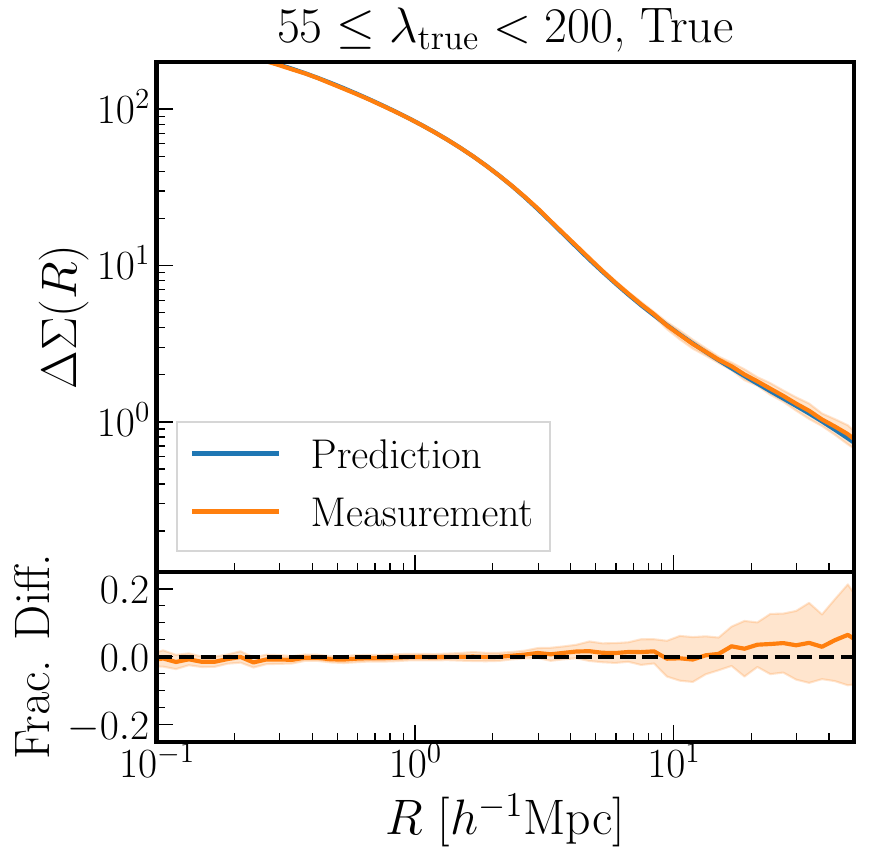}
    
    \vspace{0.5cm}
    
    \includegraphics[width=0.24\textwidth]{meas_pred_obs_20to30.pdf}
    \includegraphics[width=0.24\textwidth]{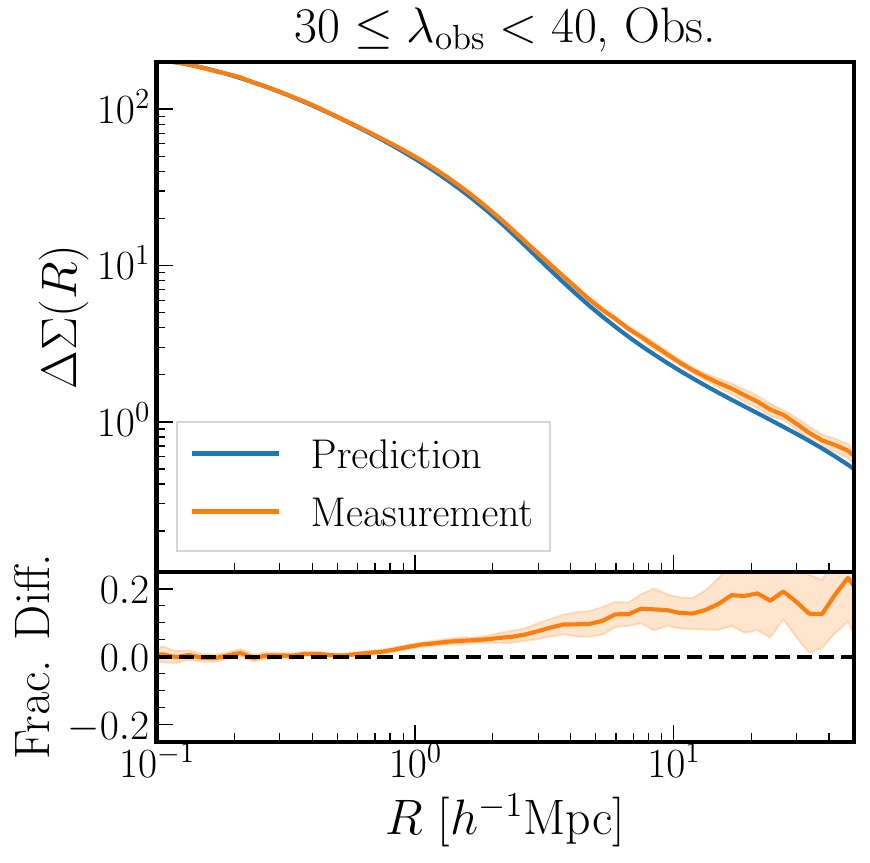}
    \includegraphics[width=0.24\textwidth]{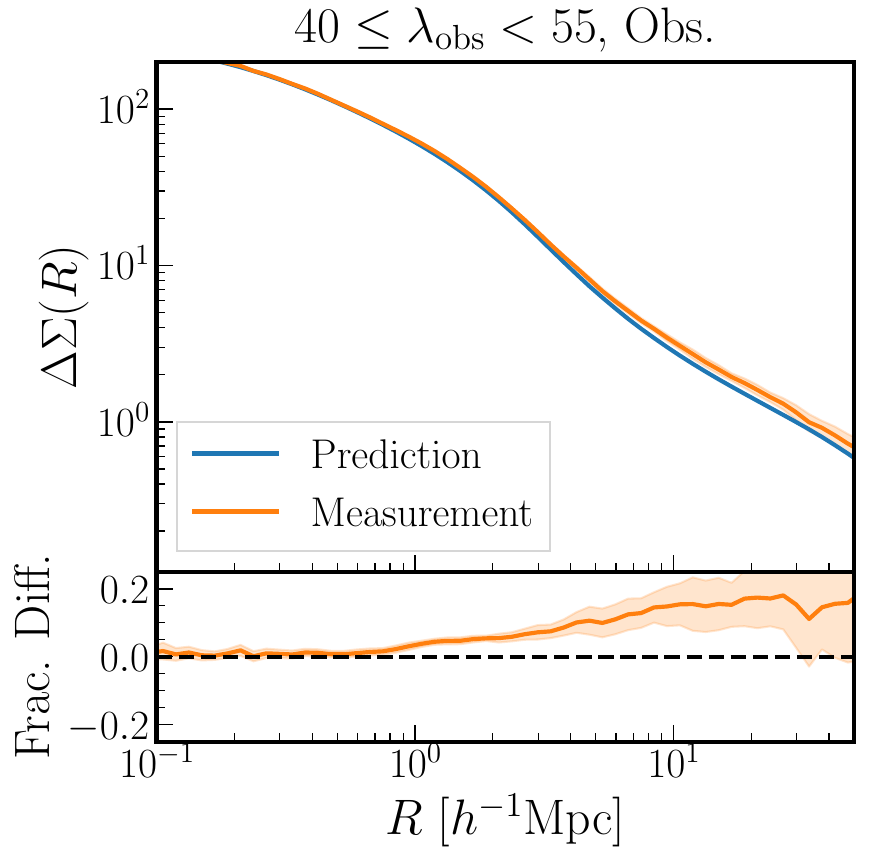}
    \includegraphics[width=0.24\textwidth]{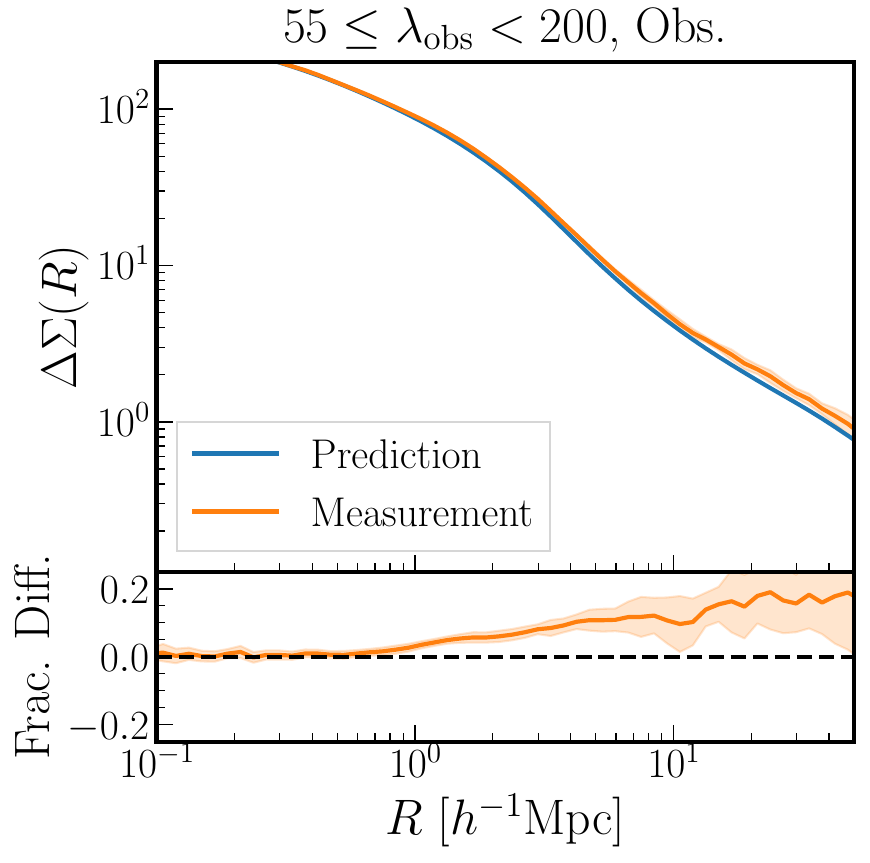}
    
    \vspace{0.5cm}

    \includegraphics[width=0.24\textwidth]{meas_pred_pure_20to30.pdf}
    \includegraphics[width=0.24\textwidth]{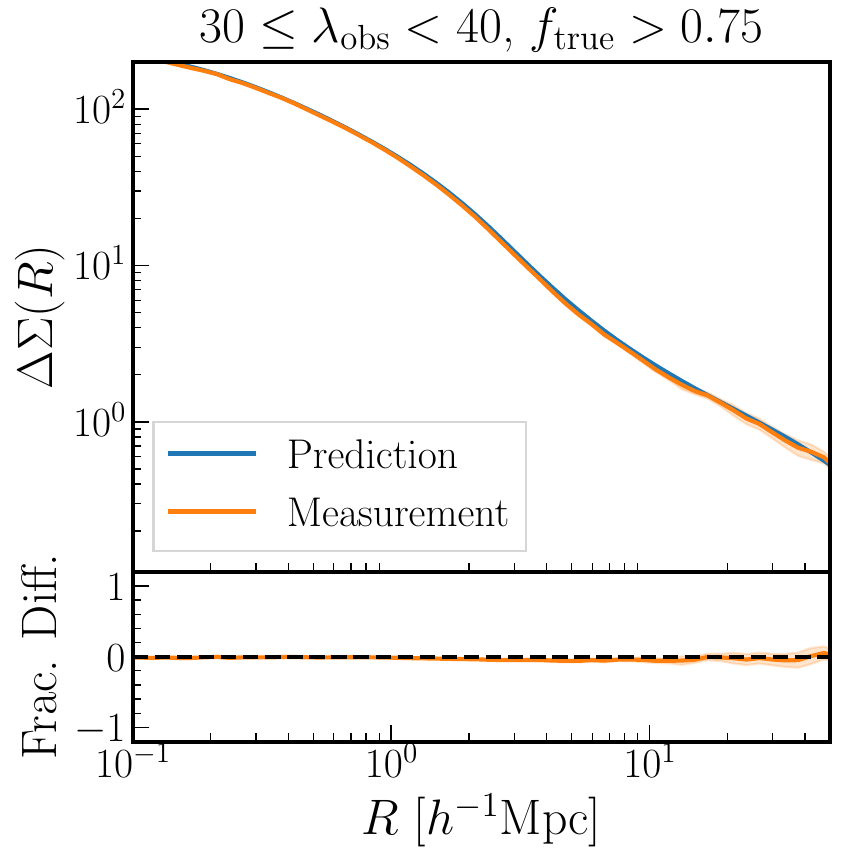}
    \includegraphics[width=0.24\textwidth]{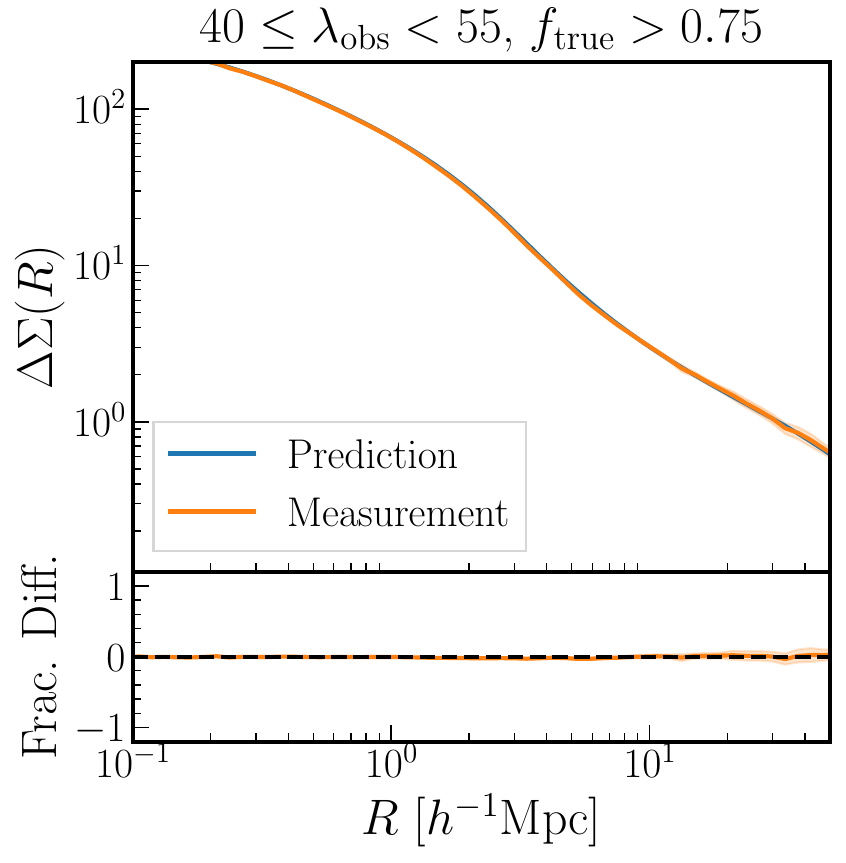}
    \includegraphics[width=0.24\textwidth]{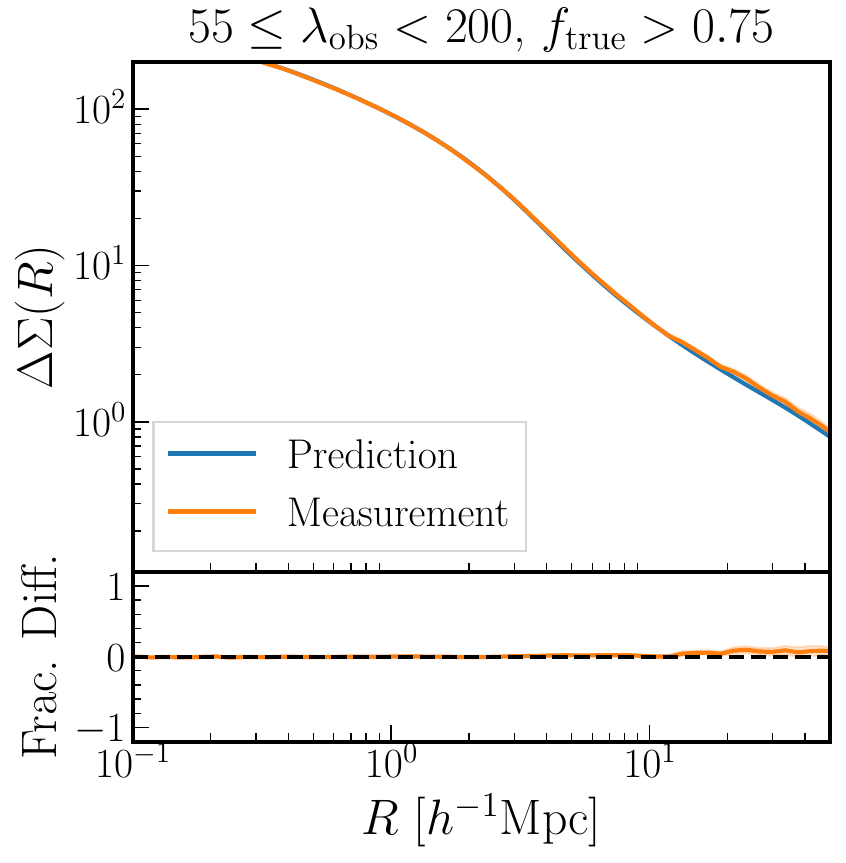}
    
    \vspace{0.5cm}
    
    \includegraphics[width=0.24\textwidth]{meas_pred_proj_20to30.pdf}
    \includegraphics[width=0.24\textwidth]{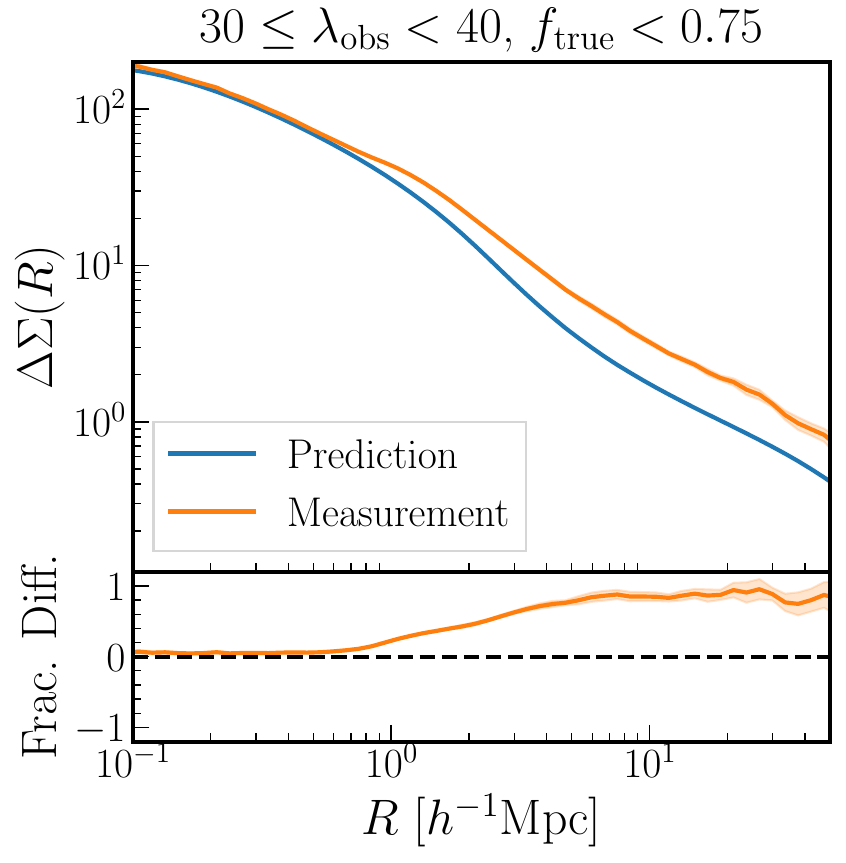}
    \includegraphics[width=0.24\textwidth]{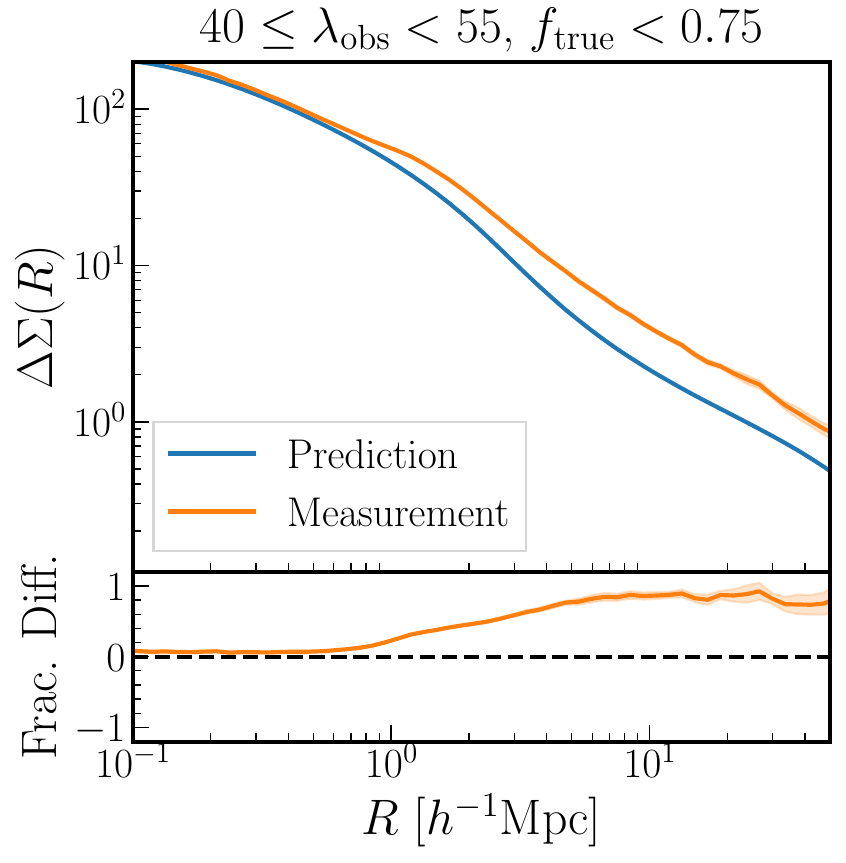}
    \includegraphics[width=0.24\textwidth]{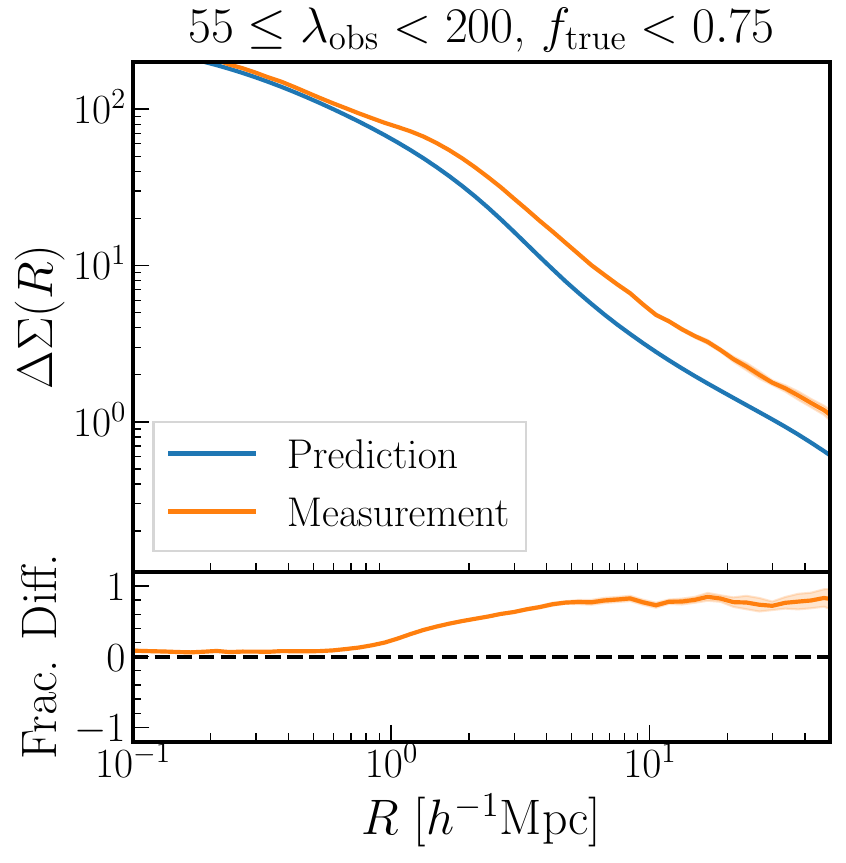}
    \caption{\label{fig:all_rich1} Similar plots to Figs.~\ref{fig:lensing_pred1} and \ref{fig:ftrue_lensing}, but for all richness bins. 
        }
\end{figure*}

\begin{figure*}
    \centering
    \includegraphics[width=0.24\textwidth]{wp_meas_pred_true20to30.pdf}
    \includegraphics[width=0.24\textwidth]{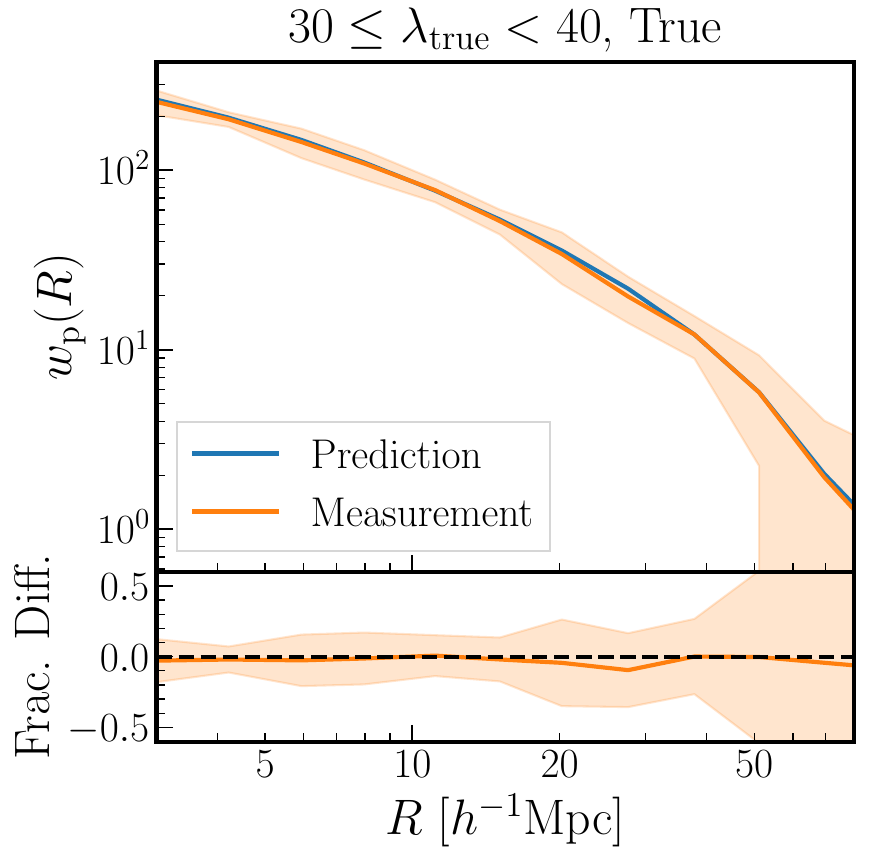}
    \includegraphics[width=0.24\textwidth]{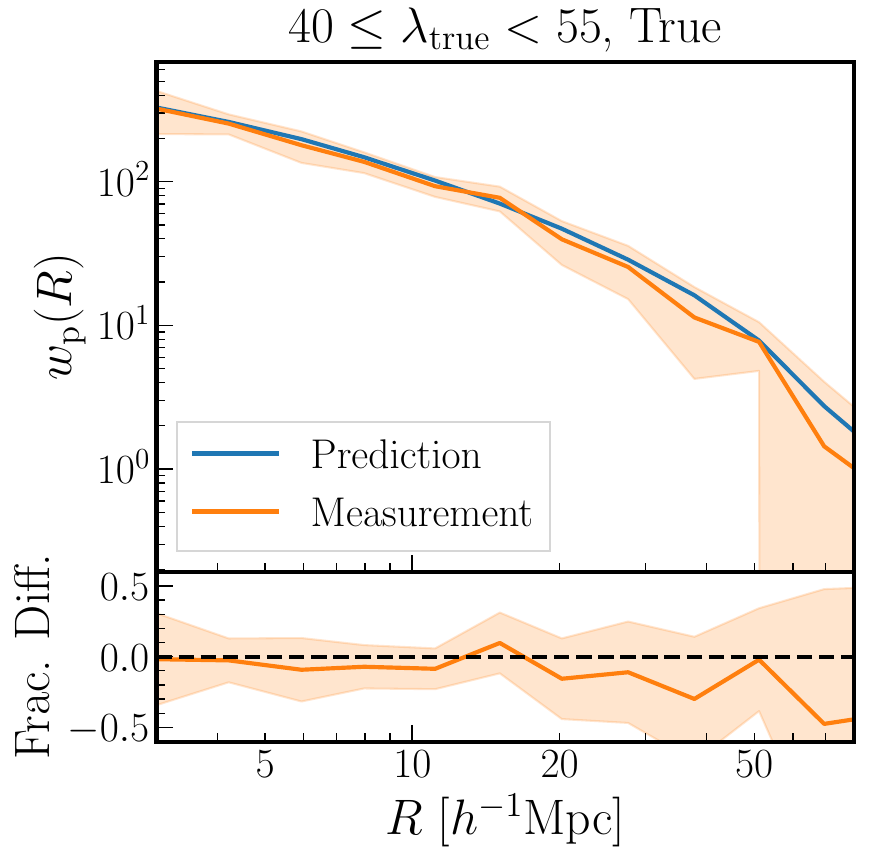}
    \includegraphics[width=0.24\textwidth]{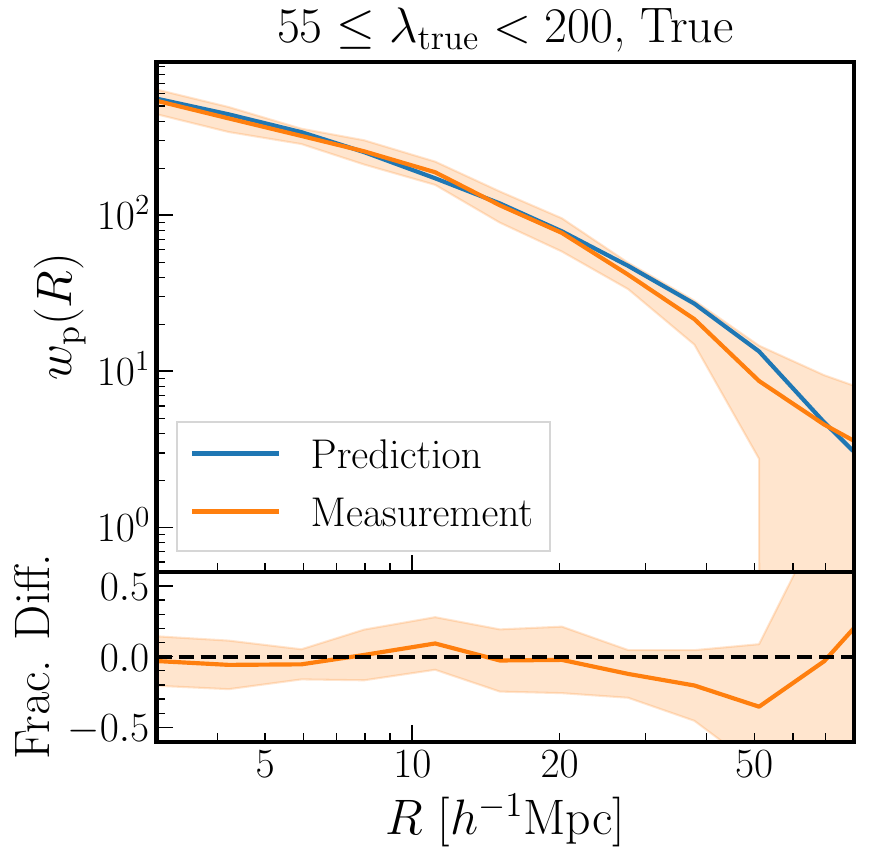}
    
    \vspace{0.5cm}
    
    \includegraphics[width=0.24\textwidth]{wp_meas_pred_obs_20to30.pdf}
    \includegraphics[width=0.24\textwidth]{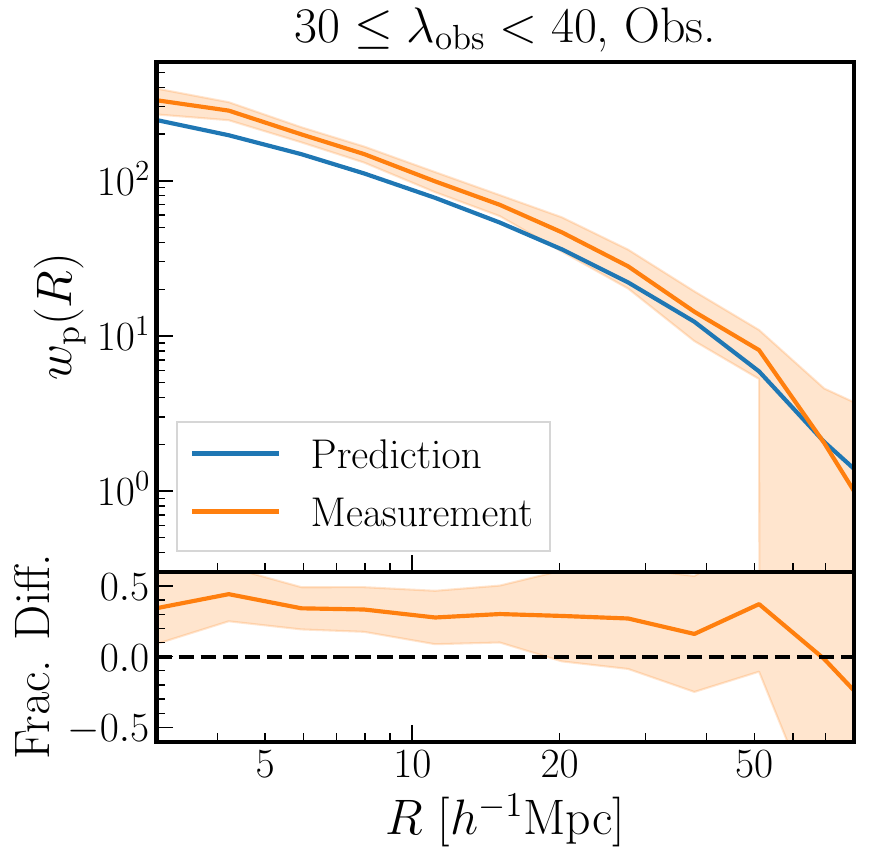}
    \includegraphics[width=0.24\textwidth]{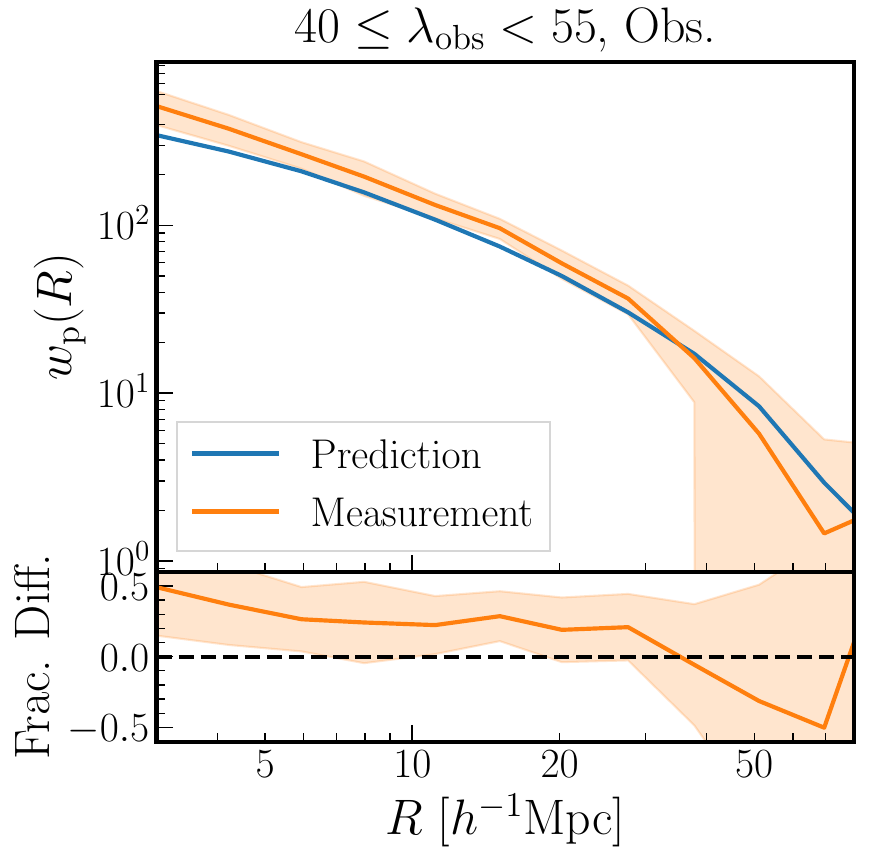}
    \includegraphics[width=0.24\textwidth]{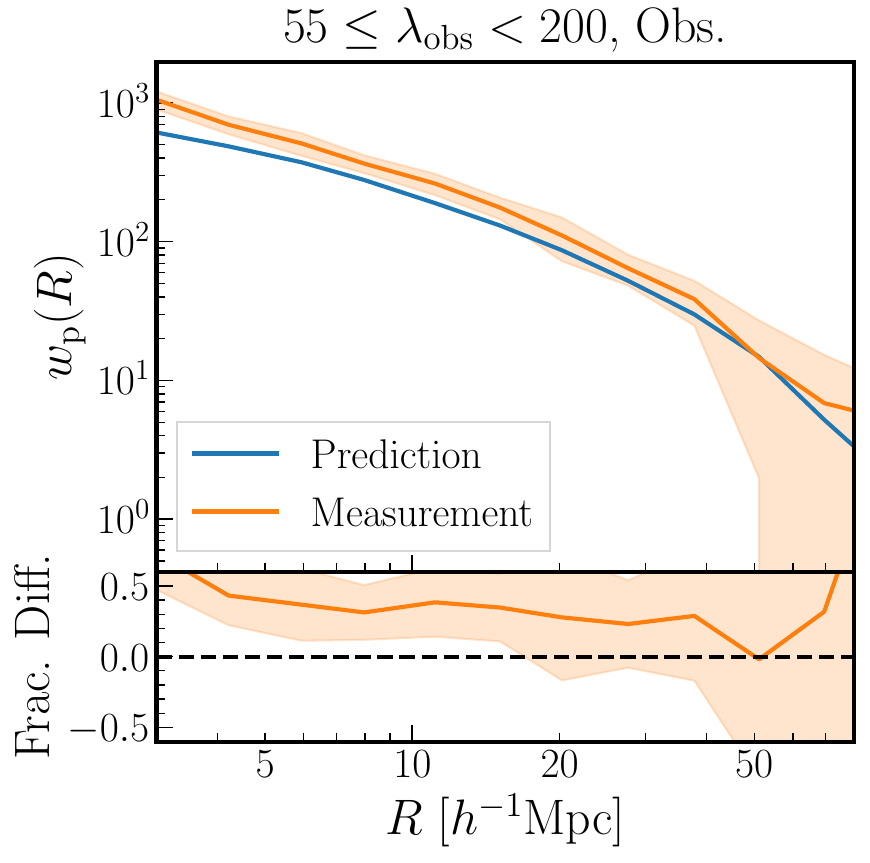}
    
    \vspace{0.5cm}

    \includegraphics[width=0.24\textwidth]{wp_meas_pred_pure_20to30.pdf}
    \includegraphics[width=0.24\textwidth]{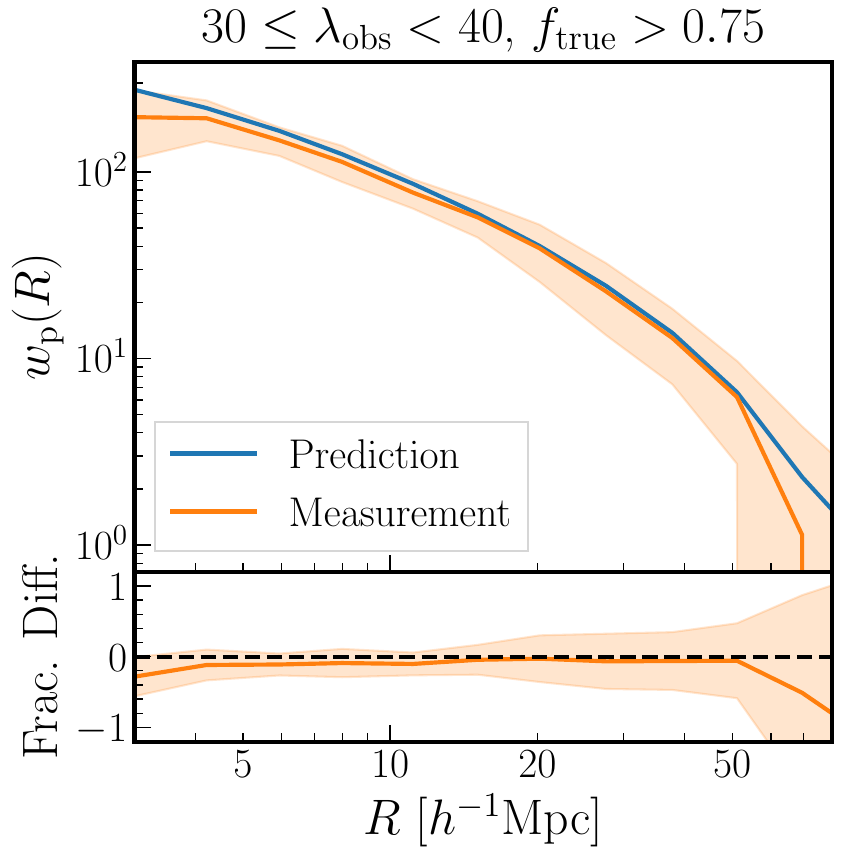}
    \includegraphics[width=0.24\textwidth]{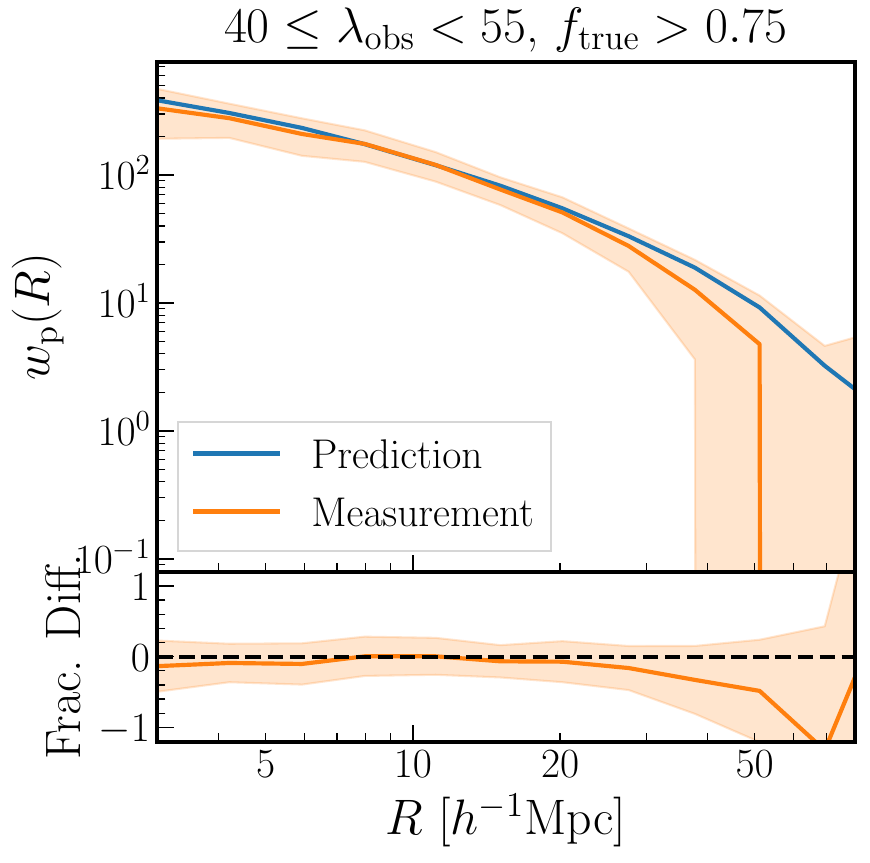}
    \includegraphics[width=0.24\textwidth]{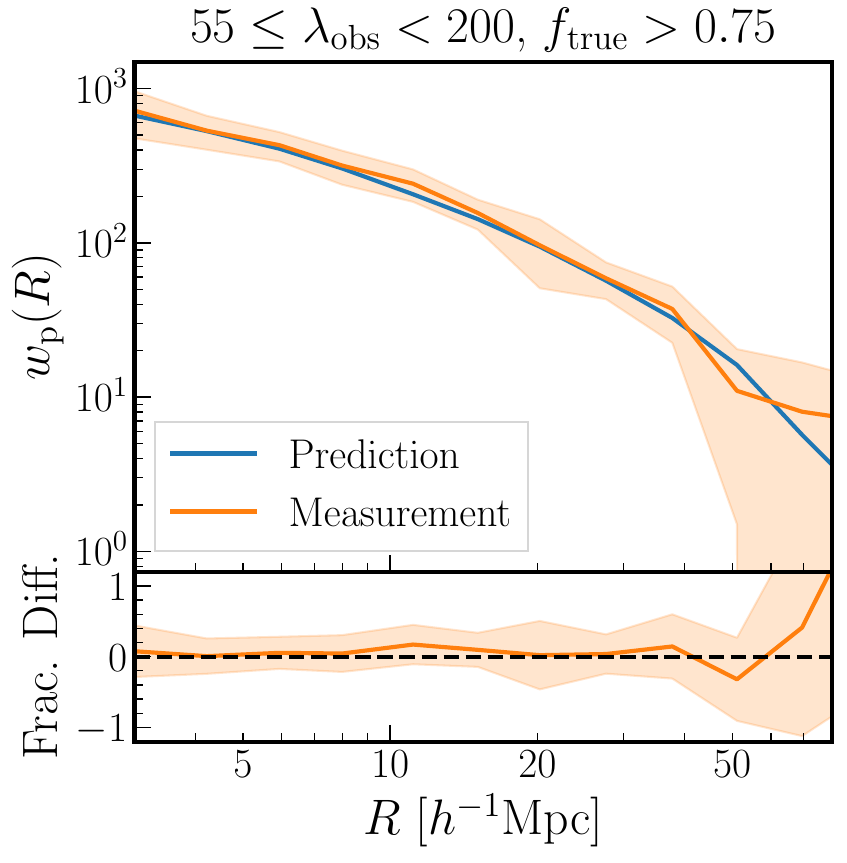}
    
    \vspace{0.5cm}
    
    \includegraphics[width=0.24\textwidth]{wp_meas_pred_proj_20to30.pdf}
    \includegraphics[width=0.24\textwidth]{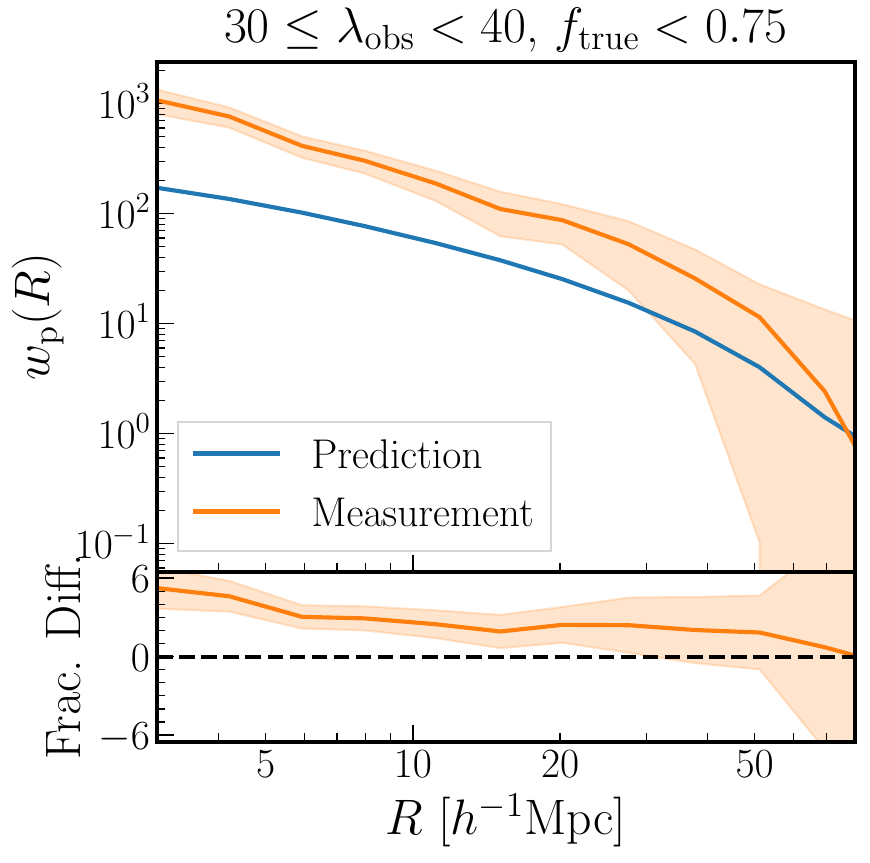}
    \includegraphics[width=0.24\textwidth]{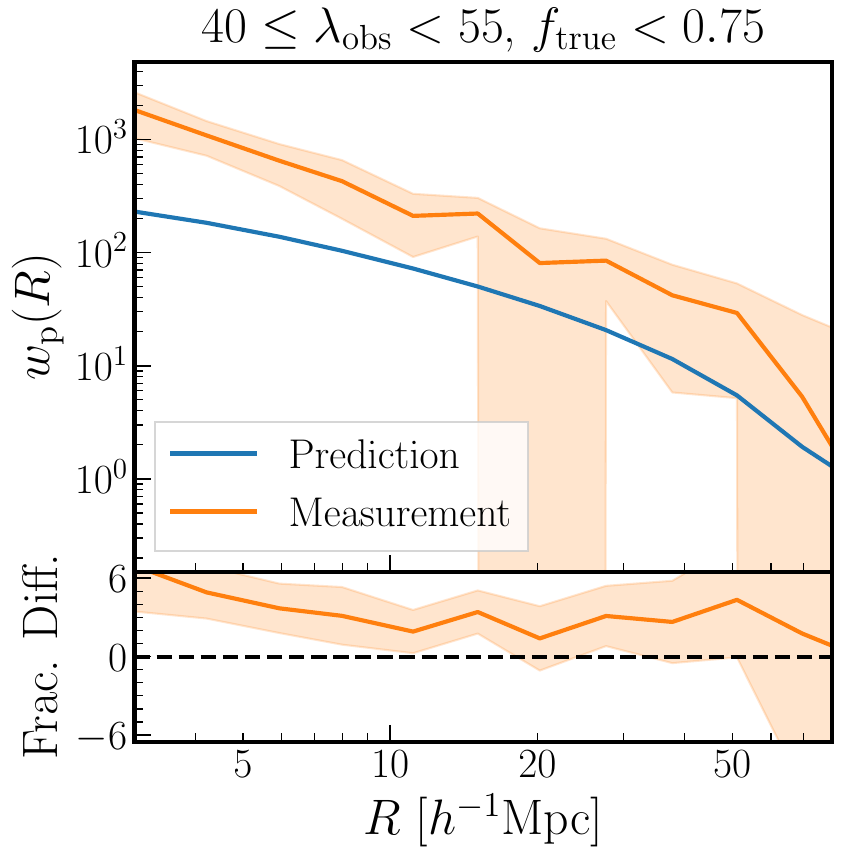}
    \includegraphics[width=0.24\textwidth]{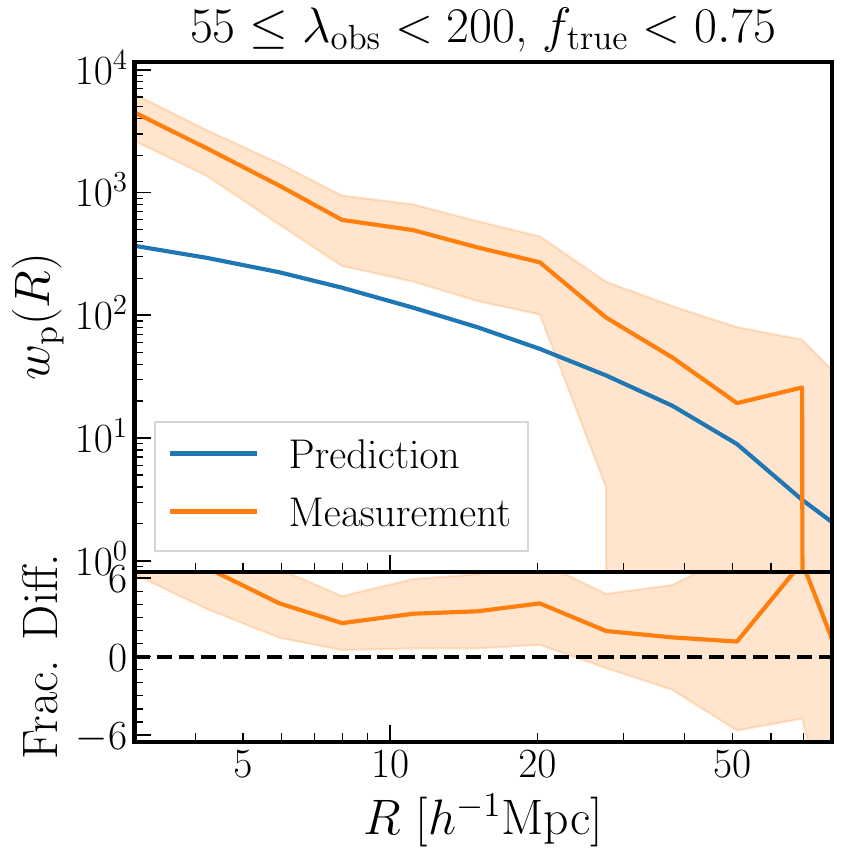}
    \caption{\label{fig:all_rich_auto} Similar plots to Figs.~\ref{fig:auto_pred1} and \ref{fig:ftrue_auto}, but for all richness bins. 
        }
\end{figure*}

For completeness of our discussions, we show in Figs.~\ref{fig:all_rich1} and \ref{fig:all_rich_auto} the impact projection effects on the lensing profiles and the cluster clustering for all richness bins, similar to Figs.~\ref{fig:lensing_pred1}, 
\ref{fig:auto_pred1}, \ref{fig:ftrue_lensing} and \ref{fig:ftrue_auto}. The fractional differences for the projected samples look remarkably similar to each other for all richness bins. 

\bibliographystyle{mnras}
\bibliography{refs} 

\bsp	
\label{lastpage}
\end{document}